\documentclass[aps,floatfix,prx,superscriptaddress,reprint,showpacs,10pt,preprintnumbers,longbibliography]{revtex4-2}

\usepackage[utf8]{inputenc}
\usepackage[pdftex]{graphicx}
\usepackage{float}
\usepackage{amssymb}
\usepackage{amsmath}  
\usepackage{amsfonts}
\usepackage{amsthm}
\usepackage{dsfont}
\usepackage{array}
\usepackage{bm}
\usepackage{mathrsfs}
\usepackage{pifont}
\usepackage{multirow}
\usepackage{upgreek}
\usepackage[dvipsnames]{xcolor}
\usepackage[pdftex,
            pdftitle={Materials Science Working Group - White Paper},
            pdfauthor={Authors},
            bookmarks,
            colorlinks,
            linkcolor=black,
            citecolor=mymagenta,
            menucolor=black,
            urlcolor=myblue,
            plainpages=false,
            pdfpagelabels,
            hypertexnames=false]{hyperref}
\usepackage{verbatim}
\usepackage{slashed}
\usepackage{bm}
\usepackage{braket}
\usepackage{booktabs}
\usepackage{multirow}
\usepackage{physics}
\usepackage{bm}
\usepackage{bbm}
\usepackage{tabularx}
\usepackage{makecell}
\usepackage{longtable}
\usepackage{textcomp}
\newcommand{\ce}[1]{$\mathrm{#1}$}
\usepackage[acronym,symbols,nogroupskip,nomain,nonumberlist,nopostdot,toc]{glossaries}
\definecolor{mymagenta}{RGB}{200, 0, 100}
\definecolor{myblue}{RGB}{45, 48, 146}

\graphicspath{{figures/}}

\makeglossaries
\newacronym{qcsc}{QCSC}{Quantum-centric Supercomputing}
\newacronym{qpu}{QPU}{quantum processing unit}
\newacronym{ucc}{UCC}{Unitary Coupled Cluster}
\newacronym{casscf}{CASSCF}{complete active space self-consistent field}
\newacronym{dmrg}{DMRG}{density matrix renormalization group}
\newacronym{mrci}{MRCI}{multireference configuration interaction}
\newacronym{dmet}{DMET}{density matrix embedding theory}
\newacronym{pde}{PDE}{partial differential equation}
\newacronym{hpc}{HPC}{high performance computing}
\newacronym{qma}{QMA}{Quantum Merlin Arthur}
\newacronym{qpe}{QPE}{Quantum Phase Estimation}
\newacronym{fem}{FEM}{finite element method}
\newacronym{qram}{QRAM}{quantum random access memory}
\newacronym{car}{CAR}{canonical anticommutation relations}
\newacronym{ccr}{CCR}{canonical commutation relations}
\newacronym{jw}{JW}{Jordan-Wigner}
\newacronym{bk}{BK}{Bravyi-Kitaev}
\newacronym{lcu}{LCU}{linear combination of unitaries}
\newacronym{ic}{IC}{informationally incomplete}
\newacronym{lasscf}{LASSCF}{localized active space self-consistent field}
\newacronym{vqe}{VQE}{Variational Quantum Eigensolver}
\newacronym{cnot}{CNOT}{Controlled NOT gate}
\newacronym{mps}{MPS}{Matrix Product States}
\newacronym{peps}{PEPS}{Projected Entangled Pair}
\newacronym{ttn}{TTN}{Tree Tensor Networks}
\newacronym{mera}{MERA}{Multi-scale Entanglement Renormalization Ansatz}
\newacronym{zne}{ZNE}{Zero-Noise Extrapolation}
\newacronym{pec}{PEC}{probabilistic error cancellation}
\newacronym{rdm}{RDM}{reduced density matrix}
\newacronym{fci}{FCI}{Full Configuration Interaction}
\newacronym{hhl}{HHL}{Harrow-Hassidim-Lloyd}
\newacronym{mcpdft}{MC-PDFT}{multiconfiguration pair-density functional theory}
\newacronym{gpu}{GPU}{graphics processing unit}
\newacronym{fpga}{FPGA}{field programmable gate array}
\newacronym{asic}{ASIC}{application-specific integrated circuit}
\newacronym{vqa}{VQA}{variational quantum algorithm}
\newacronym{varQITE}{varQITE}{variational Quantum Imaginary Time Evolution}
\newacronym{qcnn}{QCNN}{Quantum Convolutional Neural Network}
\newacronym{qcbm}{QCBM}{Quantum Circuit Born Machine}
\newacronym{povm}{POVM}{Positive Operator Valued Measure}
\newacronym{msw}{MSW}{Mikheyev-Smirnov-Wolfenstein}
\newacronym{cms}{CMS}{Compact Muon Solenoid}
\newacronym{qbm}{QBM}{Quantum Boltzmann Machine}
\newacronym{qrom}{QROM}{Quantum Read-Only Memory}
\newacronym{qdc}{QDC}{Quantum Data Center}
\newacronym{qec}{QEC}{quantum error correction}
\newacronym{qem}{QEM}{quantum error mitigation}
\newacronym{wms}{WMS}{workload management systems}
\newacronym{dua}{DUA}{Data Use Agreements}
\newacronym{isa}{ISA}{Instruction Set Architecture}
\newacronym{mpi}{MPI}{Message Passing Interface}
\newacronym{openmp}{OpenMP}{Open Multi-Processing}
\newacronym{sisd}{SISD}{single instruction, single data}
\newacronym{simd}{SIMD}{single instruction, multiple data}
\newacronym{simt}{SIMT}{single instruction, multiple threads}
\newacronym{mimd}{MIMD}{multiple instruction, multiple data}
\newacronym{sdk}{SDK}{software development kits}
\newacronym{dag}{DAG}{directed acyclic graph}
\newacronym{wmfs}{WFMS}{workflow management system}
\newacronym{cvtn}{CVTN}{Continuous Variable Tensor Network}
\newacronym{dft}{DFT}{density functional theory}
\newacronym{dmft}{DMFT}{dynamical mean-field theory embedding}
\newacronym{qdet}{QDET}{quantum defect embedding theory}
\newacronym{pes}{PES}{potential energy surface}

\begin{document}

\newpage

\title{Quantum-centric Supercomputing for Materials Science:\\ \vspace{0.2cm}
A Perspective on Challenges and Future Directions \vspace{1.2cm}}

\author{Yuri Alexeev}
\affiliation{Argonne National Laboratory, Lemont, IL 60439, USA}

\author{Maximilian Amsler}
\affiliation{Corporate Sector Research and Advance Engineering, Robert Bosch GmbH, Robert-Bosch-Campus 1, D-71272 Renningen, Germany}


\author{Marco Antonio Barroca}
\affiliation{IBM Research, Rio de Janeiro, 20031-170, RJ, Brazil}
\affiliation{Centro Brasileiro de Pesquisas F\'isicas, Rio de Janeiro, 22290-180, RJ, Brazil}

\author{Sanzio Bassini}
\affiliation{CINECA, via Magnanelli 6/3, 40033 Casalecchio di Reno, BO, Italy}

\author{Torey Battelle}
\affiliation{Arizona State University, Tempe, AZ, USA}

\author{Daan Camps}
\affiliation{National Energy Research Scientific Computing Center, Lawrence Berkeley National Laboratory Berkeley, CA, USA}

\author{David Casanova}
\affiliation{Donostia International Physics Center (DIPC), 20018 Donostia-San Sebasti\'an, Euskadi, Spain}
\affiliation{Ikerbasque, Basque Foundation for Science, 48009 Bilbao, Spain}

\author{Young Jai Choi}
\affiliation{Department of Physics, Yonsei University, Seoul 03722, Korea}

\author{Frederic T. Chong}
\affiliation{University of Chicago, Chicago, Illinois, USA}

\author{Charles Chung}
\affiliation{IBM Quantum, IBM T.J. Watson Research Center, Yorktown Heights, NY 10598, USA}

% ...

\date{\today}

\begin{abstract}
Computational models are an essential tool for the design, characterization, and discovery of novel materials. Computationally hard tasks in materials science stretch the limits of existing high-performance supercomputing centers, consuming much of their resources for simulation, analysis, and data processing. Quantum computing, on the other hand, is an emerging technology with the potential to accelerate many of the computational tasks needed for materials science. In order to do that, the quantum technology must interact with conventional high-performance computing in several ways: approximate results validation, identification of hard problems, and synergies in quantum-centric supercomputing. 
In this paper, we provide a perspective on how quantum-centric supercomputing can help address critical computational problems in materials science, the challenges to face in order to solve representative use cases, and new suggested directions.
\end{abstract}

\maketitle

\newcommand{\cmt}[1]{\textcolor{red}{#1}}

\newpage
\section{Introduction}
\label{sec:intro}
Materials science use cases are being investigated as some of the first applications of quantum computing to show practical quantum advantages. The strongest motivation for this belief is that a great deal of materials science problems are quantum in nature~\cite{feynman1982simulating}. 
A computational advancement in the field will have consequences in many areas, from scientific exploration to industrial advancements and sustainability.

Currently, high-performance supercomputing centers dedicate a substantial amount of resources to computations in the materials science space~\cite{ALCF_MaterialsScience,austin2020nersc,Connor2023instcomp,DIPC_HPC}. A fundamental issue is that ab-initio computations of materials, in the full configuration interation (CI) limit, suffer from an exponential or factorial growth of the computational resources in the system size~\cite{ciapprox-book,ciapprox-1,doi:10.1021/acs.jctc.7b00725,ciapprox-2,ciapprox-3,ciapprox-4}. Therefore, memory and processing times requirements rapidly become intractable, as the size of the quantum systems increase.
To tame this dire scaling, various approximations are employed to reduce classical computational costs, however, even these either fail -- for highly correlated materials -- or become intractable for sufficiently large and general systems~\cite{solomonik2014massively,Kim_2018,supermatchem-1,supermatchem-2,supermatchem-3,supermatchem-4,supermatchem-5,supermatchem-6,supermatchem-7,supermatchem-8,supermatchem-9,10.1063/1.3659143}.

Quantum computers present an attractive alternative, since many quantum algorithms avoid incurring in the exponential memory overheads of classical computations of quantum matter. For practical use cases, quantum computers may be expected to operate embedded in classical \gls{hpc} environments, so-called \emph{\gls{qcsc}}. Intensive classical processing is required before, after, and concurrently with these quantum computations, for a number of reasons: alleviating the workload of quantum computers, integration into existing classical HPC algorithms, signal extraction from noisy quantum devices, and exploiting fault-tolerant codes to the best extent.

 In this paper, we motivate the need for quantum-centric supercomputing in materials science research and industrial development. We also identify the challenges ahead to achieve practical quantum advantage for materials science use cases and propose directions to address them. It is important to note that this work is intended as a perspective and not as an exhaustive review of the existing literature. More specifically, our discussion will be centered around addressing several particular aspects:

\emph{Key Algorithms.} We identify quantum algorithms for materials science applications, in terms of practical applications and potential for quantum advantage. 
    
\emph{Implications on the design of quantum-centric supercomputing architectures.} We discuss the requirements that arise for quantum-centric supercomputing architectures. This includes evaluating the computational and operational demands, scalability, and integration hurdles when deploying these algorithms in coupled quantum-classical HPC environments.
    
\emph{Materials Science Use Cases.} We highlight specific use cases in materials science where quantum and HPC algorithms can be most effectively utilized. 
We select the algorithms and use cases according to three principles, which are necessary for quantum advantage:
\begin{enumerate}
    \item the use case must be classically hard in some limit;
    \item the use case and algorithm considered must be amenable to execution on a noisy or fault-tolerant quantum computer, depending on the scenario;
    \item the use case represents an interesting problem in materials science.
\end{enumerate}

Our discussion is organized as follows.
Section~\ref{high_level_algo_summary} highlights some key  existing algorithms relevant to materials science.
Section~\ref{sec:DataPreprocessing} exposes classical computations in quantum-HPC workflows, keeping key algorithms and use cases as a reference.
In Sec.tion~\ref{sec:Queuing}, we detail of the challenges for hybrid quantum-classical workload management and quantum-HPC integration, and in Sec.~\ref{sec:ProgrammingModels} give a perspective on the main components of a programming model for such systems.
Section~\ref{sec:ClassicalSimulationOfQSystems} summarizes the state-of-the-art in classical simulation of quantum circuits, which is a necessary crucial consideration in evaluating the potential for quantum advantage.
Finally, Sec.~\ref{sec:UseCaseDiscovery} gives an overview of potential use cases in materials science, bringing together considerations and threads from previous sections.

\section{Fundamental quantum algorithms}
\label{high_level_algo_summary}

In this section we summarize quantum algorithms that can be used for materials science applications. We consider algorithms that simulate quantum systems and mention quantum computing approaches to simulate \gls{pde}. Although the simulation of quantum systems is the most natural application of quantum computers in the materials science space, we mention  quantum algorithms for PDEs since they can be useful for some well-crafted use cases. Within the simulation category, we discuss the time dynamics, ground states, open quantum systems and finite-temperature properties. 

Simulation of dynamics generated by a Hamiltonian of interest is one of the most natural applications for quantum computers.
Quantum algorithms for approximating time evolution have been studied extensively, and so here briefly describe a few key methods.
In recent years the development of \emph{quantum signal processing} culminated in an algorithm called \emph{qubitization}~\cite{low2019hamiltonian}, the complexity of which is measured by the number of queries to an oracle that encodes the Hamiltonian.
For a target evolution time $t$ and an error tolerance $\epsilon$, qubitization makes only $O(t + \log 1/\epsilon)$ queries to the oracle, which is asymptotically optimal. However, the construction of the Hamiltonian oracle generally involves ancillary qubits, making qubitization difficult to implement in near-term and early fault-tolerant devices. 

In this regime, \emph{Trotterization}~\cite{lloyd1996universal,childs2019theory} (see also Sec.~\ref{trotter_sec}), the original quantum algorithm for simulating dynamics, remains one of the leading candidates for early applications, because its low overhead permits fine-grained tradeoffs between circuit depth and accuracy.
Asymptotically, the gate complexity of the $p$-th order Trotterization scales as $O(t^{1+1/p}/\epsilon^{1/p})$, which can be made close to that of qubitization at large $p$.
Additionally, the error of Trotterization depends on the commutativity of the terms in the Hamiltonian, allowing the algorithm to take advantage of the structure of the Hamiltonian.
The final example we mention is the \emph{HHKL} algorithm~\cite{hhkl2018}, which achieves nearly optimal scalings for Hamiltonians that are geometrically local (or even with power-law decaying interactions~\cite{tran2019locality}) on some lattice.
HHKL can be considered a hybrid between Trotterization and the high-accuracy methods that exploits commutation between spatially-separated Hamiltonian terms.

Perhaps the most paradigmatic type of quantum simulation for materials is the approximation of ground states and low-energy states in general.
While such problems are \gls{qma}-hard~\cite{kitaev02,kempekitaevregev04} and thus believed to be intractable in worst cases, their broad utility across physics, chemistry, and materials science has made them the objects of intense study in the hope that for physically interesting cases, they may be manageable.
The most heavily studied algorithm for simulating ground states in the near term is the \gls{vqe}, e.g.,~\cite{du2010nmr,lanyon2010towards,peruzzo2014variational,wang2015quantum,omalley2016scalable,shen2017quantum,paesani2017bayesian,kandala2017hardware,hempel2018trappedion,santagati2018witnessing,dumitrescu2018atomicnucleus,kokail2019selfverifying,kandala2019errormitigation,ganzhorn2019gate,sagastizabal2019experimental,mccaskey2019quantum,smart2019quantum,nam2020trappedion,liu2021representation,wang2021resource,liu2023analytic,zheng2023speeding,wang2023ever}), which is based on optimizing a parameterized wavefunction prepared on the quantum computer, and is well-adapted to near term due to the possibility of implementing it with low-depth circuits. Similar considerations apply to the use of finite-depth quantum approximate optimization algorithms for finding the ground state of frustrated materials \cite{doi:10.1073/pnas.2006373117,lotshaw2023simulations}.
However, VQE has a number of drawbacks, mostly related to the difficulties of establishing convergence of the optimization and of accurately measuring the target energy to be estimated, which can result in long runtimes or getting stuck in local minima. 

If we consider fault-tolerant architectures, \gls{qpe}~\cite{kitaev1995phaseestimation} and related algorithms (e.g.,~\cite{lin2020nearoptimalground,lin2022heisenberglimited,dong2022groundstate}) promise to offer ground state simulation with finite success probability, given access to initial reference states with sufficiently high overlaps with the true ground state.
A good initial state assumption can be traded off by assumptions on the mixing time of a dissipative Lindbladian associated with the problem~\cite{dingchenlin23}.

In the near term, quantum Krylov algorithms~\cite{mcclean2017subspace,colless2018computation,parrish2019filterdiagonalization,motta2020qite_qlanczos,takeshita2020subspace,huggins2020nonorthogonal,stair2020krylov,urbanek2020chemistry,cohn2021filterdiagonalization,yoshioka2021virtualsubspace,seki2021powermethod,cortes2022krylov,klymko2022realtime,baek2023say,tkachenko2022davidson,lee2023sampling,zhang2023measurementefficient,kirby2023exactefficient,shen2023realtimekrylov,kirby2024analysis,motta2023subspace} and related techniques~\cite{shen2023estimating} have recently emerged as a promising family of ground state simulation methods since they fall somewhere between VQE and QPE: many variants~\cite{parrish2019filterdiagonalization,motta2020qite_qlanczos,stair2020krylov,urbanek2020chemistry,cohn2021filterdiagonalization,epperly2021subspacediagonalization,seki2021powermethod,cortes2022krylov,klymko2022realtime,tkachenko2022davidson,zhang2023measurementefficient,lee2023sampling,kirby2023exactefficient,shen2023realtimekrylov,kirby2024analysis} have provable convergence~\cite{epperly2021subspacediagonalization,kirby2024analysis,kirby2023exactefficient,zhang2023measurementefficient,motta2023subspace} subject to similar assumptions to QPE, but the circuits they involve are only moderately more complex than those in VQE.

For some Hamiltonians, in particular spin and fermionic lattice models often studied in condensed matter and materials science, some variants of quantum Krylov methods could be amenable to noisy quantum devices.
For a recent review of quantum Krylov algorithms, see~\cite{motta2023subspace}.
 
Quantum computers also enable the simulation of open quantum system dynamics. Open systems more accurately model physical phenomenon in nature and their study is particularly relevant in fields such as condensed matter, material science, and quantum chemistry \cite{olmos2012facilitated, may2023charge, nitzan2006chemical, kastoryano2023quantum,ding2023single, cubitt2023dissipative, hubisz2021quantum,schlimgen2022quantum}. The simulation of open systems requires both coherent evolution under a Hamiltonian and dissipative processes that capture the interactions with the environment. Formally, the evolution of a state under an open-system dynamics is governed by the Lindblad quantum master equation \cite{lindblad1976generators}.

Quantum algorithms for simulating Hamiltonian dynamics have been extended to simulate Lindbladian dynamics. A naive approach is to explicitly simulate the environment~\cite{wang2011quantum, kliesch2011dissipative,ding2023simulating}. Trotterization can be used to simulate the dynamics of $k$-local Lindbladians~\cite{kliesch2011dissipative} and non-Markovian systems in~\cite{sweke2016digital}. Simulation of sparse and non-local Lindbladian dynamics requires the implementation of sparse
Stinespring isometries~\cite{childs2016efficient}. When Lindbladians can be written as a linear combination of Pauli operators, one can employ a variant of linear combination of unitaries~\cite{cleve2016efficient}. Another promising approach is to implement imaginary-time evolution to approximate the dissipative dynamics~\cite{kamakari2022digital}. Other emerging quantum algorithms to simulate open system dynamics include qubitization~\cite{pocrnic2023quantum}, wave matrix Lindbladization~\cite{patel2023wave,patel2023wave2}, and partial probabilistic error cancellation~\cite{guimaraes2023noise}. 

In addition to estimating observables on ground states, expectations with respect to finite-temperature states are also critical for studying a variety of phenomena including phase transitions. In contrast to preparing ground states, which may encode a computationally intractable optimization problems \cite{kempekitaevregev04}, thermal state preparation is generally computationally easier as long as the temperature is sufficiently high.
There is a range of quantum algorithms that aim to accelerate estimation of thermal properties of quantum systems. One approach \cite{poulinwocjan09}, analogous to the ground-state projection method used by phase estimation, leverages Grover-assisted rejection sampling to either cool an infinite-temperature state to a given finite temperature \cite{chowdhurysomma16}, or to use two finite-temperature systems to simulate a single system at half the temperature \cite{cotler18}. 

A different approach for thermal state preparation involves simulating the physical process of thermalization. Early work in this direction involves a discrete-time Metropolis-Hastings algorithm over the eigenstates of the Hamiltonian that is guaranteed to converge on the thermal state \cite{temme09}. Quantum techniques for achieving quadratic speedups in classical Markov Chain Monte Carlo have been adapted to these quantum analogues \cite{yungguzik10, wocjantemme21}. More recent work involves the simulation of continuous-time quantum Markov processes known as Davies generators, a kind of open quantum system that is regarded to emulate thermalization in nature \cite{davies76}. A recent line of work aims to develop quantum algorithms that efficiently simulate these open systems despite their long-range interactions \cite{rallwangwocjan21, chenkastoryanobrandaogilyen23, chenkastoryanogilyen23}. Some of these proposals aim for minimal quantum resources \cite{dingchenlin23}, and others explore connections between thermalization and the fundamental capabilities of quantum computation \cite{chenhuangpreskillzhou23}.

In addition to the Schr\"odinger equation, there are also applications of other PDEs in materials science, which include Maxwell's equations for deriving electromagnetic properties, the heat equation, and PDEs involved in the characterization of structural properties. These frequently rely on the \gls{fem}.

There are several proposals for quantum algorithms for PDEs, usually based on the quantum linear systems algorithm \cite{harrow2009quantum}. Boundary value problems analyzed via finite element methods often result in large linear systems of equations \cite{Rao}. Their matrix elements can frequently be derived analytically which may allow sidestepping the need for a \gls{qram}. Existing literature has shown that quantum algorithms for finite element methods exhibit polynomial speedups over their classical counterparts \cite{montanaropallister16}. These speedups have been studied for applications in solving the heat equation \cite{lindenmontanaroshao20} and electromagnetic structure problems \cite{zhangfengzhang21}.

There is also a growing body of literature on quantum algorithms for initial value problems of both linear \cite{berrychildsostranderwang17, fanglintong23} and nonlinear PDEs \cite{krovi22}. These methods highlight that a variety of dynamics can be encoded into the Schr\"odinger equation \cite{jinniuliu22}, subject to certain restrictions pertaining to normalization \cite{anliuwangzhao22} and chaos \cite{lewiseidenbenznadigasubasi23}.

\section{Classical information processing in quantum-classical workflows\label{sec:DataPreprocessing}}

\subsection{Mapping to qubits}
\label{mapping_to_qubits}

Simulation of quantum systems is a promising use case for both noisy and fault-tolerant quantum computers.
In order to simulate a physical quantum system, the governing physical Hamiltonian, for example, that of a fermionic or bosonic system, must be appropriately mapped to a qubit Hamiltonian such that the quantum computer faithfully simulates the physical system of interest.
Note an appropriate mapping preserves the operator algebra: the physical operators that comprise the physical Hamiltonian and the qubit operators that comprise the qubit Hamiltonian must obey the same algebra.
Fermions and bosons follow their respective algebras, commonly known as canonical \gls{car} and \gls{ccr}.
Qubits are the so-called hard-core bosons~\cite{wu2002qubits}. 

Plentiful choices of the aforementioned kind of mappings exist as briefly described below.  
Which mapping to use may thus be determined by a practical consideration, such as the computational cost of the resulting simulation.
A good choice of mapping will likely depend on the specifics of a given physical Hamiltonian.
In this subsection, we highlight some of the existing techniques and challenges around this problem.

Bosons have in common with qubits their commutation relations (bosonic wavefunctions are symmetric), but while a qubit has only two states, a bosonic mode has infinitely many.
Any digital simulation of bosonic modes requires cutting off the dimension at some finite value, typically much larger than two.
Once a cutoff is applied, however, the choice of mapping is essentially a decision about how to encode a nonnegative integer representing the occupation of the mode into qubits.
One could for example use unary or binary representations, with the former yielding local representations of creation and annihilation operators with linear qubit cost in the occupation cutoff, and the latter yielding log-local representations of creation and annihilation operators with only logarithmic qubit cost in the occupation cutoff.
This is a significant trade-off, but one that is at least well-understood~--- see for example~\cite{somma2005quantum,macridin2018electronphonon,mcardle2019digital,sawaya2019quantum,kan2021lattice,kreshchuk2022quantum,kan2022simulating}.

Fermionic wavefunctions, on the other hand, are antisymmetric, which means that in second-quantization their creation and annihilation operators anticommute on different modes.
Since spatially disjoint qubit operators commute, mapping fermions to qubits requires care.
Also known as bosonization in the physics community (in the sense of qubits sharing the commutation relations of bosons), a collection of qubit operators can be used to embed a single fermionic creation or annihilation operator to preserve the fermionic algebra between the embedded fermionic operators.
The first such mapping is the \gls{jw} transformation~\cite{jordanwigner1928}, and more recently this area has seen extensive research~\cite{bravyi2002fermionic,verstraete2005mapping,seeley2012bravyikitaev,bravyi2017tapering,setia2018bksf,steudtner2018fermions,steudtner2019fermions,wang2021resource,derby2021compact,derby2021alternative,kirby2022fermiontoqubit,chien2022optimizing,wang2023ever}, with each of these references discussing new fermion-to-qubit mappings.

Some of alternative mappings may be related to the JW transformation by a unitary transformation since conjugation by a unitary preserves the entire algebra. Others, especially those designed for particular fermionic lattices, may preserve the algebra between the relevant operators only, implied by the given lattice.
A subset of unitaries, such as Cliffords or linear-reversible circuits, can be used to describe other well-known mappings including \gls{bk}~\cite{bravyi2002fermionic} or parity.

The art of finding and choosing a good mapping is centered around the computational resource savings.
This is because different mappings can result in vastly different quantum circuits; e.g., with respect to the number of qubits or the number of quantum gates, when combined with both problem-specific and problem-agnostic circuit optimization techniques.
In the near term, circuits will necessarily have limited width and depth due to noise, so providing a mapping that is efficient in both dimensions is crucial for reliable execution.
Indeed, while the parity transform has been used to reduce the qubit counts~\cite{bravyi2017tapering} and the BK transform has been shown to admit the asymptotically optimal operator locality for generic fermionic Hamiltonians~\cite{bravyi2002fermionic}, customized mappings that take advantage of the input problem specifics while working in concert with circuit compilation tools, e.g., ~\cite{nam2018automated,nam2020trappedion}, have been reported to result in substantially smaller circuits~\cite{wang2021resource,wang2023ever}.

Other mappings that aim to exploit limited connectivity in fermionic lattices to localize qubit operators over the limited qubit connectivity offered by some quantum computers have also been explored~\cite{wang2021resource,derby2021compact,derby2021alternative,miller2023bonsai,chien2022optimizing,chen2023equivalence}.
With very few exceptions, e.g.,~\cite{kirby2022fermiontoqubit}, most fermion-to-qubit mappings constitute mappings from products of Majorana operators to Pauli operators.
Many interesting and useful configurations have been deployed within this framework.
The space of mappings of this type is well defined~\cite{chien2022optimizing}, although not necessarily easy to search over.
In this sense future work in constructing new useful mappings of this type constitutes a collective exercise in combinatorial optimization, which remains a useful and interesting activity, especially in the near-term regime where these details of efficiency can make or break an algorithm.
The broader challenge for the field is finding ways to push beyond this paradigm, to either find new kinds of useful mappings that are not somehow equivalent to mapping Majoranas to Paulis, or to otherwise leverage our understanding of the landscape of mappings to some novel useful effect.

\subsection{Hamiltonian encodings for dynamics simulation}

The intractability of simulating dynamics of generic quantum systems using known classical methods has driven the development of quantum computation since its inception~\cite{feynman2018simulating}, and it is the branch of quantum simulation that is most strongly believed to offer a quantum advantage~\cite{lloyd1996universal}.
Hamiltonian dynamics, or the simulation of time evolution operator, $e^{-iHt}$, for a given Hamiltonian, $H$, is a core component in many quantum algorithms. Generally, the Hamiltonian is represented as a sum, $H=\sum_i H_i$, of operators which are diagonalizable using simple unitary transformations, $H_i = U_i D_i U_i^\dagger$; e.g., exact squares of one-electron fermionic operators for quantum chemistry calculations, weighted sum of fully commuting Pauli operators, or one-sparse operators, whose dynamics admit more straightforward simulation.
The choice of these operators and their ordering, as well as the analysis of their sparsity and commutation relations, which is usually done classically, can greatly impact the efficiency of the algorithm~\cite{martinez2trotter}. Depending on the decomposition, a variety of well-known techniques, such as Trotterization and linear combination of unitaries, can then be used to compile the dynamics operator into fundamental gate operations.

\subsubsection{Trotterization}
\label{trotter_sec}

The basic idea of Trotterization is to break down the evolution of a given quantum system into a series of smaller, more computationally manageable steps. For a given Hamiltonian operator $H=\sum_i H_i$, the first-order Trotter approximation is given by the product formula
\begin{equation}
    \left(\prod_{k=1}^m e^{-iH_k\frac{t}{n}}\right)^n = e^{-iHt} + O\left(\frac{m^2t^2}{n}\right).
\end{equation}
As shown, the error vanishes for large $n$ (number of Trotter steps) or small time.
Higher-order product formulas are readily derivable but become increasingly complex to implement. In practice, the choice of which order, $p$, would best suit a particular application is application-specific.
For example, the second-order Trotter-Suzuki formula
\begin{equation}
\left(\prod_{k=1}^me^{-iH_k\frac{t}{2n}}\prod_{k=m}^1e^{-iH_k\frac{t}{2n}}\right)^n=e^{-iHt}+O\left(\frac{m^3t^3}{n^2}\right)
\end{equation}
is a preferred choice for some lattice-gauge theory simulations~\cite{shangnan2021quantum,kan2022simulating} due to its favorable asymptotic scaling in large parameters such as the number of levels used to simulate bosons, while admitting a relatively simple implementation detail. The fourth-order formula was shown to admit the best error bounds in intermediate-sized system simulations using several dozen qubits for the Heisenberg Hamiltonian~\cite {childs2018toward,nam2019low}.

Trotterization introduces errors in the simulation as shown by the above examples.
At the highest level, this error depends on the order $p$ of the product formula used and scales asymptotically as $O(t^{p+1}/n^p)$ for $n$ Trotter steps~\cite{lloyd1996universal,berry2007efficient}.
However, this error is coming from non-commuting terms in the Hamiltonian decomposition, which suggests that tighter error bounds may be available when the specific structure of commutation is leveraged.
In addition, in quantum computation, Trotterized time evolutions are always applied to some state, in which case one would ultimately care about the error in the resulting approximate time-evolved state rather than the operator error in the Trotter approximation itself.
Research into providing tighter bounds and optimizing this error (e.g. by grouping and reordering of terms) is important and ongoing~\cite{tranter2019ordering,childs2019theory,martinez2trotter}.

While Trotterization is most typically applied to Pauli decompositions of Hamiltonians, in principle, it can be applied to any decomposition into terms whose time evolutions can be simulated directly.
For example, methods for simulation of sparse Hamiltonians usually involve decomposition into one-sparse terms~\cite{berry2007efficient,berry2012black,berry2015hamiltonian,berry2015simulating,berry2019qubitization}, and Trotterization of these is the approach for example in~\cite{berry2007efficient}.
To our knowledge, there has been no systematic study of the Trotter error of these that goes beyond the asymptotic expressions obtained via tail bounds in~\cite{berry2007efficient}.

In the case of Pauli decompositions
\begin{equation}
    H=\sum_i c_iP_i
\end{equation}
for Pauli operators $P_i$ and real coefficients $c_i$, an upper bound on the number of steps can be given in terms of the $\ell_1$-norm of the coefficients, $|||H|||=\sum_i|c_i|$.
Such bounds, however, ignore commutativity between terms and tend to grossly overestimate the number of Trotter steps required to approximate the Hamiltonian evolution to a certain error.
Indeed, in the trivial case where all terms in the Hamiltonian representation commute pairwise, a single Trotter step is enough to represent the Hamiltonian evolution exactly.
A recent study on the theory of Trotter errors~\cite{childs2019theory} provides more accurate bounds on the number of steps for a $p$th order formula in terms of norms of nested commutators $\sum_{\imath_1 \cdots i_{p+1}}||[H_{i_{p+1}}, [\cdots[H_{i_2}, H_{i_1}]]\cdots]||$.  Computing the precise commutator-scaling is exceptionally useful for reducing the number of Trotter steps, and, consequently, the quantum resources required, without sacrificing accuracy. It is, however, computationally expensive, as the computational cost, $O(N^{p+1})$, scales polynomially with the number of terms $N$ in the Hamiltonian.
Classical workflows (e.g. ordering of the Hamiltonian terms to reduce circuit depth) and their efficient implementation (e.g. leveraging high-performance parallel computing for accurate commutator-scaling evaluation) will play an important role in reducing the quantum resources required for practical applications and shortening the timeline to practical quantum advantage. 

One final point to note is that the above discussion focuses on generic Hamiltonian simulation and the analysis of the Hamiltonian operator. In practice, especially when targeting early quantum applications, one is interested in the accurate evolution of a given Hamiltonian, for a given time, with a given initial quantum state. The details of choosing the appropriate Trotterization (and, more generally speaking, the Hamiltonian simulation approach) will be very much application-specific, and, when compared to results obtained by only analyzing the Hamiltonian operator, further reduction of the estimated resources will most certainly be achieved.

\subsubsection{Linear Combination of Unitaries}

Another approach to the simulation of $e^{-iHt}$ involves block-encoding the Hamiltonian as a part of a larger unitary, $U(H)$. One of the most typical block-encodings is accomplished by writing $H$ as a linear combination of unitaries, $H = \sum_n c_n U_n$, and building $U(H)$ using $U_P = \sum_n \mathcal{P}_n\otimes U_n$, where $\{\mathcal{P}_n\}$ is a complete set of orthogonal projectors, $\mathcal{P}_n\mathcal{P}_{n'} = \mathcal{P}_n \delta_{nn'}$ built in the ancilla qubit space.
The computational cost of the algorithms based on \gls{lcu} depends on the $\ell_1$-norm of $c_n$ coefficient vector, $\lambda = \sum_{n} |c_n|$~\cite{LCU_childs}.
Thus, much effort has been directed to reducing the $\ell_1$-norm by various selections of $U_n$ operators~\cite{PRXQuantum.2.030305,loaiza2023LCU1,loaiza2023LCU2,loaiza2023LCU3}.
It is also possible to show using triangle inequalities for norms that the lowest possible $\ell_1$-norm for LCU decompositions is $\lambda\ge (E_{\max}-E_{\min})/2$, where $E_{\max/\min}$ are highest/lowest eigenvalues of the Hamiltonian in the entire 
fermionic Fock space or qubit Hilbert space~\cite{loaiza2023LCU1}.
Thus, in order to obtain $\ell_1$-norm lower than $(E_{\max}-E_{\min})/2$ one needs to modify the Hamiltonian.

Often in many-body problems, one is interested in eigenstates and eigenenergies in a particular symmetry subspace (e.g., with a fixed number of electrons for molecules). For quantum dynamics, if the initial state lies within a particular symmetry subspace, it continues to evolve there since the Hamiltonian preserves symmetries.
In both cases, to reduce $\ell_1$-norm of the LCU decomposition, one can create an effective Hamiltonian that acts identically to the original in the particular symmetry subspace of interest, but has a lower spectral range, $(E_{\max}-E_{\min})/2$. In~\cite{loaiza2023LCU2}, this was done by employing the form of the effective Hamiltonian, $\hat H_{\rm eff} = \hat H - \hat O_{1e}(\hat N_e - N_e\hat I) + c(\hat N_e^2 - N_e^2\hat I)$, where $\hat O_{1e}$ is an arbitrary one-electron operator, $\hat N_e$ is the operator of the number of electrons, $N_e$ is the number of electrons in the symmetric subspace of interest, and $c$ is a constant. Since finding the true spectral range of $\hat H_{\rm eff}$ is difficult, to optimize $\hat O_{1e}$ and $c$ in $\hat H_{\rm eff}$ one can use the simplest LCU decomposition in terms of Pauli products. $\hat O_{1e}$ and $c$ can be optimized to minimize the $\ell_1$-norm of Pauli product decomposition of $\hat H_{\rm eff}$.
As a direction for further improvement, simulating low-energy states or dynamics via LCU would benefit from either removing from consideration or shifting energies of highly excited states with the right number of electrons so that these states would not limit the $\ell_1$-norm of LCU. 

Generally, the Pauli products ($\hat P_k$) are not the only unitaries that one can use for the LCU decomposition. A simple example of improving Pauli products is by grouping them into sets of anti-commuting Pauli products. If $\hat A = \sum_k c_k \hat P_k$, where $c_k$ are real constants and $\{\hat P_k,\hat P_{k'}\}=0$ then $\hat A$ is proportional to the Hermitian unitary $\hat R = (1/\bar{c})\sum_k c_k \hat P_k$, with $\bar{c}=\sqrt{\sum_k c_k^2}$. It is shown in Ref.~\cite{loaiza2023LCU1} that such grouping of Pauli products in a qubit Hamiltonian will always reduce the $\ell_1$-norm with respect to the original Pauli product decomposition. There are other forms of unitaries that can be obtained from various fermionic tensor decompositions of the electronic Hamiltonians, for example, tensor hypercontraction or double factorization~\cite{PRXQuantum.2.030305,loaiza2023LCU1}. To compare different LCU decompositions, one needs to account for three factors: 1) classical cost, 2) $\ell_1$-norm, 3) quantum resources for implementing (e.g., the number of ancilla qubits, T-gate count) of each time step. Note that $\ell_1$-norm mainly affects the size of the time step that one can take, while the third aspect concerns the quantum resources required for each time
step.

\subsection{Quantum measurement}

Many quantum algorithms involve obtaining numerical values of observables, such as energies and dipole moments, as expectation values of their associated operators.
Valuable use cases could lead to observables on 50 to 100 qubits, which when decomposed into Pauli operators can easily involve millions of terms; for instance, electronic structure Hamiltonians are characterized by approximately $N^4$ such Pauli products.
Measuring all these Pauli products separately will take a prohibitively long time even on superconducting architectures, which have relatively rapid measurements.
Grouping Pauli products is a natural approach to reduce the number of individual operators to measure.
Here, we consider various grouping approaches to measure an expectation value of a single operator, like a Hamiltonian, for an arbitrary wavefunction.
The generalization of this problem is measuring the expectation values of a set of operators for a set of wavefunctions.
Such generalization poses additional optimization issues that can be important for some applications~\cite{Choi:ExcSt}, but for the sake of brevity, we focus on the single operator, single wavefunction case for the remainder of this subsection.

All measurement schemes we describe can be seen as yielding estimators of the Hamiltonian expectation value by combining results of projective measurements of some Pauli operators.
One can think of these as either measurements of generic Pauli operators or as measurements of the Ising Hamiltonian terms (i.e., diagonal Pauli operators) with respect to various states related by local basis transformations.
Commonly examined physical Hamiltonians, like those for electronic structures, are not readily convertible to Ising form; for instance, through local unitary transformations. Therefore, they must be decomposed into simpler parts.
A simple illustration of this idea is an observable $\hat H = \sum_n \hat H_n$ that is a sum of Hamiltonian fragments $\hat H_n$ that can be transformed to the Ising form with some unitary rotations: $\hat H_n = \hat U_n^\dagger \hat Z_n \hat U_n$, where $\hat Z_n$ is a polynomial of Pauli $\hat z$ operators. Then
\begin{equation}
    \bra{\psi}\hat H\ket{\psi} = \sum_n\bra{\psi}\hat H_n\ket{\psi} = \sum_n\bra{\hat U_n \psi}\hat Z_n\ket{\hat U_n\psi}.
\end{equation}
Hence,
\begin{equation}
    \mathit{\mathbb{E}}(\hat H) = \sum_{n,k}\frac{Z_n^{(k)}}{M_n},
\end{equation}
where $Z_n^{(k)}$ are the results of measurements of $\hat Z_n$ in the state $\ket{\hat U_n\psi}=\hat U_n\ket{\psi}$, and $M_n$ are the numbers of measurements for each fragment.   

Within this general scheme, a variety of expectation value estimation methods have been proposed.
The methods can be categorized based on the following criteria:
\begin{enumerate}
    \item Measurement Approach: Measurements within the same Hilbert space as $\hat H$ versus in an extended space using ancilla qubits (e.g., as for performing positive operator-valued measures);
    \item Functional relations: Relations between estimators for measurable operators and that for the total Hamiltonian, $\mathit{\mathbb{E}}(\hat H) = f[\mathit{\mathbb{E}}(\hat H_n)]$ (e.g., linear combinations or more complex functional relations\cite{Zapata:Work}); 
    \item Unitary Transformations: The types of unitary transformations that are needed to bring the measurable operators to the Ising form; 
    \item Estimator Error Scaling: The error scaling with the number of measurements is either $1/\sqrt{M}$ (shot-noise limit) or $1/M$ (quantum limit, i.e., Heisenberg scaling). Most near-term methods, which require low-depth measurement unitaries (i.e., $\hat U_n$) achieve only the $1/\sqrt{M}$ scaling. Achieving Heisenberg scaling requires much deeper circuits that are likely not feasible before fault-tolerance. 
\end{enumerate}

One of the estimation techniques that has become popular recently is classical shadow tomography~\cite{Preskill:ShadowTom}.
In the framework described above, the unitary transformations $\hat U_n$ are chosen randomly from the Clifford group (later, other groups were considered as well~\cite{RubinMiyake,ogorman2022fermionic}).
Applying random Clifford transformations, $\{\hat U_k\}$, as an extension to a state preparation circuit allows one to effectively measure parts of the Hamiltonian that would be transformed to Ising forms $\hat Z_k$ if $\{\hat U_k\}$ were applied to the Hamiltonian:
\begin{equation}
    \hat U_k \hat H \hat U_k^\dagger = \hat Z_k + \hat R_k,
\end{equation}
where $\hat R_k$ is the non-Ising part).
This shows that the information about the expectation value of $\hat H$ is only obtained from the Ising parts because the expectation value of $\hat R_k$ on any product state resulting from the wavefunction collapse is zero.
This insight spurred various efforts to enhance the sampling convergence by biasing selection of $\{\hat U_k\}$ to those that increase the norm of $\{\hat Z_k\}$ parts~\cite{huang2020predicting,Huang_Preskill:2021,Hadfield_Mezzacapo:2022,Hadfield:2021,lukens2021bayesian,shlosberg2023adaptiveestimation,koh2022classical,elben2023randomized,hillmich2021decision}.
A recent paper~\cite{dutt2023practical} has benchmarked these randomized measurement methods against a collection of molecular electronic structure Hamiltonians.

A more general view of this approach involves the use of adaptive, \gls{ic} \gls{povm}~\cite{garciaperez2021learning,acharya2021informationally}. This framework allows the optimization of the measurement for both the observable (e.g., $H$) and the state iteratively while collecting data. Moreover, the same mathematical framework can describe various physical implementations of the measurement~\cite{fischer2022ancillafree} and naturally allows for the modelization of noise in the measurement~\cite{glos2022adaptive}. Finally, it is worth mentioning that classical shadows and IC-POVMs allow for the estimation of multiple observables and for further post-processing, which can include noise mitigation methods~\cite{filippov2023scalable}.

To analyze the efficiency of various estimators one needs to take into account several factors: 1) classical optimization cost for finding optimal measurable operators; 2) quantum resource overhead due to the need for extra unitary transformations (e.g., $\hat U_n$); and 3) measurement overhead due to the number of measurements needed to achieve a desired accuracy in the expectation value. In the near-term, the overhead related to additional unitaries is crucial for overall feasibility.
For example, techniques using fermionic $\hat U_n$ \cite{Huggins_Babbush:2021,motta2021low,yen2021cartan,choi2023fluid} could be less efficient than those using qubit-based Clifford transformations \cite{choi2022improving,yen2023deterministic} even though the former require lower numbers of measurements for a given accuracy.
Additional options for mitigating circuit and measurement errors become possible using symmetries of measurable fragments.
Selecting more symmetric fragments allows one to improve statistics by removing the results that violate expected symmetry constraints -- e.g., the number of electrons -- which provides an error mitigation protocol~\cite{Cai_2021,cai2023quantum,cohn2021quantum}.

A crucial question for ongoing research is which types of physical Hamiltonians admit feasible measurement schemes in the near term.
Spin models typically partition into a few natural measurement bases; e.g., a Heisenberg model typically contains terms of the form $XX$, $YY$, $ZZ$, and $Z$, which form three local measurement bases (all $X$, all $Y$, and all $Z$).
On the other extreme, molecular electronic structure Hamiltonians have several terms that scale $\sim N^4$ for $N$ orbitals. It is unclear whether any known measurement scheme can reduce the resulting extremely large measurement counts to something tractable on near-term devices for instances large enough to offer the potential for quantum advantage.
Hence an important question is whether there exist intermediate problems where measurement techniques beyond simple grouping -- as in spin models -- can be leveraged to achieve tractable measurement counts.
For fermionic Hamiltonians, this is closely connected to choices of fermion-to-qubit mapping, as discussed in Sec.~\ref{mapping_to_qubits}.
A potential example is the Fermi-Hubbard Hamiltonian, where it is possible 
to partition the Hamiltonian into 5 operators consisting of commuting Pauli products, and each operator is converted into Ising form by at most one layer of Bell measurement circuit (\gls{cnot} + H)~\cite{PhysRevB.102.235122}. By estimating symmetries such as parity, these Bell measurement circuits allow one to mitigate circuit errors~\cite{PhysRevApplied.14.014059}.
Another excellent study on the topic as well as other elements of the materials simulation stack can be found in Ref.~\cite{clinton2022towards}.

\subsection{Problem-level preprocessing and optimization}

Achieving quantum advantage on pre-fault-tolerant quantum computers is likely to require optimization at all levels of the algorithmic stack.
This includes optimization of the input problem itself, as well as optimization that takes the input instance into account.
Since the family of possible input problems to quantum algorithms for materials science is too large for an exhaustive survey here, we instead provide two examples of this kind of problem-level preprocessing.
Part of the goal of this is simply to emphasize that this step is almost certain to be a necessary part of any serious attempt to achieve quantum advantage on a noisy quantum computer.

Our first example is active space selection (see also Sec.~\ref{embedding_sec}).
A fundamental problem in chemistry and materials science is the eigenvalue problem, where the eigenstates of the Hamiltonian describing the system are to be solved.
Diagonalizing the Hamiltonian of the full system is, however, too expensive and oftentimes unnecessary.
In many systems of practical interest, e.g., point defects in wide bandgap semiconductors~\cite{wolfowicz2021quantum} and catalysts on surfaces or interfaces~\cite{gujarati2023quantum}, electronic excited states of extended molecules~\cite{Sarkar2023}, a certain subset of electrons and orbitals in the molecule or solid is more relevant to the problem than the rest.
Therefore, a so-called active space can be defined, leading to an effective Hamiltonian. Solving this effective Hamiltonian (on a quantum computer) is much easier than solving the full Hamiltonian, while the essential properties of the system are retained. Generally speaking, defining active spaces requires identifying a chemically active site on the system. Efforts have been put into developing automated active space selection (see next paragraph), but a universal approach is so far lacking. In practice, prior knowledge of the system studied and the chemical intuition of the researcher is usually relied upon. For example, spin defect systems, e.g., the nitrogen-vacancy center in diamond and an electronic defect in MgO~\cite{maze2011properties,mitra2021excited, haldar2023local, Verma2023}, are heterogeneous materials promising for the realization of quantum communication. They have a defect center hosted in a periodic solid, whose local environment is vital to its electronic, optical, and mechanical properties~\cite{doherty2013nitrogen}. Therefore, a proper active space should include localized orbitals around the defect center~\cite{ma2020quantum, sheng2022green}. Other examples are pseudotetrahedral organometallic complexes containing chromium(IV) and aryl ligands which have been experimentally identified as promising molecular qubit candidates~\cite{Sauza-delaVega2022, Bayliss2020}.

Very often, local perturbations in extended systems such as spin defects described above or gas adsorption on solid-state catalysts are multiconfigurational, which means that several electronic states are degenerate in energy.
The accurate description of such systems require multireference methods. Among the most common ones are the \gls{casscf} method~\cite{CASSCF0,CASSCF1,CASSCF2}, \gls{dmrg}~\cite{White1992}, \gls{mrci}~\cite{Buenker1974}, and multireference perturbation theories.
However, these methods scale poorly with system size and exponentially with the size of the active space and thus are impractical for use in extended systems.

An ab-initio quantum embedding technique, \gls{dmet}~ \cite{DMET,DMET_mol,DMETpractical,5years_of_DMET}, and its periodic implementations provide a framework that allows users to effectively reduce an extended system to a finite embedding subspace~\cite{pham2019periodic,cui2020efficient}.
The user selects a fragment space characterized by orbitals obtained by using orbital localization schemes. Electrons in such orbitals are generally entangled to the environment, so it is not possible to define a wave function in the fragment orbitals alone. However, the entangled part of the environment can be obtained approximately using the Schmidt decomposition of a single-determinant whole-system approximate wave function (usually the Hartree--Fock wave function) \cite{DMETpractical}. This involves the diagonalization of the environmental block, $D_{\text{env}}$, of the one-body \gls{rdm}:

\begin{equation}
D_{\text{env}} = \mathbf{U} \lambda \mathbf{U}^\dagger
\label{eq:SVD}
\end{equation}

Here, $\lambda$ represents a diagonal matrix containing the eigenvalues $\lambda_i$ for $i = 0, 1, \ldots, N_{\text{env}}$, and where $N_{\text{env}}$ indicates the count of environmental orbitals. The columns of the unitary matrix $\mathbf{U}$, corresponding to non-zero and non-two $\lambda_i$ values, identify the entangled bath orbitals. Localized fragment orbitals and entangled bath orbitals together define a subspace of the molecule, no larger than twice the size of the original fragment, which is unentangled at the level of the underlying single-determinantal level of theory and within which any quantum chemistry method can be applied by treating all other orbitals as part of a frozen core.

In the case of spin-defects, the fragment active space is usually selected by including orbitals centered around the defect, whereas in the case of gas adsorption, the orbitals centered around the gas molecules and the surface atoms they are bound to are considered.
Methods like CAS-DMET and NEVPT2-DMET~\cite{mitra2021excited, haldar2023local, Verma2023}  utilize this framework and provide users the capability of modeling multireference ground and excited states arising from such local perturbations in extended solids with high accuracy.
In CAS-DMET, the embedding subspace constructed using Eqn.~(\ref{eq:SVD}) is subjected to a CASSCF calculation which further reduces the effective active space and accounts for the most relevant static correlation.
The remaining dynamic correlation is captured by multireference perturbation theories like NEVPT2~\cite{NEVPT2-0,NEVPT2-1,NEVPT2-2} as is done in NEVPT2-DMET.

A more cost-effective method compared to NEVPT2-DMET to include electron correlation is DME-PDFT~\cite{Mitra2023}.
In this approach, the 1- and 2-RDMs generated from a CAS-DMET calculation are used to compute the densities and on-top pair densities, which are then used in a subsequent 
\gls{mcpdft} calculation~\cite{LiManni2014, mcpdft,Zhou2022}.
MC-PDFT offers a method for integrating multiconfiguration wave function theory and density functional theory, allowing for the comprehensive treatment of both near-degeneracy correlation and dynamic correlation in strongly correlated systems. It proves to be a more cost-effective alternative to multireference perturbation theory, multireference configuration interaction, or multireference coupled cluster theory. Moreover, MC-PDFT demonstrates superior accuracy for various properties compared to Kohn–Sham density functional theory.
MC-PDFT computes the total energy as a functional of the electron density ($\rho$) and on-top pair density ($\Pi$) obtained from a multiconfiguration wave function.

The MC-PDFT energy is expressed as:
\begin{multline}
    E_{\text{MC-PDFT}}  = V_{NN} + \sum_{pq} h_{pq} D_{pq}\\
     + \frac{1}{2} \sum_{pqrs} g_{pqrs} D_{pq} D_{rs} + E_{ot} [\rho,\Pi],
\end{multline}
where \( V_{NN} \) is the nuclear-nuclear repulsion energy, \( h_{pq} \) and \( g_{pqrs} \) are one- and two-electron integrals, \( D_{pq} \) are the elements of the one-electron RDM, and the on-top energy \( E_{ot} \) is a functional of the density \( \rho \) and the on-top pair-density \( \Pi \).
It has been used in combination with various multireference wave functions like the generalized active space wave function~\cite{Ghosh2017}, DMRG~\cite{Sharma2019, Sharma2020} and it has been used to compute the total energy of strongly correlated systems starting from the 2-electron RDM evaluated on the quantum device.

In the context of DMET, PDFT is especially attractive because the on-top energy is expressed in terms of electronic density on a quadrature grid using semi-local (and therefore linear-scaling) functionals which are agnostic to the origin of the underlying density matrices. This formalism thus models electron correlation of all electrons regardless of any underlying embedding methodology. Therefore, DME-PDFT is formally less sensitive to the size of the embedded fragment than methods such as NEVPT2-DMET, which can only model electron correlation within the embedded subspace of fragment and bath orbitals~\cite{Mitra2023}.

A promising step towards heterogeneous catalysis applications is the recent study of adsorption with memory-efficient DMET algorithms which, when combined with multireference electronic structure solvers, can be extended to the bond-breaking phenomenon on surfaces~\cite{Mitra2022}.
A computationally affordable alternative to CASSCF for modeling larger active spaces is the \gls{lasscf} method~\cite{Hermes2019, Hermes2020}.
LASSCF divides a given active space into localized non-interacting subspaces, connected through a mean field, and has the ability to capture strong electron correlation within these subspaces without encountering the computational cost associated with CASSCF.

The LASSCF wavefunction is represented as an antisymmetrized product of the \gls{fci} wavefunctions of these individual localized subspaces and the single determinantal wavefunction of the inactive space, which remains delocalized over the entire molecule.
This is mathematically represented as:
\begin{equation}
\Psi_{\text{LAS}} = \bigwedge_{K} (\Psi_{A_K}) \wedge \Phi_D
\end{equation}

where \( \Psi_{A_K} \) denotes the many-body (generally FCI) wavefunction of the \( K \)th localized subspace, and \( \phi_D \) denotes the single-determinantal wavefunction of closed-shell occupied inactive orbitals.
The energy of the system \( E_{\text{LAS}} \) is obtained through variational optimization and is expressed as:
\begin{equation}
    E_{\text{LAS}} = \langle \psi_{\text{LAS}} | \hat{H} | \psi_{\text{LAS}} \rangle
\end{equation}
where \( \hat{H} \) denotes the molecular Hamiltonian.

Although LASSCF treats strong correlations within these localized smaller active spaces, it only accounts for interactions between localized fragments at a mean-field level.
Methods such as State Interaction (LASSI) on a classical computer~\cite{Pandharkar2022} or the variational unitary coupled cluster singles and doubles (UCCSD) on a quantum computer (LAS-UCC)~\cite{Otten2022} model correlation between these fragments using LASSCF as a reference wave function.
In the LAS-UCC method, the multiconfigurational subspace wavefunction $\Psi_{A_K}$ for each fragment is prepared using either QPE or Direct Initialization (DI), as outlined by D'cunha \textit{et al.}~\cite{Dcunha2023StatePreparation}.
The inter-fragment correlations are then captured through a standard VQE optimization process using the \gls{ucc} ansatz.

Despite these recent advances in computational methods for modeling electronic structures, significant research challenges exist, particularly involving the treatment of ``larger" active spaces and the inherent poor scaling of multireference perturbation theories. Numerous research efforts have been directed towards the treatment of larger active spaces using classical electronic structure techniques like DMRG, selected CI, and quantum Monte Carlo methods. However, expanding the active space relies on physically motivated approximations to balance computational cost~\cite{ciapprox-1,Holmes2016, Stein2019, Booth2009,ciapprox-2}. The basic philosophy in all these approaches is focusing on the most important configurations while eliminating less significant ones, based on physically motivated criteria.
Quantum computing holds the promise of effectively modeling these larger active spaces accurately, especially as the number of available qubits increases and noise mitigation techniques become more powerful.
Embedding techniques such as LAS-UCC may help accelerate this process.
While perturbation theories scale poorly on classical hardware, recent studies on formulating quantum implementations of perturbation theories\cite{Li2023} highlight the potential for creating multireference quantum algorithms, including NEVPT2-DMET.
With more qubits being available, the primary research challenge still remains to determine the optimal methods for quantum implementation that maintain a favorable balance between computational cost and the desired accuracy.

Our second example of problem-aware preprocessing is VQE ansatz design~\cite{peruzzo2014variational}.
VQE (and its variants) is the most widely used algorithm on quantum computers for the eigenvalue problem, although its experimental applications to date have been limited to classically-simulable demonstrations; see, e.g.,~\cite{du2010nmr,lanyon2010towards,peruzzo2014variational,wang2015quantum,omalley2016scalable,shen2017quantum,paesani2017bayesian,kandala2017hardware,hempel2018trappedion,santagati2018witnessing,dumitrescu2018atomicnucleus,kokail2019selfverifying,kandala2019errormitigation,ganzhorn2019gate,sagastizabal2019experimental,mccaskey2019quantum,smart2019quantum,liu2021representation,liu2023analytic,zheng2023speeding}.
The core of the VQE algorithm is the parameterized quantum circuit $U(\Vec{\theta})$, usually called the ansatz, whose output is a manifold of states that should contain an approximation to the ground state of the system.
The ansatz can be evaluated according to two key attributes: expressivity and practicality.
The former measures how well the ansatz can approximate the solution to the problem to a certain accuracy, while the latter takes circuit depth, the difficulty of optimizing those parameters, and the qubit connectivity of the underlying hardware into consideration.
Quantum advantage in VQE requires identifying an ansatz that satisfies both criteria, i.e., it is a feasible circuit on a given device, it is trainable at scale on a classically hard problem, and it can represent a low-energy state with sufficient accuracy to surpass classical methods.

Two typical examples of ansatzes are the UCC ansatz~\cite{peruzzo2014variational, shen2017quantum,nam2020trappedion,anand2022quantum} and hardware efficient ansatzes~\cite{kandala2017hardware}.
The UCC is chemically-inspired and thus very expressive for most chemistry and materials, but usually entails a large number of gates and is likely beyond near-term hardware.
The hardware-efficient ansatzes are problem-agnostic and designed specifically to use the native gates and connectivity of a given device, but may be impractical to optimize due to the barren plateau problem~\cite{mcclean2018barren}, and also may fail to be sufficiently expressive at low depth, lead to symmetry breaking, and non-smooth potential energy surfaces \cite{d2023challenges}.

Efforts have been put into designing new ansatzes to strike a balance between these two dimensions and unify the advantages of UCC and hardware efficiency~\cite{ryabinkin2018qubit,elfving2021simulating,motta2023bridging,tang2021qubit}.
The qubit coupled-cluster ansatz~\cite{ryabinkin2018qubit, ryabinkin2020iterative, huang_quantum_2023} and ADAPT-VQE~\cite{tang2021qubit} are among such examples, where a pre-screening process is proposed to select certain Pauli operators to generate the gates in the ansatz, thereby reducing the circuit depth and number of parameters.
These methods that trim the number of variational parameters are promising, but there is still a lack of evidence for the scalability of parameter training.
An exception may be HMP2~\cite{wang2021resource}, where second-order perturbation theory is employed to reduce the number of variational parameters, while providing real-time, quantum computation-based methodologies to identify additional Pauli operators to include in the VQE progression.
These advancements indeed merit further research despite outstanding challenges due to the simplicity of \gls{vqa} in all other respects (particularly in circuit complexity) and the extensive success of classical variational algorithms.

Another challenge with ADAPT-VQE-like algorithms lies in the larger measurement overhead due to the selection of Pauli operators, which can however be mitigated by the use of IC-POVM measurements~\cite{nykanen2023mitigating}.
While it is important to design an ansatz structure that is most appropriate for a target problem, appropriately initializing the ansatz parameters is an equally critical task when it comes to the trainability of the resulting optimization.
This is because random parameter initialization can be extremely far from the optimal parameters, and the ability for VQE to converge to the optimal parameters (and thereby find the ground state) will heavily depend on the optimizer and the effects of noise.
Identifying good classical initialization can therefore considerably improve the accuracy and speed of VQE convergence.
One promising method is CAFQA~\cite{ravi2023cafqa}.
The CAFQA ansatz is built with only Clifford gates and is therefore classically simulable in polynomial time.
A discrete optimizer is used to search through different possible Clifford gate parameters until the optimizer converges to a set of parameters that minimizes the VQE expectation (to the best possible extent within the Clifford space).
This `CAFQA' state is then used to initialize traditional VQE.
There is an opportunity to explore other forms of classical simulation to bootstrap VQE; near-Clifford circuits and tensor network approaches are promising directions~\cite{preopt-1,preopt-2}.
All these methods could benefit from problem-specific optimization.
For instance, restricting the CAFQA Clifford search to a state space that is expected to have some overlap with the ground state can improve initialization speed and accuracy.

\subsection{Classical data loading}
\label{ssec:classical-data-loading}

At the highest level, quantum computation involves the following steps. First, a classical description for a specific problem is established. This description is input into a quantum circuit. After executing the quantum circuit, we retrieve classical measurement results as the solution to the problem. Within this workflow, an essential challenge is the data input. A robust quantum algorithm must explicitly detail this input technique and the associated hardware.

For some types of use cases, e.g., quantum simulation algorithms, the problems themselves are quantum mechanical in nature, and often this permits direct mappings to quantum circuits, such as those discussed in Sec.~\ref{mapping_to_qubits}. In more general cases, including when the inputs are classical data, scientists who focus on theoretical complexity analysis often abstract such input methods as an `oracle' and investigate how many times this oracle must be applied, using this as a basis to estimate the algorithm's complexity. However, creating a physical realization of a quantum oracle is also a fundamental aspect of realizing quantum computation.

QRAM~\cite{giovannetti2008quantum} serves as the physical embodiment of the quantum oracle. Unlike traditional RAM, where a single piece of data can be swiftly loaded into the central processor, QRAM allows for multiple classical data items to be loaded concurrently in superposition. This unique capability positions QRAM as a bridge between classical and quantum computing realms. More than that, QRAM can function as a rapid and versatile quantum oracle, effectively addressing the data-input bottleneck prevalent in numerous quantum applications. QRAM is also a fundamental assumption for many quantum algorithms. For instance, many quantum machine learning algorithms~\cite{aaronson2015read,biamonte2017quantum} may require QRAM as a prerequisite to potentially achieve quantum speedups, although some algorithms have been discovered to achieve the speedups without~\cite{niroula2021quantum}. Currently, given the significant challenges in physically constructing QRAM, both theoretical and experimental research on QRAM is an active field of study.

For quantum simulation and material sciences, QRAM could primarily be used to provide fast, efficient, and reliable initial states and Hamiltonian encoding. QRAM architectures could provide useful interfaces, benefiting practical use cases of quantum simulation to large-scale, challenging problems in material science.

\subsubsection{QRAM construction}
\begin{figure*}
    \centering
     \includegraphics[width=0.45\textwidth]{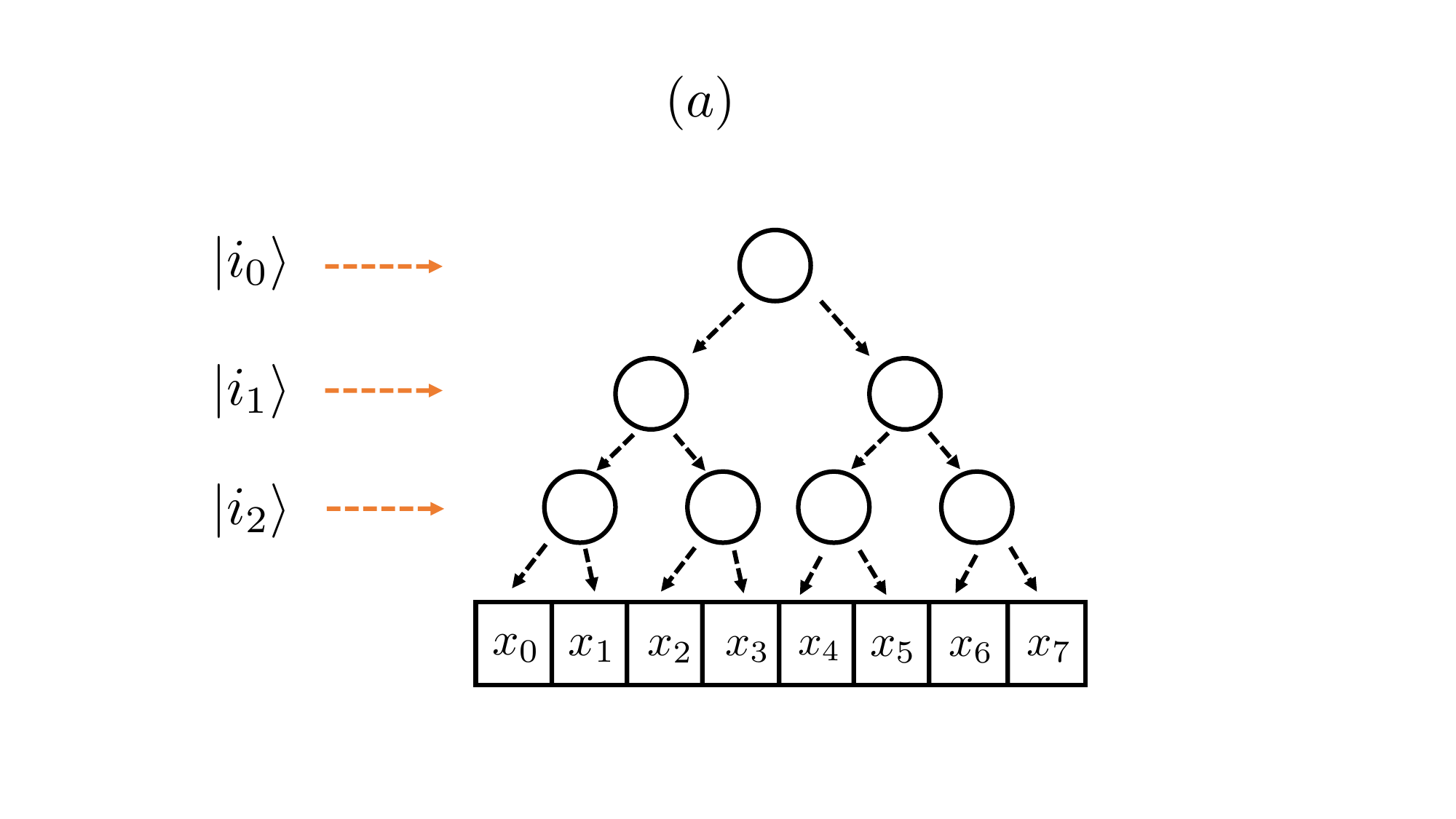}
     \includegraphics[width=0.45\textwidth]{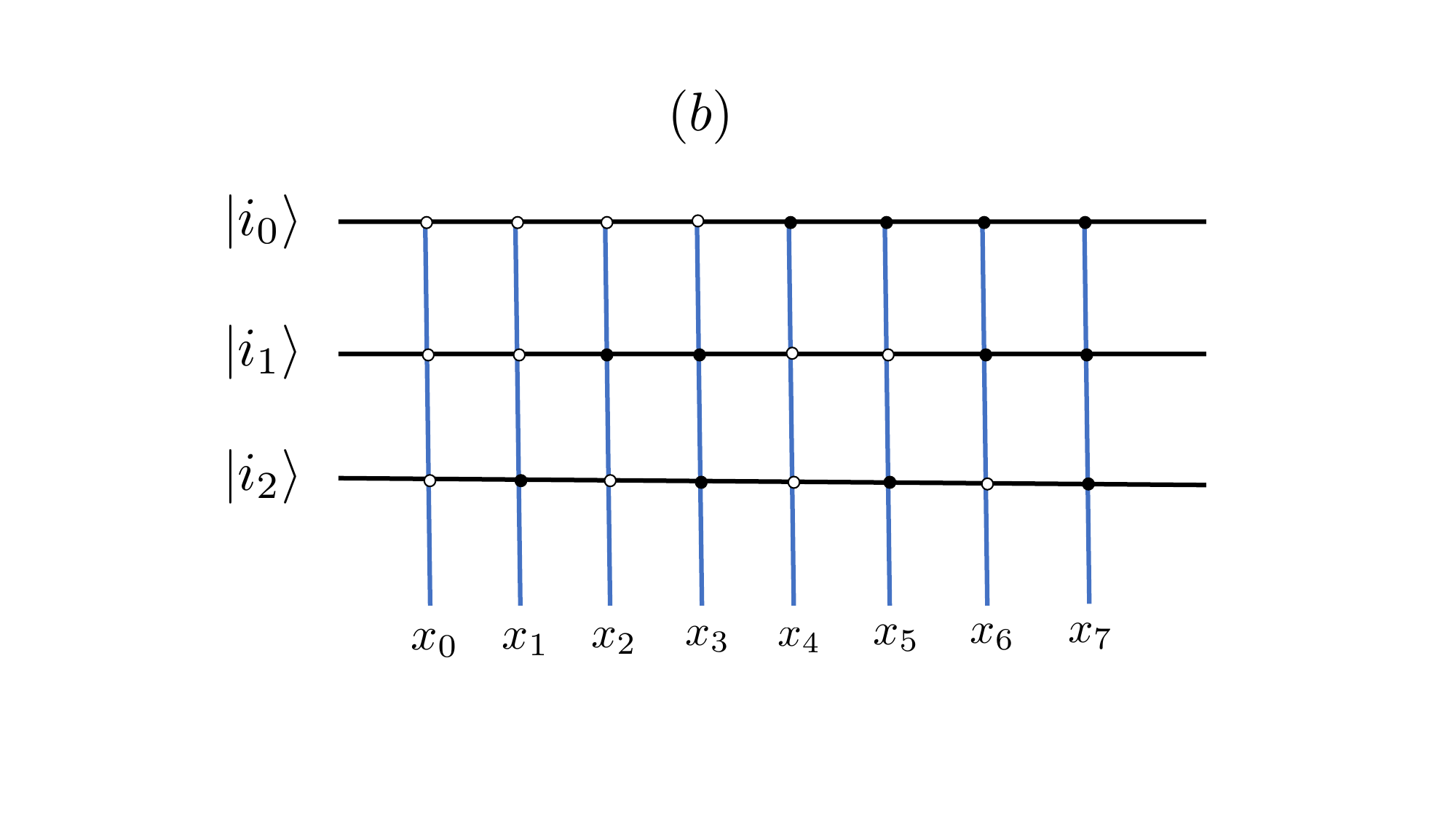}
     \includegraphics[width=0.6\textwidth]{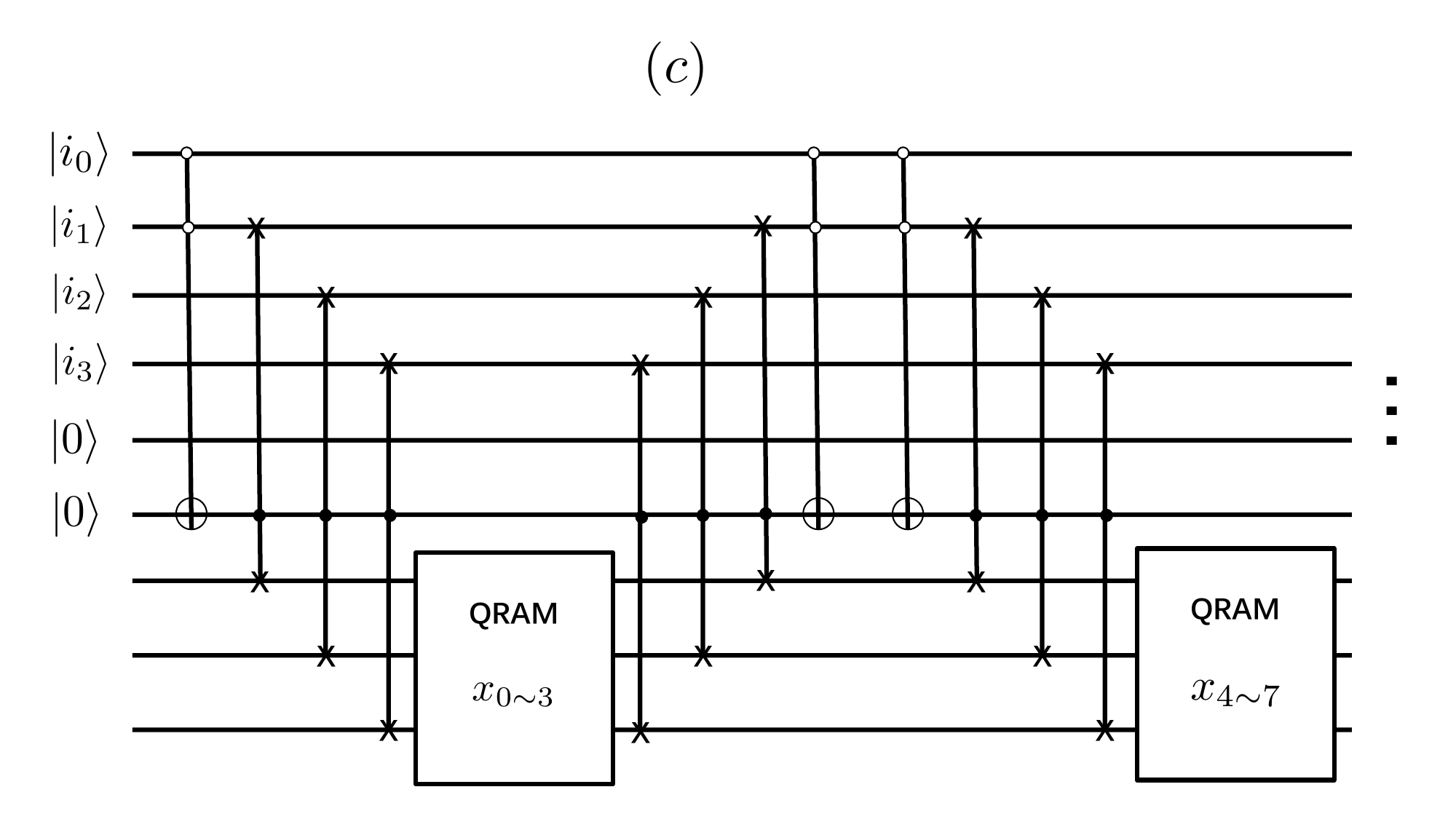}
    \caption{$(a)$ An illustration of QRAM made by quantum routers. $(b)$ An illustration of QROM. $(c)$ Hybrid QRAM-QROM architecture.}
    \label{fig:qram}
\end{figure*}

QRAM could be considered as the following unitary,
\begin{align}
\left|\psi_{\text {in}}\right\rangle=\sum_{i=0}^{N-1} \alpha_i|i\rangle^A|0\rangle^B \stackrel{\text { QRAM }}{\longrightarrow}\left|\psi_{\text {out}}\right\rangle=\sum_{i=0}^{N-1} \alpha_i|i\rangle^A\left|x_i\right\rangle^B
\end{align}
starting from an input state $\ket{\psi}_{\text{in}}$ in superposition, one can get an output state $\ket{\psi}_{\text{out}}$. We assume that the data vector $x_i$ has $N$ components. In order to construct QRAM, quantum routers are important ingredients. One can write quantum routers as quantum circuits made by a sequence of controlled-SWAP gates. A QRAM can be assembled using quantum routers~\cite{nielsen2010quantum}, as illustrated in Figure \ref{fig:qram}, where routers are organized in a binary tree formation. This basic QRAM configuration made by quantum routers is termed fanout QRAM, and it allows for data storage of $O(N)$ elements in a time complexity of $O(\log N)$.

On the other hand, the bucket-brigade architecture presented in~\cite{giovannetti2008quantum} serves as a nuanced version of the fanout design, distinguished by dynamic routing of address qubits: address qubits are dynamically introduced into the tree during a query. If an address qubit comes across a router in the $\ket{0}$ or $\ket{1}$ state, it routes to the left or right, respectively. However, upon encountering a router in the $\ket{W}$ state (each router could incorporate a third state labeled $\ket{W}$, for ``wait,'' so it might be a qutrit system), the router and incident mode states are interchanged, setting the router's state to $\ket{0}$ or $\ket{1}$ based on the initial incident address. Such designs are more robust to decoherence~\cite{giovannetti2008quantum}. 

QRAM stands for a type of extremely fast quantum memory that may be better suited for handling large-scale data. On the other hand, when the amount of data is relatively small, one might use what's known as \gls{qrom} to upload quantum data, for tasks like simulating certain Hamiltonians~\cite{babbush2018encoding}. Designs that combine QRAM and QROM are referred to as hybrid architectures~\cite{hann2021resilience}. Since QROM requires $O(N \log N)$ time and $O(\log N)$ space, hybrid architectures might represent a space-time trade-off for specific hardware conditions (see Figure \ref{fig:qram}).

QRAM might potentially have numerous applications if it is realized on a large scale. Here we summarize some potential applications of QRAM.
\begin{itemize}
\item Quantum algorithms. Many quantum algorithms based on quantum machine learning~\cite{wittek2014quantum,biamonte2017quantum} might require QRAM as a quantum oracle for providing potential exponential quantum speedups. Refs.~\cite{jiang2022quantum,liu2022quantum} discuss how quantum algorithms could provide hardware requirements to QRAM architectures. Ref.~\cite{liu2023data} discusses how QRAM could bring potential enhancement for quantum simulation algorithms in a data center setup. For instance, one could design QRAM circuit for Hamiltonian simulation of quantum chemistry or quantum materials similar to~\cite{babbush2018encoding}, which is an efficient way for uploading Hamiltonian data towards quantum computers. 

\item \gls{qdc}. Nowadays, data centers are important businesses in the digital society, and it is natural to have a quantum version of data centers in the coming era. Refs.~\cite{liu2023data,liu2023quantum} propose a theory of QDCs, where QRAM and quantum networks are combined to provide an elementary definition of QDC. These works show that QDCs with several applications in QRAM as fast quantum memories, will provide fast, secure, and accurate applications in quantum computing, communication, and sensing. 
\item Quantum communication and sensing. Quantum Private Queries~\cite{giovannetti2008queries} and the associated blind quantum computing~\cite{giovannetti2013efficient} could be flagship applications of QRAM in quantum communications, where the users could ask data centers to provide specific services without privacy leakage. Moreover,~\cite{liu2023data} provides a quantum data compression algorithm based on variants of QRAM, an application helpful for distributive quantum sensing applied to, for instance, quantum telescopes~\cite{gottesman2012longer}. 
\end{itemize}

Fault-tolerant designs of quantum random access memory are investigated by~\cite{di2020fault}, showing that QRAM with full fault tolerance might be very challenging on a large scale. On the other hand, noisy QRAM might still be error-resilient for generic setups of decoherence and noisy channels~\cite{hann2021resilience}.
In fact, though fanout QRAM designs might be vulnerable to decoherence, rendering them non-scalable, the bucket brigade QRAM architecture is known for its high noise resilience~\cite{giovannetti2008quantum, hann2021resilience}. Specifically, the infidelity of a query grows only logarithmically with memory size~\cite{hann2021resilience}. Yet, past studies leaned on specific noise models, raising questions about the practical advantages in real-world implementations. Ref.~\cite{hann2021resilience} delved into the effects of decoherence on QRAM in a comprehensive manner. A significant takeaway is the affirmation that the logarithmic infidelity scaling persists across varied error channels, including challenges like depolarizing noise and coherent errors. The research identifies limited entanglement among the memory components as the chief reason behind this noise resilience. Interestingly, this insight also suggests potential architectural simplifications without compromising noise resilience, serving an interesting future research question. This understanding implies that with contemporary hardware, QRAM might be potentially realized in real-world noisy environments, facilitating high-fidelity queries even with early fault-tolerant technologies. Additionally, when quantum error correction is integrated, the bucket brigade architecture continues to show promise, enhancing hardware efficiency and fortifying against logical errors~\cite{giovannetti2008quantum, hann2021resilience}.

\subsubsection{Experimental efforts towards QRAM}
Aside from theoretical studies, significant experimental progress has been made towards realizing QRAMs. In one of the original papers on QRAM~\cite{giovannetti2008architectures}, two different fanout designs are presented together with the bucket-brigade designs, quantum optical fanout and phase gate fanout, where hybrid systems of photons and trapped atoms are used. Ref.~\cite{jiang2019experimental} realizes a quantum memory of 105 qubits carried by 210 memory cells in a macroscopic atomic ensemble, with demonstrations of storage of optical qubits into these memory cells and their readout, which might be a significant step towards full quantum random access memory with superpositions of addresses.

Refs.~\cite{hann2019hardware,hann2021resilience,connorthesis} propose a hardware-efficient QRAM design with Hybrid Quantum Acoustic Systems. More precisely,~\cite{hann2019hardware,hann2021resilience,connorthesis} introduce a methodology for quantum computation utilizing multi-modal quantum acoustic systems and, based on this framework, suggest a streamlined approach for QRAM construction. Quantum data is preserved in high-Q phonon modes. Interactions between these modes are deliberately crafted by applying off-resonant stimuli to a transmon qubit. Contrasted with previous suggestions that emphasize direct qubit excitation, it is claimed that their approach can significantly enhance gate fidelity, especially for enduring acoustic modes.

Ref.~\cite{chen2021scalable} introduces a QRAM construction approach using a photonic-integrated-circuit (PIC) architecture combined with solid-state memories. Additionally,~\cite{chen2021scalable} offers a novel scheme rooted in quantum teleportation, further expanding its application in quantum networks. Both of these methods effectively carry out the primary QRAM functions: quantum state transfer and quantum routing. These implementations are grounded in existing components, including electro-optic modulators, a Mach-Zehnder interferometer (MZI) network, and nanocavities linked to artificial atoms for spin-based memory operations.

\begin{figure*}
    \centering
     \includegraphics[width=0.6\textwidth]{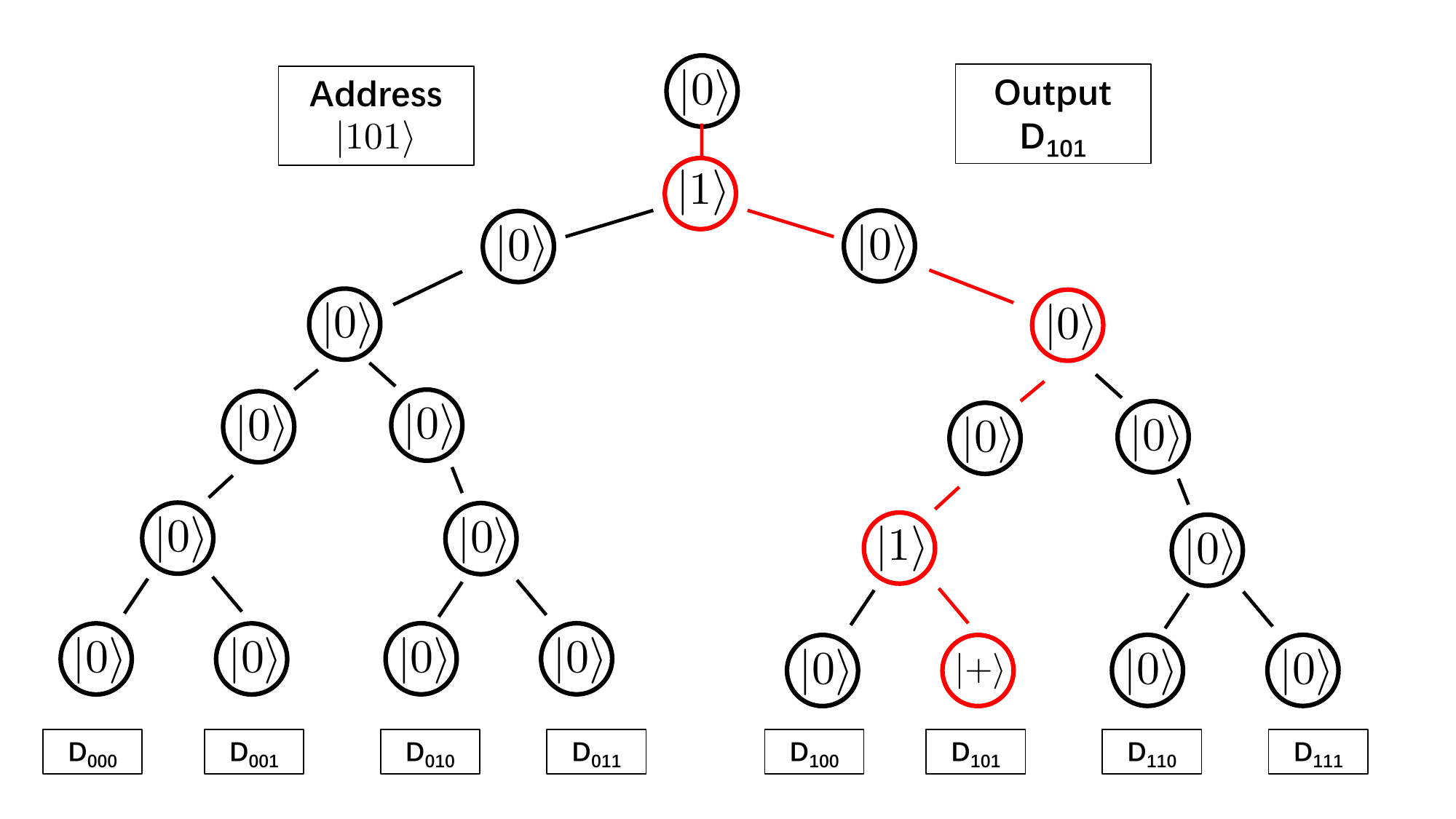}
     \caption{QRAM as a binary tree of $\text{Q}^2$ routers. Here, $\text{Q}^2$ router refers to the router with both control and signal states being quantum states.}
     \label{fig:challenges}
\end{figure*}

\begin{figure*}
    \centering
    \includegraphics[width=0.45\textwidth]{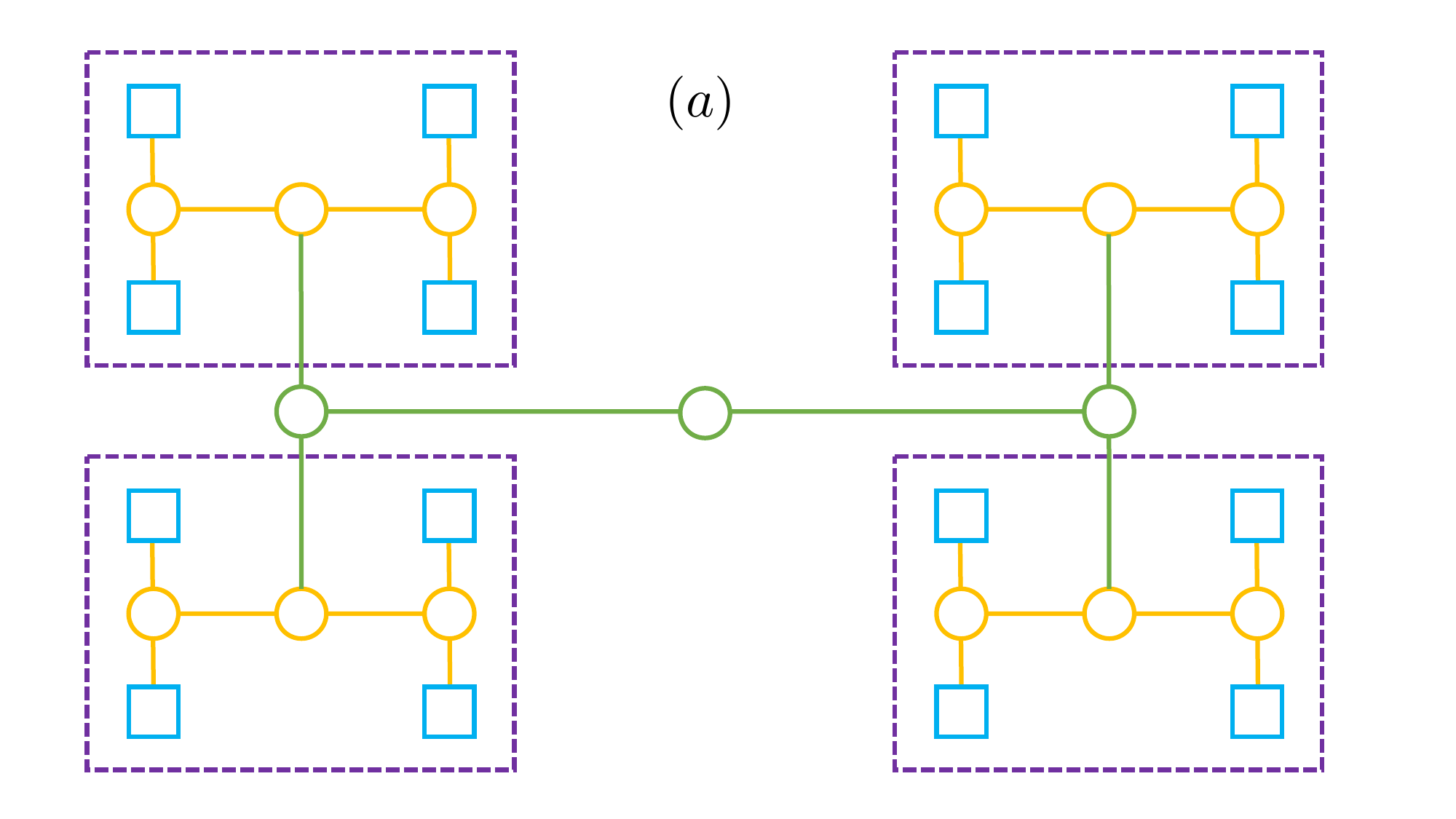}
     \includegraphics[width=0.45\textwidth]{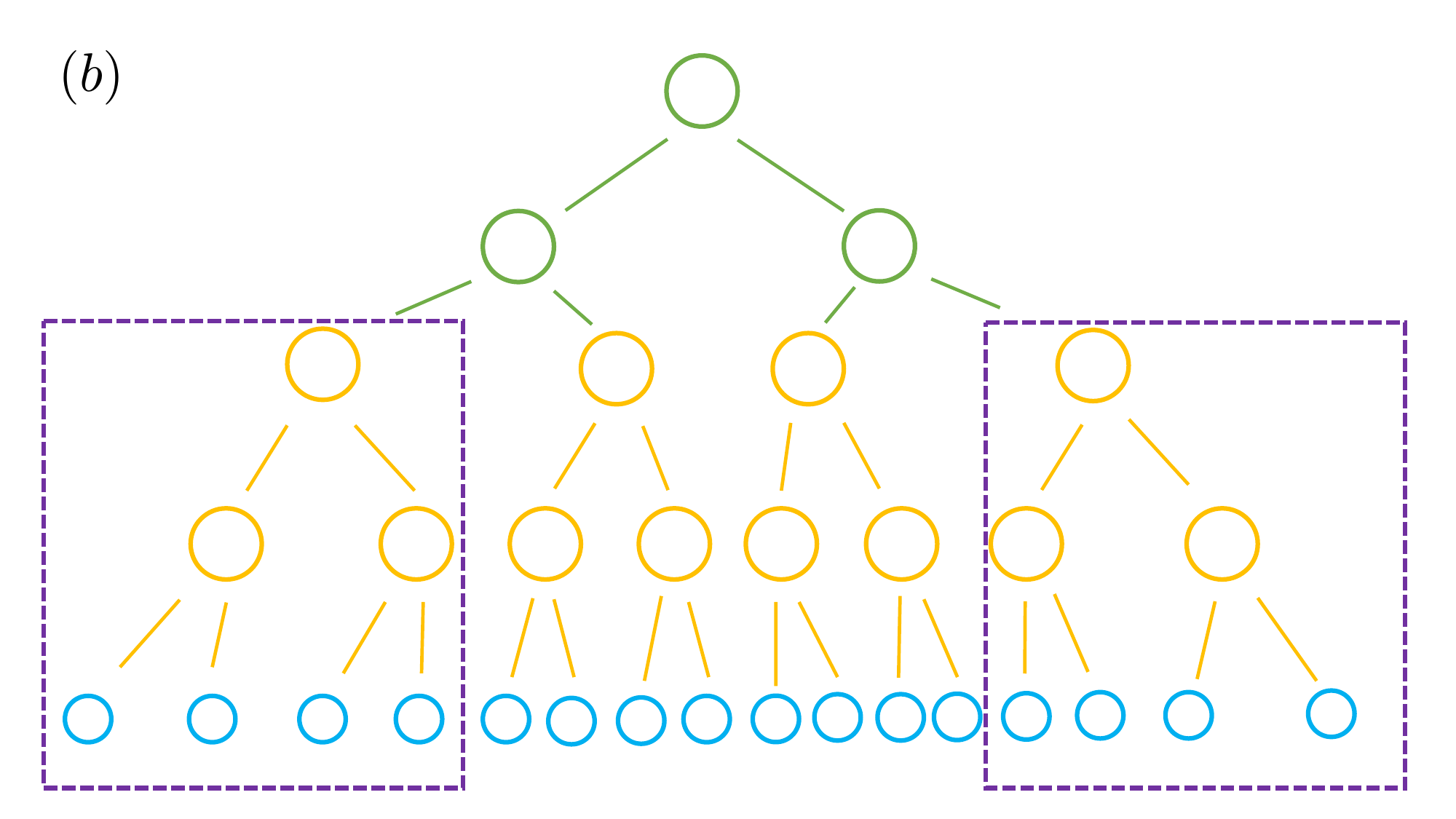}
     \caption{$(a)$ Self-similar fractal structure of the H-tree designs (see~\cite{xu2023systems}). One can use the self-similar structure to create a size $2^{n+2}$ QRAM from 4 units of size $2^n$ QRAM. $(b)$: An example of depth-4 tree with $2^4=16$ leaves. }
     \label{fig:challenges2}
\end{figure*}

Constructing a large-scale, fault-tolerant QRAM presents significant challenges, which could be summarized as the following.
\begin{itemize}
\item Scalability: The number of qubits required for a QRAM scales linearly with the data set size. As data sets grow, so does the qubit demand. Moreover, larger data sets necessitate more substantial QRAM hardware with higher circuit depth and more sophisticated designs.

There are various ways to address the scalability issue by hardware co-designs of QRAM. For instance, we can utilize $\text{Q}^2$ routers as foundational components for constructing QRAM (Here, $\text{Q}^2$ router refers to the router with both control and signal states being quantum states). This modular design allows us to assemble QRAM by interlinking various modules, as depicted in Figure \ref{fig:challenges}. Alternatively, a 2-D integrated QRAM can be achieved using \emph{H-tree} patterns, either through the teleportation protocol as outlined in~\cite{xu2023systems} or by employing long wires (see Figure \ref{fig:challenges2}. Note that this design is beyond~\cite{xu2023systems} by implementing long-range coupling wires, within the capability of IBM devices. Moreover, the design is compatible with a \emph{single-layer} architecture without crossing coupling wires). For a scalable QRAM, extensive range connectivity might be bounded by some principles of physics as discussed in~\cite{wang2023fundamental}. 

\item Efficiency: To ensure QRAM's efficiency and reliability, the quantum hardware must be both highly compact (fractal or self-similar structures in the design of QRAM with more efficient usages of space like~\cite{xu2023systems}) and highly coherent. In addition to superconducting circuits, it is also intriguing to explore more compact phononic circuits for QRAM like~\cite{hann2019hardware}.

\item Error Correction: As QRAM size expands, errors become inevitable. Thus, strategies for reliable data loading despite these errors are essential. Standard quantum error correcting codes might be applicable to QRAM. However, developing quantum error correction techniques that are specifically adapted to QRAM~\cite{di2020fault,jaques2023qram} remains an open research area. On the other hand, error resilience of designs of QRAM~\cite{hann2021resilience} might provide possible resolutions. Moreover, error resilience is a feature of bucket-brigade QRAM, which shows that errors only scale poly-logarithmically with the system size (see above). This indicates that QRAM can already be a useful application for some non-error-corrected quantum computers.

\item Need for Specialized Solutions: Given its unique purpose and architecture, QRAM's scalability issues should be approached with its specific needs in mind. Unlike universal quantum computers, QRAM doesn't need a universal set of quantum operations, which could be an avenue for simplifying its architecture. Moreover, some classical optimization could be helpful to enable more sophisticated QRAM designs. For instance,~\cite{hann2021resilience} designs a noisy QRAM simulator in classical devices, which might be for further explorations of QRAM designs and simulation in a classical space.

\end{itemize}

\subsection{Circuits, transpilation, and architecture}

Just as in classical computing, every quantum computation that is to be carried out on a quantum computer needs to be broken down to a netlist of native operations that admit a direct implementation on a quantum computer. One of the most widely-used representation of such a netlist is the so-called quantum circuit, which is to be transpiled for a target quantum computer backend. Different quantum backend hardware may support different qubit-to-qubit connectivity, which limits the interactions between the qubits that may be implemented natively. 

To this end, quantum applications in materials science are no exceptions. Indeed, quantum circuits for Heisenberg Hamiltonian simulation~\cite{childs2018toward,nam2019low}, relevant for many-body localization and room-temperature superconductivity, and electronic structure calculations for small molecules~\cite{nam2020trappedion,wang2021resource,wang2023ever}, have been optimized for both pre-fault-tolerant and fault-tolerant settings before. Other examples include particle physics simulations~\cite{shaw2020quantum,kan2021lattice,kan2022simulating}, where fault-tolerant cost, including the number of non-Clifford quantum gates and qubit counts, have been minimized. In all these examples, it is critical that the resulting netlist of instructions to be carried out by a quantum computer is optimized, targeting {\it time to solution}: a key metric in all types of computation, including quantum. Case in point, if time to solution is of no matter, the computational advantages offered by quantum computing is moot and classical computing suffices, and a high-fidelity operation of a quantum computer would become unnecessary since repeated executions of computation can be performed until an acceptable answer is obtained. The important roles that circuits, transpilation, and architecture will continue to play in leveraging the computational power enabled by quantum computation then form the central topics of this section.

\subsubsection{Connectivity and Architecture}
\label{sec:CNA}

We begin by highlighting key topics around qubit connectivity, a fundamental aspect of a \gls{qpu} that dictates the efficiency, scalability, and feasibility of the quantum devices. We first consider SWAP networks, highlighting their role in enabling Hamiltonian simulation in quantum systems with limited connectivity. The discussion then moves to distributed quantum computing, emphasizing its importance in scaling quantum computing beyond the capabilities of current QPUs. Quantum circuit cutting is also examined for its practicality in decomposing complex quantum circuits into smaller segments, facilitating distributed processing.

\begin{figure}[htbp]
    \includegraphics[width=\linewidth]{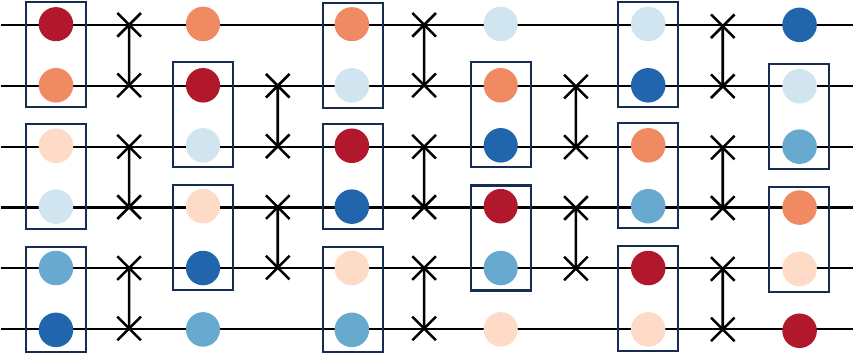}
    \caption{Example SWAP network for $n=6$ qubits. In $O(n)$ steps, each of the $O(n^2)$ pair of qubits (colors) performs an interaction, even with just linear connectivity.}
    \label{fig:swap_network}
\end{figure}

\paragraph{SWAP networks} One of the tools that can be leveraged to effectively implement Hamiltonian simulation primitives on hardware with minimal (linear) connectivity is the SWAP network~\cite{kivlichan2018quantum, tomesh2021coreset,hashim2022optimized}. Through a SWAP network, depicted in Figure~\ref{fig:swap_network}, all $\binom{n}{2} \in O(n^2)$ pairs of (commuting) interactions between $n$ qubits can be accomplished in just $O(n)$ circuit layers, even on a quantum computer architecture that admits nearest-neighbor qubit-to-qubit connectivity. With the exception of an all-to-all architecture that admits a single-instruction implementation of commuting, overlapping two-qubit gates~\cite{grzesiak2020efficient,grzesiak2022efficient,bravyi2022constant}, the asymptotic $O(n)$ scaling matches that of an architecture that allows for random-access two-qubit interactions that may be implemented in parallel so long as the interactions do not overlap. In the context of the Materials Science applications in this paper, any ``dense'' interactions with commuting terms can have a natural implementation on an all-to-all architecture (c.f. Refs.~\cite{nam2019low,wang2021resource,kan2022simulating,wang2023ever}) or on a limited-connectivity architecture with SWAP networks. Reducing the overhead in moving the quantum information around for optimal implementation of a given set of two-qubit interactions remains an area of ongoing research.

\paragraph{Distributed/modular}
Just like classical HPC, distributed quantum computing is necessary for the continued scaling of quantum computing. While monolithic QPUs continue to make significant progress in both sizes and qubit qualities, there remains a huge gap between available QPUs and practical benchmarks. To make matters worse, developments from classical computing demonstrate that the computing and memory capacity gaps between large-scale industrial use cases and single computing cores only get larger, despite having more powerful monolithic cores. To combat this increasing gap, almost all industry-scale classical applications today such as Large Language Model (LLM) training, and even many personal-scale workloads such as intensive gaming, happen in parallel. In fact, distributed computing is only more imperative for quantum computing as single-core scalability poses significant engineering challenges.

There have been distributed quantum computing proposals from both hardware and software researchers. IBM, Quantinuum, and IonQ all announced plans to develop modular QPUs with interconnects to enable distributed quantum computing.
Ref.~\cite{ang2022architectures} models hardware performances of potential superconducting modular QPUs.
Ref.~\cite{wu2022collcomm} proposes compilation algorithms for distributed QPUs with EPR pair connections. 
Ref.~\cite{Zhang_2022} proposes ClusterVQE which enables domain-specific optimizations to map applications from a target domain (here, VQA) to distributed QPUs.
Further research across the stack from applications to devices is necessary to enable distributed quantum computing with tolerable overheads.

\paragraph{Circuit cutting}
Quantum circuit cutting offers an orthogonal software path to enable distributed quantum computing to hardware modular QPUs. Cutting qubit wires~\cite{peng2020simulating} or quantum gates~\cite{piveteau2023circuit} breaks down a large quantum circuit into multiple smaller subcircuits. Multiple less powerful QPUs run the smaller subcircuits in parallel. Classical computing eventually reconstructs the original circuit output. End-to-end implementations~\cite{tang2021cutqc, tang2022scaleqc} demonstrate cutting up to $200$-qubit benchmarks to distribute onto multiple QPUs available nowadays, which are otherwise intractable for purely quantum computing or classical simulations. The primary challenges in applying quantum circuit cutting at practical scales stem from the demands of classical post-processing: firstly, the output length of an n-qubit circuit is $2^n$, which rapidly strains classical memory and processing time for large quantum circuits; secondly, the classical post-processing effort scales exponentially with the number of required cuts, $K$, further constraining the runtime.

Despite its ability to execute previously intractable workloads, there is still a gap between the computational capacity of circuit cutting now and future practical quantum workloads. The first theory proof~\cite{peng2020simulating} proposed an exponential classical reconstruction cost. Instead, subsequent works~\cite{tang2022scaleqc} drastically reduce the overhead to sub-exponential but remain prohibitive in the worst case. Filling the gap calls for further developments in classical reconstruction techniques, algorithm-backend co-design, and quantum computing cloud backend system adaptations~\cite{serverless}.

In the next two paragraphs, we shift attention to elements beyond QPUs, addressing key developments in memory and qubit-state storage. We delve into QRAM and QROM, emphasizing how compiler strategies like the Gray Code can enhance memory efficiency. Further, we examine multimode cavity systems, an innovation in quantum memory, providing effective storage solutions for qubit states while also introducing distinctive integration and scalability challenges. How these memory elements are to be eventually integrated with a to-be-developed compiler to enable efficient quantum computing is an open area of research.

The QRAM and QROM memory primitives, discussed earlier in Sec.~\ref{ssec:classical-data-loading}, can be optimally implemented with appropriate compiler techniques. As one example, consider the Gray Code, a sequence of the $2^n$ n-bit bitstrings such that only 1 bit changes between sequential neighbors, for instance, 000, 001, 011, 010, 110, 111, 101, 100. Integrating awareness of the Gray Code in a compiler can significantly reduce gate counts for memory implementations. Additional compilation techniques for these memory primitives, including routing and mapping, are discussed in prior work~\cite{di2021improving, xu2023systems,wang2023fundamental}.

Multimode cavities with tens of modes and photon lifetimes on the order of milliseconds pose as efficient mediums to store multiple qubit states. When used as quantum memory or registers, they can allow for effectively extending the lifetimes of qubits. These cavities can be coupled with one physical qubit, to allow for swapping of a qubit state in and out of the cavity to perform multi-qubit operations with other qubits. 
While integrating multimode cavities reduces swap-distance to enable multi-qubit gates for a larger number of qubits, it also imposes serialization of swaps of qubit states from a single cavity posing additional compilation challenges. In terms of physical integration of these cavities to superconducting circuits, there remain challenges in terms of reducing 3D-cavity size or utilizing 2D resonators while maintaining long lifetimes of stored qubits~\cite{chakram2021seamless}.

\subsubsection{Circuit optimization}
\label{sec:Circ}
\paragraph{Sequences of Pauli rotations} Most standard implementations of Hamiltonian propagators rely on the implementation of a sequence of Pauli rotations, often coming from a Trotter expansion.
In the near-term setting, a sequence $R_{P_k}(\theta_k)\cdots R_{P_1}(\theta_1)$ is usually implemented via an alternating sequence of Clifford circuits and single-qubit rotations.
This prompts the natural problem of finding the best possible sequence of interleaved Clifford circuits and single-qubit layers implementing some goal sequence of Pauli rotations.
In some cases, the ordering of the rotations may also be relaxed, leading to even further optimizations.
Some work already tackles this synthesis problem both for all-to-all architectures and for restricted coupling maps~\cite{childs2018toward,nam2019low,nam2020trappedion,cowtan2020generic,wang2021resource, li2021software, martiel2022architecture,wang2023ever}.
This problem is tightly related to the \emph{parity network synthesis problem} introduced in~\cite{amy2018controlled}, for which many architecture-aware heuristics have been developed~\cite{vandaele2022phase,meijer2020architecture}. It is still an open question to formalize the notion of Pauli networks in the presence of ancillas and measurements. 

\begin{figure*}
    \includegraphics[width=0.45\textwidth]{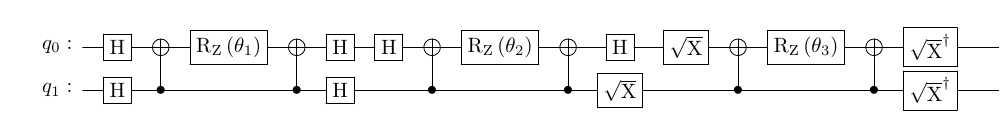}\\
    \includegraphics[width=0.45\textwidth]{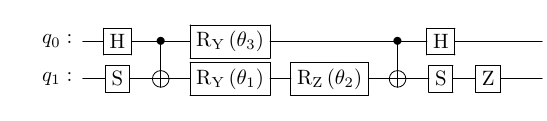}
    \caption{Two circuits implementing the same sequence $R_{\operatorname{YY}}(\theta_3)\cdot R_{\operatorname{XZ}}(\theta_2)\cdot R_{\operatorname{XX}}(\theta_1)$.}
    \label{fig:enter-label}
\end{figure*}

\paragraph{Feedforward} One of the key emerging tools for efficient compilation is dynamic circuits that make mid-circuit measurements that feedforward into subsequent operations. While in principle any quantum circuit can be implemented without midcircuit measurements, due to the \textit{principle of deferred measurement}, there are several scenarios where the dynamic approach significantly lowers circuit depth or circumvents topology constraints. For example, controlled rotation gates can be implemented with a smaller number of non-Clifford gates using feedforward~\cite{jones87novel,nam2020approximate}, which are beneficial for optimizing Heisenberg hamiltonian simulations~\cite{nam2019low}. Another example may be that, while a $n$-qubit GHZ state requires $O(n)$ layers with linear qubit connectivity (and $O(\log n)$ layers with all-to-all connectivity), the state can be produced in constant $O(1)$ depth with feedforward midcircuit measurements. More generally, any Clifford circuit can also be implemented in constant depth with midcircuit measurements on linear nearest-neighbor connectivity. In the context of materials science applications, this is useful for primitives such as the CNOT ladder used in Hamiltonian simulation. Provided with the already identified utility of feedforward in optimizing quantum circuits, further developments in the use of feedforward to optimize additional elements of quantum circuits are anticipated.

\paragraph{Pre-fault tolerant vs. Fault-tolerant} The central task of a compiler is to translate a human-readable code to a machine code that can then be executed by a computer using its native operations. As such, the compiler plays a critical role in both the accessibility of the software development environment for the end users and the efficiency of the machine code that faithfully implements the code that the users want to execute using the least amount of computational resources. Specifically to the latter, an optimizing compiler can be developed, which takes into account various computational operations, possibly having varying cost~\cite{nam2018automated,nam2019low,wang2021resource}. An example applicable to quantum computing may be in the context of the pre-fault tolerant vs. fault-tolerant quantum computers, where the former typically enjoys faster and higher-fidelity single-qubit operations compared to entangling operations, whereas the latter admits significantly less expensive implementations of Clifford operations over non-Clifford counterparts, e.g., single-qubit T gates. A powerful compiler would then enable an end user to develop quantum software at ease -- hopefully in most cases independent of the backend details -- while behind-the-scenes custom-optimizing for different backend quantum computer hardware, including the availability of fault tolerance.

\paragraph{Approximate compilation} Oftentimes, the freedom to implement target unitaries only \textit{approximately} is a powerful tool that can lead to significant cost reductions. For example, this technique has enabled improvements in quantum volume~\cite{cross2019validating}, qudit SWAP gates~\cite{campbell2023superstaq}, bosonic interactions~\cite{shaw2020quantum,kan2021lattice,kan2022simulating}, and even the leading proposal for Shor's Algorithm~\cite{gidney2021factor}, in which exact addition is replaced by approximate addition. Indeed, the widely-used quantum Fourier transform is also implemented approximately in practice~\cite{coppersmith2002approximate,nam2018automated}, and in most quantum simulation algorithms an integral component is implementing a time evolution operator approximately. Beyond approximating unitaries, approximate compilation has also been employed in conjunction with knowledge of the input state, where one quantum circuit may be replaced with another as long as their actions on the input state is similar (even if the actions on some other states might be different). Given the important role that approximate methods play in classical computing -- consider for example numerical integrators or polynomial-time approximation scheme for NP-hard or \texttt{\#}P-complete problems -- fruitful research outcomes that leverage approximate compilation, broadly defined, are expected. Adding to the flavors of approximate compilation research works mentioned earlier, possible, non-limiting research avenues may include (c.f.~\cite{campbell2019applying,pabst2022quantum,shaydulin2023evidence,dalzell2023quantum} and references therein) continuing to quantify existing approximate methods in classical computing for a speed up (e.g., Grover-like or potentially more) or finding entirely novel approximation schemes for those problems known to be hard to solve for quantum computers (e.g., QMA-complete).

\subsection{Real-time classical processing of quantum information (within coherence time)}
\label{realtimeQuantumClassical}

In this model of hybrid computation, the classical part is embedded or coupled with the quantum program and runs simultaneously with the quantum operations within the coherence time.
This can significantly reduce the data-movement between the classical and quantum processors while providing the desired flexibility and short timing in controlling the quantum state.
With mid-circuit measurement~\cite{corcoles2021exploiting, govia2022randomized}, classical computation can use previous quantum operation outcomes to profile the quantum execution status, predict and even adjust the quantum operations for future iterations or steps.
Some examples are \gls{qec}~\cite{brown2023advances}, progressive projection filtering~\cite{ge2019faster, stetcu2023projection} and random walk phase estimation~\cite{granade2022using}.
The key point is that the classical computation has to be performed while the quantum state is still coherent.
Given the short coherence time for near-term devices, constraints on the latency of the classical computation can be stringent, possibly demanding a strongly coupled classical hardware accelerator -- e.g., \gls{fpga}, \gls{gpu}, or \gls{asic} -- physically in place where classical control is happening.

\subsection{Error mitigation}

The output of quantum devices is inevitably altered by noise throughout a computation. While QEC can theoretically be used to remove error from any computation once error rates of the physical devices are pushed below the fault-tolerance threshold, in the near term various techniques can be used to reduce the effect of noise on computation, generally referred to as \gls{qem}~\cite{kandala2019errormitigation,maksymov2023enhancing,cai2023quantum, stein2023q}. When a noisy quantum device is used to compute expectation values of quantum observables, this in general results in a biased estimator. The goal of QEM is to minimize the bias and variance in expectation values from several different runs of the noisy quantum circuit using classical post-processing.

One method that can in principle recover the ideal noiseless expectation values is \gls{pec}~\cite{vanderberg2023probabilistic}.
PEC is based on the inversion of a learned noise model of the device through sampling from a distribution of noisy circuits related to the noise model.
One can prove that the bias of the resulting estimator vanishes as the quality of the learned noise model improves (provided the noise does not change between the noise learning and the actual experiment), but the sampling overhead is exponential in the size of the circuit, with the base of the exponential being related to the noise rate.
While PEC becomes challenging for larger experiments, various techniques can mitigate its cost such as the use of tensor-network-based post-processing \cite{filippov2023scalable} instead of the circuit sampling, which reduces the sampling overhead quadratically, at the cost of more classical computational power. Moreover, the sampling overhead can be contained by improving the accuracy in estimating the noise factors.

An alternative approach to QEM is called \gls{zne}. ZNE is based on the collection of noisy expectation values for various amplified values of the noise parameter.
The expectation value at the zero-noise level is then extrapolated using a linear, polynomial, or exponential fit.
ZNE can be implemented using analog (relying on pulse stretching~\cite{kim2023scalable}) or digital amplification methods (such as sub-circuit repetition~\cite{shehab2019toward,giurgicatiron2020digital,majumdar2023best}) that do not require a precise characterization of the noise, or using probabilistic error amplification (PEA) \cite{kim2023evidence}, which achieves the desired noise factor by randomly drawing circuits from a distribution implementing a rescaled Pauli noise model learned from the device.
ZNE alleviates sampling overhead of PEC, however does not guarantee an unbiased estimator.
Its success depends on a proper choice of the fitting method (methods for systematically choosing the most appropriate one have been proposed, e.g.,~\cite{majumdar2023best}), amplification factors for the noise (e.g., too large amplification will destroy the signal, too small amplification will make the extrapolation less precise), as well as the amplification method itself. Recently, ZNE has also been extended to quantum error correction~\cite{wahl2023zero}.

One more technique to boost quantum fidelity through post-processing is Quancorde~\cite{ravi2022boosting}, which uses Clifford ``canary circuits'' (which are classically simulable but also resemble the target application structure and thus suffer similar structural noise impact) to order an ensemble of devices or qubits/mappings approximately along the direction of increasing fidelity of the target application.
One then estimates the correlation of measurement outcome probabilities with this ordering and uses this correlation to weight the noisy probability distribution; correct measurement outcomes are expected to have a higher correlation with the ensemble order, and thus their probabilities are boosted, while those of incorrect outcomes are suppressed.

QEM has become an integral part of quantum simulations with near-term devices and, together with the constant improvement of the hardware, has allowed for impressive experiments reaching the boundary of what is computable using brute-force classical calculations~\cite{kim2023evidence}.
It will enable us to push the capabilities of the upcoming generations of hardware up to the transition to fault-tolerance, where it can still complement QEC for various use cases.
One potential avenue towards increasing the effectiveness of QEM is to develop either applications that are well-adapted to certain types of error mitigation, or mitigation techniques that exploit information about the specific application.

\section{Workload management\label{sec:Queuing}}

\subsection{Overview\label{sec:OverviewQueuing}}

We cover here the different aspects of the management of heterogeneous workloads involving quantum and high-performance classical resources. Whereas both technologies have an extensive set of functional tools for resource access and management, scheduling, and data handling, there is still need for substantial work towards extending the elasticity of these tools to operate in a hybrid quantum-classical environment and to cater for the needs of the different user profiles. 

We consider and discuss the specificities of different access models, with classical and quantum resources operating in co-located environments or remotely, and explore the impact that different integration timescales, given the disparity in speed and bandwidth between quantum and classical, have on such operational models.

Finally, we describe the current state of the middleware layer for quantum and classical integrations across many different vendors in the industry and give some brief overview of their challenges and involved complexities.

Computational workloads can be categorized broadly into two classes:
(1) trivial (or ``embarrassingly parallel'') tasks, where each task executes independently of all others using its own set of dedicated computational resources;
and (2) non-trivial interoperating sets of tasks, potentially sharing an allocated set of computational resources, where the outcome of a subset of tasks can affect the behavior of another. For HPC systems to handle the various requirements of such workloads, sophisticated \gls{wms} and job schedulers enable efficient resource allocation, initiate and monitor workload execution, and govern a queue of pending work requesting resources.

In QCSC systems, computational workloads can also be defined based on Quantum-HPC integration types, such as HPC-for-Quantum, Quantum-about-HPC, Quantum-in-HPC scenarios, involving interplay between quantum and classical tasks. Consistency in Quantum-x-HPC or HPC-x-Quantum seems desirable and possible. Quantum-about-HPC workload involves quantum tasks that require HPC resources for pre- and post-processing. In this integration type, the quantum tasks are independent of the HPC tasks, and the HPC resources are used to support the quantum tasks. Quantum-in-HPC workload involves HPC tasks that require quantum resources for acceleration, where the quantum tasks are tightly coupled with the HPC tasks, and the HPC resources are used to support the quantum tasks.

Finally, HPC-for-Quantum workload involves tight integration with quantum resources in real-time, within the coherence time of the quantum system \ref{realtimeQuantumClassical}. The classical resources are used to support the quantum tasks, and the classical tasks are tightly coupled with the quantum tasks~\cite{saurabh2023conceptual}. Circuits with classical control often referred as \textit{dynamic circuits}.

As discussed above, integrating quantum systems into classical HPC facilities will add complexity to managing and scheduling quantum-based workloads due to the disparity between quantum and classical processing times, sometimes requiring dynamically allocating and deallocating resources within and among jobs.

In this section, we discuss in detail challenges related to classical-quantum workload management, job scheduling, queuing, and operation modes of these jobs.
We review these concepts and current solutions in the classical setting, and give recommendations for
on-premises and cloud-based integration of quantum computational resources with HPC systems.

We differentiate between (i) on-premises integration, which involves directly incorporating quantum computational resources within a local HPC system infrastructure, and (ii) cloud-based integration, which entails accessing quantum computational resources remotely through cloud services.
Scenario (i) suits workloads requiring tight integration between quantum and classical resources (e.g., Quantum-in-HPC and HPC-for-Quantum workloads). Further, it is ideal for sensitive or proprietary workloads where data security and privacy are paramount, as the data does not leave the premises. Scenario (ii) is beneficial for workloads that can tolerate latency, e.g., Quantum-about-HPC workflows that offload encapsulated parts to a cloud quantum system.

\subsection{Resource management}

This section is structured into two main parts: exploring the resource model, which deals with the allocation and architecture of quantum computing resources, and job scheduling, which is essential for optimizing the performance and efficiency of quantum computing tasks.

\subsubsection{Resource model}

A resource model refers to the framework or method used to manage and allocate computational resources. Specifically, in this section, we explore the distinction and relationship between QPUs and qubits, the placement and location of quantum and classical resources and ownership models.

\paragraph{\textbf{QPU vs qubits}}

The introduction of QPUs marks a significant departure from the conventional approach to traditional HPC scaling. Instead of pursuing the conventional path of continually improving semiconductor processors for greater speed and density, or incorporating specialized processors like GPUs, and optimizing their coordinated use through parallelization, QPUs take a fundamentally different approach. While one might assume that QPUs merely serve as an additional component, they fundamentally alter the way data is represented, employing the language of quantum circuits governed by the principles of quantum mechanics. This constitutes a distinct computational model, diverging from all that has come before it. However, it's essential to note that QPUs are not designed for general-purpose computing tasks. Consequently, practical implementations necessitate the careful orchestration of both classical and quantum computation resources.

Of particular importance for this type of scheduling task is the fact that QPUs operate at very different time scales and bandwidths from standard HPC processes. One could argue that the quantum circuit represents the essence of QPUs as co-computing units with very specific tasks in a supercomputing environment. As the fundamental unit of quantum information, the quantum circuit is a computational routine consisting of coherent quantum operations, including unitary transformations and projective measurements, and concurrent classical computation. As such computational units, quantum circuits may require executions that span inordinate amounts of time for a single computation. For example, consider the fastest quantum hardware that exists today at scale, superconducting qubits, tackling the task of running a circuit with 100 qubits and $10^8$ $T$ gates. Using the surface code, and assuming a physical error rate of $10^{-4}$ and a syndrome cycle time of 1 $\mu$s, a single execution of that circuit, considering both magic state distillation and consumption, can take on the order of tens of minutes and up to several hours, depending on the number of state distillation factories (and therefore physical qubits) used~\cite{litinski2019game}. This circuit execution cannot be preempted, which is important for scheduling purposes. 

The above scenario represents an ideal, still far from what quantum hardware can offer today. Currently, QPUs don’t run in fault tolerant ways, can only tackle much smaller circuits, and must leverage error mitigation solutions that result in large sampling overheads. In contrast with the example above, QPUs today are extremely non-fungible, each of them with its own noise profile, and this affects user preferences and scheduling complexity. 

The introduction of this new type of resource opens a range of models of operation:

\textit{QPU-based Model}: This model resembles the current GPU model, where the entire quantum device is treated as a unified resource unit. Due to its natural inflexibility, this approach has the downside of potential device underuse when jobs only target a subset of the QPU.

\textit{Qubit-Based Model}: In this approach, the qubits themselves are treated as resources, allowing a QPU to be fully occupied simultaneously by a variety of jobs. In near-term devices, with qubits exhibiting different types and amount of noise, and in the presence of limited device connectivity, this approach has shortcomings, as there would be competition among jobs for the best or more suited resources within the QPU.

\textit{Shares Model}: This introduces a pseudo-unit of resource definition where each ``share'' represents a part of the device, a specific period of time, or the circuit size. Essentially, each share signifies a portion of the computational capacity of the resource. Can be treated as sub-category or QPU-based model \cite{qiskitRuntimeFairShareAdmin}.

\paragraph{\textbf{Quantum resource location.}}

In the landscape of HPC, the location of computing resources determines the efficiency, performance, and accessibility of computational tasks. Two primary modes of resource location, co-located and remote resources, offer distinct advantages and considerations for HPC facilities and users.

\textit{Co-located resources} refer to computational resources physically situated within the same facility or data center as the user or workload management system. There are several advantages of having co-located quantum and classical resources. Minimal network latency and high-speed data transfer capabilities, which is especially advantageous for applications where access to large amounts of (low-level) data (like counts) is required. Full control and management of co-located resources, enabling customization and configuration based on specific requirements. Data security. Co-located resources location implements on-premises integration type (i) \ref{sec:OverviewQueuing}.

\textit{Remote quantum resources}, on the other hand, are located at a physical distance from HPC classical infrastructure, often in off-site data centers, cloud environments (cloud model integration type (ii) \ref{sec:OverviewQueuing}), or partner organizations. Remote resources extend the reach and capabilities of HPC systems, but they come with unique characteristics. Elasticity is one of them, allowing for scaling up and down quantum resources based on need. Another characteristic is sharing. At this moment, it is a case when a quantum processor is shared among multiple users, which drives down the cost of execution, but also introduces variation in performance due to resource contention. Network overhead can impact data transfer rates and communication latency, especially in applications with large loads. 

\paragraph{\textbf{Quantum resource ownership.}} 
\label{sec:WM_ResourceOwnership}

High-performance computing facilities generally serve large numbers of users with varying computational resource requirements for their computational workloads (jobs) \ref{sec:DataPreprocessing}.
Users are part of user groups and projects, which have assigned queues.
A project has some resource allocation quota (e.g., $10^{6}$ node hours), whereas queues are defined by criteria such as priority and the amount of computational resources available per job (e.g., $n$ CPU cores, $m$ GPUs) to a user of that queue.
At the level of user groups and projects, the integration of quantum information processing devices with classical HPC computational resources is straightforward: access to a quantum resource could be granted to a given user assigned to a \texttt{qpu\_user} group with project quota $Q$ specifying the number of ``quantum node'' hours.

\begin{figure*}
    \centering
     \includegraphics[width=0.85\textwidth]{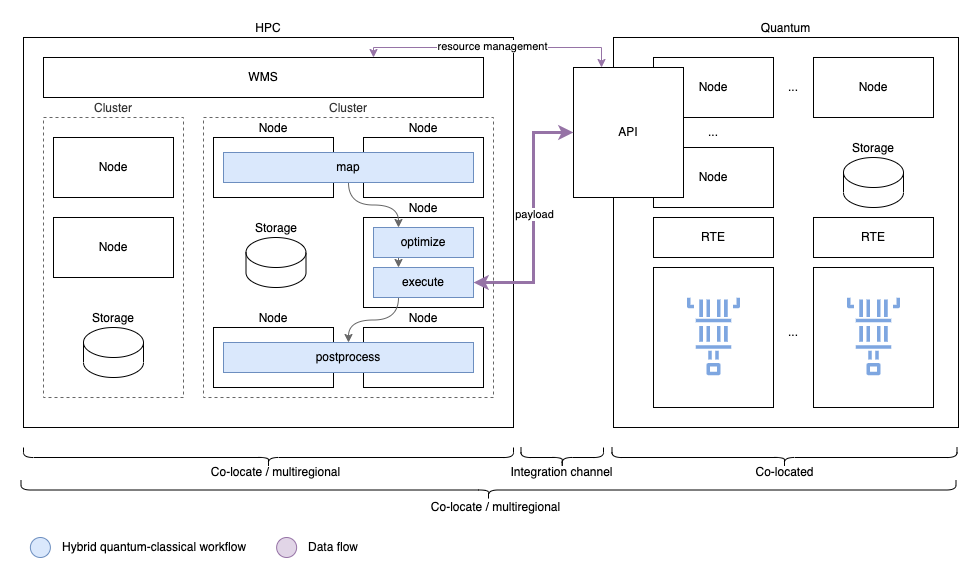}
    \caption{Quantum-centric supercomputing integration overview. Integration channel can be tight or loosely coupled, which will affect throughput and latency of transmitted payload. Real-time compute is possible on co-located quantum side; near-time compute can be executed on quantum and HPC sides; long-time compute should happen on HPC side. Depending on implementation WMS can schedule tasks on quantum side through API or direct access to classical nodes of quantum side.}

    \label{figure:HPCQuantumIntegration}
\end{figure*}

\begin{figure}
    \centering
     \includegraphics[width=\columnwidth]{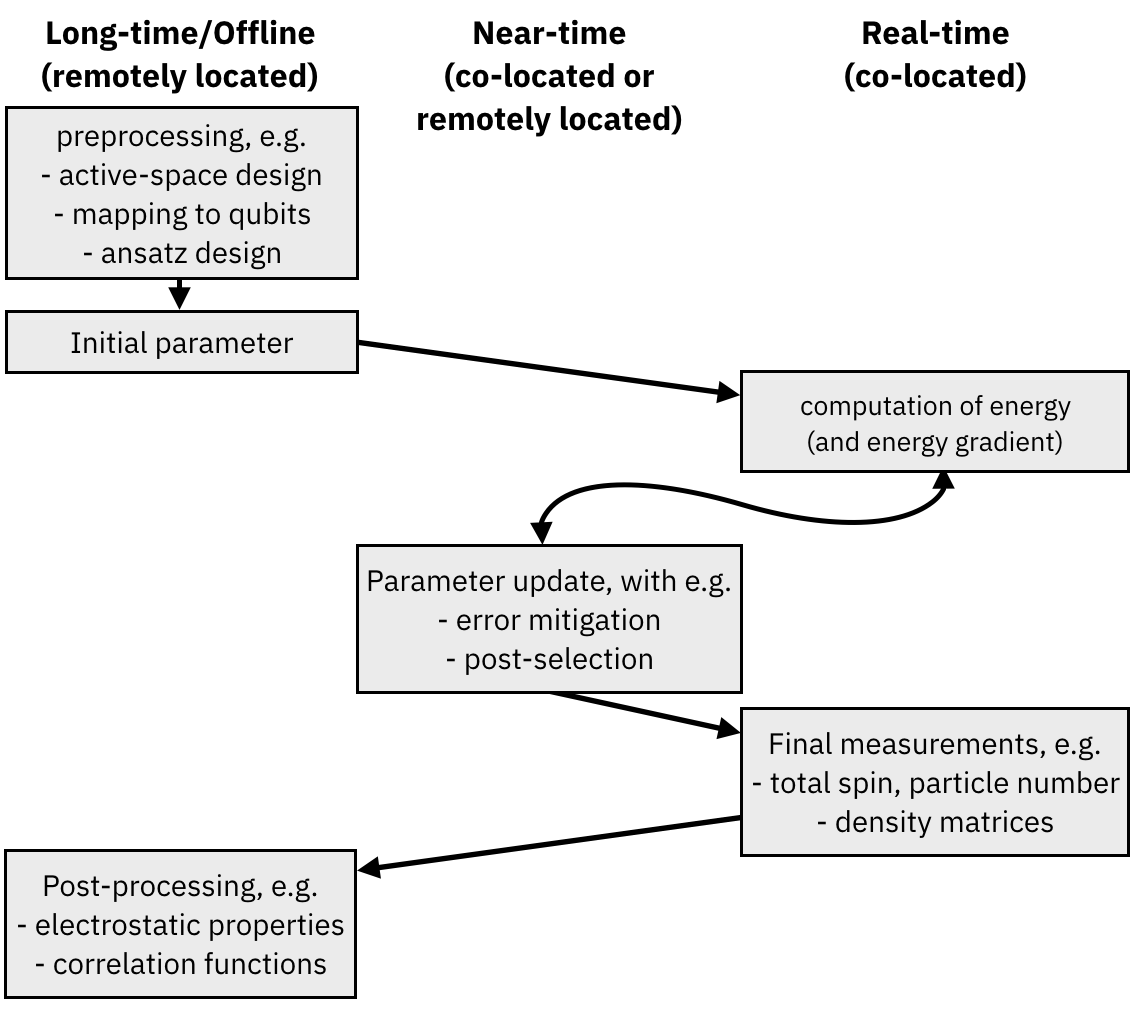}
    \caption{Integration between classical (HPC) and quantum computing resources exemplified by the variational quantum eigensolver. The steps of the calculation are represented by gray blocks, connected by arrows describing the flow of operations and arranged left/center/right for operation that require ``long-time/near-time/real-time'' interaction between HPC and quantum computers (see also main text). 
    }
    \label{figure:HPCQuantumIntegrationExample}
\end{figure}

\subsubsection{Job Scheduling}

When introducing quantum resources, one could envisage \textit{real-time} quantum nodes, where classical computations are performed within the coherence time of the qubits (e.g., for error correction studies), and \textit{near-time} quantum nodes, where higher latency classical communication suffices (e.g., variational algorithms).

Common frameworks in use today within HPC facilities and cloud platforms include SLURM~\cite{yoo2003slurm}, TORQUE~\cite{torqueengine}, Altair Grid Engine~\cite{altairgridengine}, and IBM Platform LSF~\cite{ibmlsf}, which provide highly scalable control over heterogeneous computer clusters and jobs submitted to them.
To execute a workload, a user configures their job using a \textit{submission script} based on the syntax of the associated scheduler.
Submission scripts essentially define a job:
project name, resource requirements (e.g., node count, CPU count, memory size, and wall clock limit), tasks to execute, input and output data paths, etc.
Note the governance of an HPC system (see Sec.~\ref{sec:queuing:gov}) could require all input data be readily available, e.g., on a scratch area accessible only by the user.
These scripts are submitted from a login (head) node to the WMS or job scheduler, where they are queued and finally executed on one or more compute nodes.

A queued user job is executed in the order of the project's or queue's priority as well as resource availability;
for instance, a higher-priority job whose resource requirements are met will be executed before a lower-priority job.
Resources can be exclusive or shared, so if \texttt{job\_x} does not need all cores on the node(s) to which it is assigned, \texttt{job\_y} could be assigned to the remaining cores.
However, resources allocated to a given job are fixed and, therefore, remain allocated for the duration
of the job's lifetime (and thus charging against the user/project quota), regardless of whether or not they are performing useful work.

Extensions to existing job schedulers could include quantum-specific configuration settings, e.g.,
number of circuits to be executed and depths, from which, along with architecture-dependent factors such as data loading, gate application, and readout times, runtime estimates can be made.
However, while separating classical and quantum configuration spaces may be amenable to near-term
applications (i.e., wherein classical and quantum operate cooperatively but have loose latency requirements, as in VQE),
mid-term quantum-classical applications requiring lower latencies likely will necessitate tighter integration between the two types of resources.

Additionally, quantum-centric supercomputing workload scheduling must also consider the coupling between the quantum and classical tasks and assign and co-locate tasks to the resources accordingly. Different quantum-classical coupling types represent workloads with unique requirements and challenges, such as the lack of a unified standard for accessing QPUs and expressing hybrid workloads, and the need for parallelization of quantum workloads across nodes, cores, and accelerators. For example, classical tasks tightly coupled to quantum tasks, e.g., for the HPC-for-Quantum integration type, must be co-allocated and assigned to nearby QPUs and GPUs. Additionally, scheduling of  QPU jobs must be prioritized to ensure the optimal utilization of the QPUs while minimizing the time-to-solution and energy-to-solution~\cite{saurabh2023conceptual} (section \ref{queueing}).

\paragraph{\textbf{Queueing}}
\label{queueing}

HPC environments typically offer shared resources that are made available to a large user base. To ensure the equitable and efficient distribution of computing resources, HPC systems implement queuing mechanisms. 

One of the aspects of execution management in HPC is the presence of multiple queues, each configured with specific priorities and constraints. These queues are designed to categorize and prioritize computational tasks based on the specific needs and requirements of the users or projects.

Queues can have resource constraints. Resource constraints may include limitations on the number of nodes, processors, memory, or specific software specifications available for jobs in each queue. These constraints help balance the allocation of resources across different user groups and projects while ensuring that the computing environment remains stable and efficient.  As QPUs are "rare" resources, it is beneficial to increase the priority of the "quantum" queue while maintaining constraints on execution time and classical resources. It might be also beneficial to increase the priority of classical jobs connected to the quantum queue.

Because QPUs are a new and currently limited resource type, it is beneficial to specify a separate queue for quantum jobs.

Furthermore, as quantum computing technology evolves and the number of available QPUs increases, HPC administrators will need to adapt the queuing system to accommodate this growth. This might involve not only adjusting the priority of the quantum queue but also dynamically redistributing workloads to make optimal use of the expanding quantum infrastructure. Balancing the allocation of quantum and classical resources will remain a complex task, requiring constant monitoring and fine-tuning to ensure efficient utilization of the entire HPC facility. In this context, the queuing system's flexibility and adaptability will be crucial in meeting the evolving demands of quantum computing within the HPC environment.

\textit{Cloud queues}. There is a practice nowadays of cloud vendors are providing access to quantum services that are executing quantum payload. In order to provide fair allocation of resources, cloud providers are implementing their own queuing mechanisms behind cloud endpoints. It is important to understand that cloud vendors have different job queuing and priority implementations, and all might employ a shared resource ownership model \ref{sec:WM_ResourceOwnership}. 

\textit{Qiskit Runtime service fair share queue}. Access to quantum resources are divided between hubs, groups and projects.
A \textit{Hub} represents the top level of an organization such as an academic, industry, or research partner.
A \textit{Group} represents a mid-level structure to which access shares can be allocated by the hub for one or more collections of users (projects).
A \textit{Project} represents the base-level construct to which shares are allocated from the overarching group, and to which users are directly assigned.

For each group and project, the duration of the scheduling period is used to convert shares into an equivalent amount of time that an instance would receive under ideal conditions. The ratio between the time used and the shares equivalent time is used as the basis for scheduling jobs.

Many HPC job scheduling systems employ a so-called "fair-share algorithm" designed to distribute computational resources among multiple jobs in a manner that is considered fair. The definition of "fair" can vary significantly depending on the context and the specific requirements of the system. The Qiskit Runtime fair-share algorithm takes into consideration how shares of resources are distributed across groups and projects to determine job prioritization \cite{qiskitRuntimeFairShare}.

\paragraph{\textbf{Cloud-bursting model}}

Similar to a classical cloud bursting \cite{bicer2011cloudbursting}, we propose a Quantum cloud bursting model in which an HPC facility offloads quantum computation to cloud vendors when there are insufficient quantum resources within its local infrastructure.

As with any approach, there are pros and cons of quantum cloud bursting. 
Quantum resources may be expensive to maintain locally. In such cases, offloading workloads to cloud vendors with readily available QPUs can significantly decrease maintenance. 
Certain applications require proprietary or specialized software. Acquiring and managing licenses for such software can be both costly and time-consuming. Cloud bursting allows organizations to leverage cloud vendors' pre-configured environments, providing access to the required software without local setup and licensing.
It also presents challenges related to data transfer and increased requirements for security.

\begin{table*}[t]
    \centering
    \begin{tabularx}{\textwidth}{@{}>{\bfseries}l*{3}{X}@{}}
        \toprule
        Endpoint & Request Body & Response & Description \\
        \cmidrule(r){1-4}
        \it{GET allocations/} 
            & \makecell[l]{allocation\_ids: list(str) \\  fields: list(str)}
            & allocations: list(dict)
            & Fetch resource allocations listed by id. \\
        \it{POST allocations/} 
            & specification: dict
            & allocation\_id: str
            & This allocates a set of resources from a resource request specification. \\
        \it{DEL allocations/} 
            & allocation\_id: str     
            & status: bool
            & Destroy a resource allocation and return resources to the pool of available resources. \\
        \it{GET resources/} 
            & \makecell[l]{specification: dict \\  fields: list(str)}   
            &  resources: list(dict)
            & Fetch the current state of resources on a system.  This should report status of anything that a user may be able to specify for their job placement, including QPU status and software stack status \\
        \it{GET tasks/} 
            & \makecell[l]{task\_ids: list(str) \\  fields: list(str)}
            & tasks: list(dict)
            & If a list of tasks are passed in, provide information on currently running tasks on the system.  If no tasks passed in, provide information on all running tasks.  This may constrain the results for the executing user so that it shows only their own jobs if the user is otherwise unprivileged. \\
        \it{POST tasks/} 
            & \makecell[l]{task\_information: dict \\  allocation\_id: str \\ batch\_id: str}  
            & allocation\_id: str
            & Launch a task. \\
        \it{DEL tasks/} 
            & \makecell[l]{task\_ids: list(str) \\  signal: int} 
            & status: bool
            & Send a signal to tasks for termination. \\
        \bottomrule
    \end{tabularx}
    \caption{Minimal quantum cloud bursting API specification.}
    \label{table:quantum_cloud_bursting}
\end{table*}

Figure \ref{figure:HPCQuantumIntegration} is showing how quantum cloud bursting model can be implemented via integration of Quantum cloud API and workload management system. Table \ref{table:quantum_cloud_bursting} shows a minimal API specification that cloud vendors should implement to unlock better integration of quantum and classical resources.

An illustration is given in Figure \ref{figure:HPCQuantumIntegrationExample}, using the variational quantum eigensolver as an example. Classical preprocessing (including mapping to qubits, choice of the variational ansatz, and parameter initialization) is carried out on HPC resources located remotely from the quantum computer (``long-time'' interaction). The execution of quantum circuits may require the use of real-time communication between classical and quantum units through real-time electronics of RTE (see also Figure \ref{figure:HPCQuantumIntegration}). Real-time interaction is required e.g. to implement dynamical circuits used for symmetry verification/enforcing, monitored quantum dynamics, and quantum error correction.
The update of variational parameters, which requires the calculation of energies and energy gradients typically in conjunction with readout/gate error mitigation and post-selection operations, may be carried out on HPC resources located remotely from the quantum computer or co-located with it (``near-time'' interaction).

\paragraph{Quantum execution runtime prediction}

Quantum execution runtime, determined by circuit complexity (number and type of gates, layers), can be predicted using historical data~\cite{ravi2022adaptive}. Despite potential overhead from error mitigation and circuit transpilation, accurate predictions are achievable. However, dynamic circuits with their non-deterministic workload structures pose challenges. 

Runtime prediction is crucial for effective workload management, allowing quantum resources to be optimally utilized. Incorporating runtime predictions into scheduling enhances priorities and overall workload efficiency, showcasing the valuable role of prediction in quantum computing.

Orchestrating workflows that involve the integration of both HPC and quantum systems presents a set of unique challenges, particularly in the design and management of the payload, which encapsulates the instructions and data for these complex tasks. One of the primary challenges lies in achieving interoperability between classical and quantum computing environments. Quantum algorithms and classical algorithms have different computing models and requirements. Ensuring that the payload can effectively bridge this gap, translating between classical and quantum instructions and data formats, is a non-trivial task. This translation layer must not only accommodate the distinct computational paradigms but also manage the conversion of data types and formats, all while maintaining computational efficiency.

Optimizing resource allocation is another significant challenge in the payload for orchestrating quantum-centric supercomputing workflows. Balancing the computational load between HPC and quantum systems and efficiently using both types of resources is a complex task. It involves determining when to offload tasks to the quantum processor and when to keep them on classical hardware based on factors such as quantum system availability, workload characteristics, and quantum hardware constraints. Additionally, managing the data movement between these systems while minimizing latency and maximizing throughput adds further complexity to the payload design. To address these challenges, advanced workload scheduling and resource management strategies, as this paper mentioned, are required, leveraging real-time monitoring and adaptive decision-making to ensure that HPC and quantum systems work in harmony to deliver optimal performance for complex scientific and computational tasks.

\subsection{Middleware}

Quantum computing is evolving towards modular architectures comprising multiple QPUs coupled to classical computing nodes. Moreover, emerging applications that aim to benefit from quantum acceleration involve significant classical and quantum computational components, such as different phases of data collection and streaming, parameter tuning, simulation, and analysis. Typically, such quantum and classical coupling depend upon interaction between application components within and outside the coherence time of the quantum system, i.e., on tight- and medium-coupling across multiple quantum computing units (QPUs) or loose-coupling in a workflow application~\cite{saurabh2023conceptual} (see section~\ref{sec:OverviewQueuing}). Hence, there is a need to design middleware systems that can facilitate the efficient understanding and interplay between quantum and classical components in an end-to-end workflow. Regarding quantum-centric supercomputing systems, middleware should also leverage well-established high-performance computing abstractions and must be compatible with existing HPC software stacks for managing hybrid workloads, tasks, and resources to integrate quantum computing into HPC systems seamlessly. 

This section initially describes existing hybrid platforms, runtime management, and quantum workflow frameworks instrumental in realizing quantum-centric supercomputing middleware systems. Further, we envision a conceptual middleware built upon the quantum-centric supercomputing integration limitations of existing middleware systems. 

\paragraph{\bf Existing middleware:} Recently, several tools and frameworks emerged to manage quantum and classical tasks and resources efficiently. XACC is a system-level approach to integration of quantum and conventional processing, whereby  quantum kernels are offloaded using hardware-agnostic constructs \cite{McCaskey_2020}. QCOR is a high-level language specification and associated tooling for hybrid programming \cite{10.1145/3380964,10.1145/3462670}.  NVIDIA developed the CUDA Quantum~\cite{cuda_quantum} to integrate classical and quantum computing devices. It supports programming hybrid quantum-classical applications with optimized control and communication between quantum processors and classical tasks.  It is also integrated with CUDA libraries for accelerating and scaling quantum simulations and classical HPC computations across distributed multi-node, and multi-GPU architectures. 

There also exists hybrid quantum-classical runtime systems, such as Qiskit runtime~\cite{cross2018ibm} and Braket Jobs~\cite{braket-jobs-2021}, that can be integrated into quantum software frameworks (e.g., Pennylane~\cite{bergholm2018pennylane})  and platforms. The Qiskit runtime system provides primitives for defining, scheduling, and optimizing near-time quantum-classical workloads and enables their execution both synchronously and asynchronously. On the contrary, Amazon Braket Jobs are limited to a proprietary cloud environment and execute quantum-classical tasks as braket jobs with on-demand priority access to QPUs on a variety of quantum hardware (e.g., IonQ, Rigetti, etc.). There are also emerging middleware software stacks; for instance, QCG and QCG-PilotJob deliver highly efficient services and access tools for remote job management in large-scale computing environments, including HPC and quantum environments, by adding and extending capabilities to existing queuing systems (e.g. SLURM) \cite{qcg_quantum} \cite{qcg_vvuq}.

To support end-to-end quantum workflows and the interplay between quantum and classical components, several workflow orchestration frameworks exist, such as Orquestra~\cite{zapata2021orchestra}, Covalent~\cite{covalent2023} and Tierkreis~\cite{https://doi.org/10.48550/arxiv.2211.02350}. Orquestra and Covalent frameworks provide a quantum development and execution environment for quantum and quantum-inspired workflows with support for deployment, scaling, and parallelizing workflows on classical and quantum processors. Tierkreis instead utilizes a data flow-based programming model to orchestrate hybrid quantum workflows and provides a runtime environment allowing for concurrent and asynchronous execution. Recently, another middleware tool called IBM Quantum Serverless~\cite{quantum_serverless_qsw} emerged, which combined with Qiskit Patterns, a mechanism to build quantum workflows at scale, allow users to execute hybrid workloads on cloud or on-premise infrastructure.  

\paragraph{\bf Conceptual gap and vision:}

The intersection of quantum and HPC presents unique resource management and scheduling challenges, heterogeneity in handling different quantum hardware types, and thus, a high application development complexity. Effective resource management in a quantum-centric supercomputing middleware system requires sophisticated resource allocation mechanisms and scheduling algorithms that can dynamically allocate quantum and classical resources and manage complex workloads. This allocation must consider quantum processors' distinctive computational capabilities and limitations. Different quantum technologies, such as superconducting qubits or trapped ions, have varied characteristics and requirements. Thus, the complexity of developing applications for such systems is substantial.

Despite significant advancements in integrating quantum and classical computing, current tools and frameworks have limitations. One major constraint is their nature as point solutions, often tailored to specific quantum providers or hardware. This specificity can limit their applicability and flexibility, making it challenging to adapt these solutions to different quantum computing environments or hardware platforms. Furthermore, many of these tools focus on narrow types of workloads. For instance, some are optimized for near-time variational algorithms. Others are designed for workflows, which, while effective in those contexts, may not efficiently handle real-time and near-time applications. As a result, tools and approaches are fragmented, posing challenges for developers seeking a unified, versatile platform for diverse quantum-centric supercomputing tasks.

\subsection{Governance}
\label{sec:queuing:gov}

\subsubsection{Security}

Similar to classical infrastructure, some of the security topics that must be addressed when developing hybrid quantum classical environments:
\begin{itemize}
    \item {Authentication and authorization mechanisms to restrict access to authorized users or entities.}

    \item{ Fine-Grained permission to ensure that users can only access the resources they are entitled to.}

    \item {Data encryption protocols for both data in transit and data at rest. Explore the use of post-quantum cryptography to secure data.}

    \item {Secure data transfer between classical and quantum components in a workflow, preventing data interception or tampering during the transition.}

    \item {Access monitoring to detect and respond to unauthorized access attempts or unusual behaviors.}

    \item {Auditing and logging to track job submissions, resource allocations, and user interactions, aiding in post-incident analysis and compliance.}

    \item {Data privacy of user data and compliance with data protection regulations such as GDPR \cite{GDPR2016a} or HIPAA \cite{hipaa}, especially when dealing with sensitive data in hybrid computations.}

    \item {Threat detection to proactively identify and respond to potential security incidents.}
    
    \item {Resource isolation: In the event of a security breach, ensure that affected resources are isolated and investigated to prevent further damage or unauthorized access.}
\end{itemize}

\subsubsection{Policies}

\textbf{Storage Policies:} Policies will define a data lifecycle that incorporates measures for archival, backup, and the secure erasure of data to safeguard intellectual property and sensitive information.

\textit{Storage Quotas:} Storage quotas are allocated to projects, reflecting the anticipated volume of data and the project's duration. Requests for adjustments will be entertained to these quotas in response to significant changes in data needs, subject to a review process and availability of resources.

\textit{Data Lifecycle Management:} Procedures for data storage, archival, backup, and deletion. Data retention periods will be determined  based on the data's significance, sensitivity, and ongoing relevance, ensuring essential data is preserved and non-critical data does not occupy valuable storage indefinitely.

\textbf{Queue Policies:} Ideally, a queue scheduling policy will efficiently prioritize jobs based on estimated runtime, research urgency, and users' historical consumption of resources, while also incorporating backfilling, to minimize idle resources. Regular reviews of queue performance should be undertaken to validate and adjust our prioritization strategies.

\textit{Scheduled Access:} An advance reservation system ideally would be available for projects with foreseeable intensive computing needs, ensuring the availability of critical HPC and quantum resources for time sensitive projects. Such reservation requests would need to provide technical justification and be to subject to a merit review process. 

\textit{Job Time Limits:} Reasonable runtime limits must be set for various job types to ensure equitable system access, as well as providing predictability in scheduling for users waiting for system access.

\textit{Job Interruptions Handling:} Users should be notified in advance of system maintenance where possible. The capacity to pause jobs during system maintenance or unexpected downtime,and resume jobs after would be ideal, minimizing the impact on research continuity and computational efficiency. Users should be encouraged to implement best practices for checkpointing, and saving work.

\textbf{\gls{dua}:} Users are be expected to engage in data use agreements that explicitly outline conditions for access, use, and sharing of data.

\textit{Responsibilities of Data Users:} Users are tasked with upholding data confidentiality and integrity, in compliance with security protocols.

\textit{Data Sharing and Collaboration:} Establish conditions under which data can be shared with external parties, paying special attention to the transfer of data between classical and quantum computing frameworks.

\textit{Termination and Renewal:} The terms for the termination and renewal of DUAs will be clearly communicated. In the event of termination, data will be managed according to pre-established procedures, ensuring the secure disposition or transfer of data, with particular care for sensitive quantum data.

\textit{Breach of Agreement:} Outline the consequences for any breach of DUAs, recognizing the potential for increased severity in the context of a hybrid HPC-Quantum environment.

\textbf{Learning and Resources:} Given that both HPC and quantum computing require specialized knowledge and skills, providing a robust array of training programs and resources encourages both individual and cross-domain competency.

\textbf{User Feedback:} Intelligent incorporation of user feedback ensures effective alignment with user needs and expectations, and aiding in identifying bugs and issues. User feedback provides a window into how quantum and HPC resources are actually being utilized versus initial assumptions. User feedback can also identify an accessibility or security issues, and inform what specific training and educational resources are most needed. Given the pace of evolution of the field of quantum-centric supercomputing, frequent and timely inclusion of user feedback is essential in guiding future development.  A structured approach to collect user feedback involves regular forums, surveys, and user support help tickets to ensure we capture the user experience comprehensively.

\textbf{Compliance and Audit:} Importance of maintaining rigorous compliance with regulatory standards and internal policies. Our protocols will undergo periodic reviews to align with legal and industry benchmarks. Regular internal audits will be conducted to ensure adherence to our established policies and to identify areas for improvement. These audits will also assess the precision of resource allocation and the efficacy of our data management practices.

\section{Programming models\label{sec:ProgrammingModels}}

\subsection{Introduction}

A programming model describes and provides the mechanisms for instructing how a computer system is to perform a computation, based on the execution models of specific computing hardware and programming language.
These can usually be separated because any underlying execution model for a given hardware platform may be made transparent in any given language through calls to functions through abstractions.
For the purposes of this paper, we will adopt this common viewpoint and focus on language-independent programming models supporting developers in several roles working across a layered execution model.
Likewise, we will assume independence of programming paradigm, allowing for its choice to be driven by the details of an individual project and its requirements.

\begin{figure}
    \centering
     \includegraphics[width=\columnwidth]{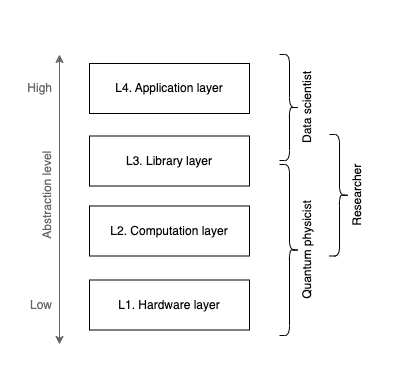}
    \caption{Programming model layers. (L1) Hardware layer: electronics, firmware \ref{sec:MemoryModel}. (L2) Computational layer: circuits \ref{sec:ExecutionModel} \ref{dataTypes}. (L3) Library layer: functions. (L4) Application layer: domain specific applications. 
    }
    \label{figure:ProgrammingModelLayers}
\end{figure}

A programming model for a heterogeneous computing system combining both classical and quantum elements can be described in terms of four layers, illustrated in Fig. \ref{figure:ProgrammingModelLayers}, comprising an application's execution and categories of developers who write code describing computational procedures across these layers.

Working up from the bottom, Layer-1, which we call the Hardware layer, is where hardware-dependent operations occur on specific devices.  On the quantum side, the programming model at Layer-1 provides access to processors and other electronics, software, and firmware which directly control the quantum device, and on the classical side it provides processor-specific instructions according to its \gls{isa}, itself an abstraction of the processor hardware.

Layer-2, which we call the Computation layer, is where low level computational flows and information  are described.  For quantum this means constructing and optimizing circuits which are then mapped onto specific quantum hardware by compilation (and transpilation), operating on quantum information and interfaces to classical information. On the classical side this means functions optimized for specific underlying processors including CPUs (of varying ISAs), GPUs via a manufacturer’s API, or other environments for specialized processors such as inference engines, processors aimed at AI problems, or custom FPGAs.

The Library or Function layer, Layer-3, is where computational capabilities in Layer-2 are abstracted into functions aimed at particular problem domains and the hardware-independent computations supporting them. These are used by application developers without necessarily being concerned with the details of the underlying execution model.

Finally, Layer-4, which we call the Application layer, is where domain-specific application developers construct programs and workflows that perform computations to solve problems relevant to domain experts, expressed in domain-relevant terms, without necessarily being concerned with the detailed mechanisms of those computations. 

\subsection{Related Work}
\label{ssec:related-work}
It's useful and instructive to consider popular existing programming models and how they relate to this layering. Apache Spark~\cite{apache-spark} is a programming model for large-scale data processing whose base programming language is Java; SYCL~\cite{10.1145/3388333.3388649}, CUDA~\cite{Sanders2010:CUDA}, and High Performance ParalleX (HPX)~\cite{heller2023hpx} are programming models developed in idiomatic C/C++, spanning Layer-2 and Layer-3.

There do exist domain-specific programming libraries such as OpenFermion~\cite{McClean2020:OpenFermion}, which facilitate the mapping of electron-structure problems to quantum circuits.  While certainly useful, it is arguable whether this approach represents a complete programming model per~se rather than a suite of helper routines for a specific workflow (module specification $\to$ integral generation $\to$ circuit mapping) at Layer-2.

 SimuQ~\cite{Peng2024:SimuQ} defines a domain-specific language for modeling arbitrary Hamiltonians, compiles these to an abstract \emph{analog} instruction set, then lowers that to a sequence of pulses rather than to a quantum circuit at Layer-2. Programming-model support for the direct expression of Hamiltonians may be beneficial in Layer-3 in defining functions for building materials-science applications.  One challenge, however, is performance scalability of the software stack.  Peng et al.~\cite{Peng2024:SimuQ}\ report SimuQ compilation times that grow rapidly with the number of sites in the Hamiltonian as well as a number of Hamiltonians that fail to map to the underlying quantum computer due to hardware constraints.

\gls{mpi}~\cite{Snir1998:MPI} and \gls{openmp}~\cite{Dagum1998:OpenMP} are open-standard specifications that aim generally for the programming of parallelized computations on heterogeneous classical computing platforms, and are examples operations spanning Layers 2 and 3.

Programming models can be directives-based, decorated, or provide extensions to their associated programming languages. In each case, a compiler capable of translating code written using a given programming model is needed to target specific backend hardware at Layer-1.
A programming model incorporating semantics and execution model for quantum computing systems will naturally extend existing models targeting classical hardware. Importantly, a quantum computing-enabled programming model should provide abstractions such that applications can be developed independently from the underlying computing model or hardware; that is, at Layer-4 they should be agnostic with respect to both classical and quantum hardware. In this way, application developers can write software that can be executed on arbitrary hardware using functions in libraries at Layer-3, with possible customizations in Layer-2 if needed, and a backend compiler to translate the higher-level code (including language extensions, etc.) to the target machine code. 

An initial prototype of such a programming model for the development of quantum computing applications is XACC/qcor~\cite{McCaskey2020:XACC,Hayashi2023}, an extensible language and classical-quantum compilation framework for Rigetti, IBM, and D-Wave quantum systems. XACC follows the co-processor model, akin to SYCL and CUDA, treating the quantum system as an computational accelerator, similar how a general purpose graphics processing unit (GPGPU) computing is used. This work paved the way for NVIDIA's open-source CUDA Quantum platform~\cite{Kim2023:CUDA-quant} which extends their existing CUDA framework, providing support for programming QPUs in addition to CPUs and CUDA GPUs.

\subsection{Execution Model}\label{sec:ExecutionModel}

An execution model describes the behavior of an application and is exposed via a programming model. The hardware, or device, architectures help to define the execution models based on their capabilities and functionalities, and the programming model presents an abstract implementation of the underlying capabilities to enable general (or expected) purpose computation. In classical computing, a well known classification of architectures includes those found in Flynn's taxonomy~\cite{Flynn1966}. In this section, we complement Flynn in the context of quantum computation and how this classification can be incorporated into a programming model for hybrid classical-quantum applications.

Flynn's taxonomy is laid out in two dimensions: the number of instruction streams or programs and the number of data streams. The \gls{sisd} model represents classical sequential machines following the von Neumann architecture. In the immediate term of quantum computing, until QPU partitioning or parallel QPU processing is possible, we foresee that SISD will be the predominant architecture. Programming languages exposing this execution model include C, C++, and Fortran, which are dominant languages within HPC. Parallel quantum computations, wherein, e.g., a large QPU can be partitioned into topologically equivalent subprocessors, \gls{simd} will be beneficial; SIMD capabilities can then be exploited to parallelize quantum computations. A recent popular addition to the SIMD family is GPUs, which manipulate multiple data through the same set of instructions executed across many processors in lock-step, known as \gls{simt}~\cite{Lindholm2008NVIDIATA}.

\Gls{mimd} is generally accepted to represent parallelism. Again, supporting an architecture within a programming model depends on the underlying hardware capabilities which are exposed to the user. Modern languages such as Erlang~\cite{10.1145/1238844}, Go~\cite{Go}, or Python~\cite{pythonWelcomePythonorg}, and recent revisions of C++~\cite{std-execution} present built-in architecture-dependent concurrency models, and parallelism is most often exploited either via extensions to traditional HPC programming models such as CUDA~\cite{CUDA}, OpenMP~\cite{OpenMP_2023}, OpenMPI~\cite{gabriel04:_open_mpi}, or via careful allocation of resources and explicit control flow declaration known as ``workload orchestration’’. Ray~\cite{rayProductionizingScaling} is an example of a parallel workload orchestrator.

In our layered vision of programming models, at Layer-1 (i.e., Hardware Layer), classical hardware shows instances of several kinds of execution models: instruction-level parallel (superscalar) processors, instances of MIMD and GPUs, address SIMD and SIMT; Tensor Processing Units, and Data Processing Units support unloading specific workloads from CPUs, ASICS are examples of programmable technologies for specialized SISD; and novel applications can be explored through the use of reconfigurable hardware such as FPGAs. On the quantum side, general-purpose quantum computers can essentially be considered SIMD machines with a computational memory in which each qubit has both independent and combined processing capabilities.

At Layer-2 (Computation Layer), the hardware architecture is abstracted into the programming model’s execution model. This, in turn, will be reflected by the programming language subject to compilation. Language-specific statements control the underlying hardware through a human-friendly syntax and developer ergonomics, presenting an abstract SISD model. It is through the careful coordination that access to parallelism and other execution models is granted to higher layers. In quantum computing, although there exist programming languages capturing quantum semantics (e.g., OpenQASM 3~\cite{openqasm3}, Q\#~\cite{qsharp}, and Quipper\cite{quipper}), the control flow of quantum circuits, circa dynamic circuits, is simple enough to be captured as data and to be translated into hardware instructions mostly directly.

Layer-3 (Library Layer) takes advantage of Layer-2's execution model libraries to access actual programming models beyond SISD variants. This layer increases the level of abstraction into libraries, frameworks, and \gls{sdk}, introducing concepts agnostic to the hardware execution model, and tools for interfacing with the abstract execution model. A classical example, in AI engineering, TensorFlow~\cite{tensorflow} exploits the concept of a graph, and implements a variant of the data-flow execution model, where operations are executed as soon as the input operands are ready. Tools such as TensorBoard~\cite{TensorBoard2023} allow for the inspection and debugging of graphs. Similarly, on the quantum side, the Qiskit SDK~\cite{qiskit} introduces concepts from quantum computing such as quantum registers, gates, and measurements, and captures the circuit execution model as a \gls{dag}. The circuit and DAG visualizers~\cite{QiskitVisualizationDAGDrawer} allow for the exploration of these quantum program representations.

Finally, Layer-4 makes use of these libraries, frameworks, and SDKs to implement domain-specific applications. Applications at this level act as coordinators, directing the workloads needed to solve specific problems. Within the quantum computing ecosystem, Quantum Serverless~\cite{serverless} can coordinate an execution workload through many clouds and quantum computing providers and thus, understand concepts coming from both the classical and quantum worlds. At this uppermost level, developers make use of the most suitable programming model based on trade-offs between speed, solution quality, and resource cost.

\subsection{Quantum Kernels}
\label{sec:QuantumKernels}

In the context of classical computing, invocation of accelerator kernels can happen in several ways: (1) host (CPU) code; (2) strictly via other kernels (device-side, kernel launch); or (3) host or kernel code. Scheduling of host code and accelerator kernels similarly can be performed in three ways: (1) implicitly, if a programming model provides runtime mechanisms for managing a task DAG; (2) asynchronously, through non-blocking calls that return immediately after being called; or (3) explicitly, via synchronization points.

A programming model should enable quantum kernels (circuits, possibly with classical control flow) to execute in parallel, synchronously and asynchronously with respect to the kernels themselves and classical computations in order to cover a wide variety of use cases. This type of execution model is akin to that of classical accelerator (e.g., GPU) programming wherein asynchronous accelerated kernels can be desirable but data dependencies permit only synchronous execution. However, a major difference between purely classical and classical-quantum hybrid execution are the latencies involved; for instance, classical accelerators interface via high-speed buses whereas today's free and publicly available quantum devices are generally cloud-based.

The rate at which quantum computations can be performed (i.e., the analog to classical CPU clock speed) is set by the qubit technology which range from $O(1)$~Hz (photonic qubits) to $O(10^6)$~Hz (superconducting qubits). Programming models must provide kernel scheduling mechanisms, or runtime, for orchestrating devices where latencies can range from $O(10^{-6})$ seconds to $O(1)$ minutes or greater. In cases where classical and quantum computations are decoupled — i.e., independent classical and quantum information processing with no communication between processes is needed — no orchestration would be required from the programming model. Indeed, classical and quantum routines or applications can be developed in separate languages and executed independently. In this section, we are concerned with ‘near-time’ hybrid computations wherein classical and quantum resources work in unison to complete a task.

Scheduling and execution of quantum kernels can be handled by a queuing system within the programming model. Similar to the SYCL programming model~\cite{10.1145/3388333.3388649}, a queue could have a one-to-one correspondence with a device; multiple queues would be necessary to work with multiple devices. Once a queue has associated to it a quantum device, that device can then be queried directly from the queue to obtain device-specific information, e.g., number of qubits and supported features. A quantum extension to the queue model would include sampling functionality to specify the number of executions (or shots) of a given kernel. Furthermore, a queue would capture the necessary semantics for execution, data input and output, etc., and provide within its API synchronous and asynchronous capabilities.

In principle, quantum accelerator kernels should be indistinguishable from classical accelerator kernels. That is, the semantics should be such that both classical and quantum kernel can expressed in the same way. For example, classical kernels generally define an ``index space’’ over which an instance of the kernel executes. Each kernel instance executes the same code but may operate on different data or traverse a different execution pathway through its code. This mechanism, coupled to embarrassingly parallel workloads, is what empowers GPUs in today’s HPCs. For a quantum device kernel, the concept of an index space can be extended to include device (qubit) partitioning wherein, e.g., a single circuit instance can be mapped multiple times to the topology of a large device and hence execute in parallel.

In addition to partitioning, various quantum ``backends’’ (i.e., control electronics, classical computational devices, etc.) will likely have different technical capabilities; e.g., control-flow, dynamical decoupling schemes, etc. Such capabilities could be configurable  so a developer may choose the schemes most appropriate for their kernel. Therefore, an API specification should permit extensibility for vendors to expose all desired functionality for their backend devices via a common plugin mechanism, allowing for interoperability between different vendor library implementations.

In HPC, vendor libraries are typically written in low-level systems languages (Layer-2) such as C and C++ which are the de facto for applications targeting a set of heterogeneous architectures on current HPC platforms. A programming model built atop these would require language extensions for allow the expressiveness necessary to write circuit applications. Language extensions also allow for a single-source programming model such that development would be streamlined and eliminate the interplay of multiple languages (and thus compilers, interpreters, etc.). Moreover, this makes classical control-flow within quantum kernels more natural for developers. An alternative to a single large-scale application is an admixture of distinct executables whose orchestration is handled by a \gls{wmfs}, see ~\cite{alsaadi2021exaworks}. The WFMS is effectively a wrapper around multiple executables — that may or may not have intercommunication — and steers their execution based on a predefined DAG or logic that enables dynamism of the overall flow.

\subsection{Memory Model}\label{sec:MemoryModel}

Quantum processors operate at a very particular level in terms of memory and, in fact, a quantum processor can be regarded as a computational memory. In contrast with typical classical formalism where operations usually happen outside dedicated memory, quantum computing operates on the data input and the lifetime of quantum information is limited to the duration of the circuit carrying out the computation, with some notable exceptions like inherently quantum memories (QRAM and QROM). These quantum memories, hitherto never realized, would naturally span layers 1 and 2, and could play a fundamental role in classical data loading for some algorithms in which the implementation of the quantum oracle is critical to retain any level of quantum advantage~\cite{mosca2008quantum}. Section~\ref{classical-data-loading} provides a thorough discussion on this subject.

On a more classical standpoint, and similar to the concept of shared memory in HPC architectures, one could envision a model whereby some memory is shared at the QPU controller level with other processing units (GPUs, CPUs) in order to exploit different aspects of the quantum pre- and post-processing tasks, as well as fast loading of quantum circuits or parts of circuits.

\subsection{Data Types} \label{dataTypes}

The integration of classical computing resources with QPUs involves necessarily some common interface for data consumption that bridges largely different DSLs. A seamless workflow that exploits both CPUs/GPUs/xPUs and QPUs will include steps to transform data types as needed by each unit. It is worthwhile to differentiate between data interpreted as instructions and other type of data like metadata or data carrying computing information. Within the scope of programming models, the quantum nature of the information existing in the QPU during the computation is unimportant. Any interaction of the QPU with supporting or co-computing units before, during, or after the quantum computation occurs largely in the form of classical information. In fact, both the inputs and outputs for quantum processors are naturally classical except for a few notable exceptions (QRAM being one example, as discussed in Section~\ref{ssec:classical-data-loading}). The natural outputs from QPUs are probability distributions, to which we have a level of access limited by our ability to sample them. Certain quantum algorithms like quantum phase estimation or Shor’s algorithm produce distributions so peaked that one or only a few samples are enough to provide the answer. Other algorithmic approaches based on heuristic methods, however, produce output distributions from which the value of certain quantum observable is computed by repeated sampling. This value can either represent the answer to the quantum subroutine, or it can be passed back to other -classical- computing resources to generate a new input to the QPU. 

Inputs to quantum processors are, at the highest level of abstraction (Layer-2), a series of operations, that can be both unitary and non-unitary, and which can be represented by a quantum circuit. These quantum circuits can then take many levels of specificity throughout the transpilation and compilation chains, from the multilevel representation that a quantum assembly language like OpenQASM provides to the binaries used by the classical control hardware to implement the quantum gates on the quantum processor. These gates, or instructions, are parameterized to get instanced by data from the algorithm input. The gates are ultimately expressed as physical pulses of energy that interact with the quantum hardware (Layer-1). Thus, these pulses ultimately contain both instruction data and computational data. It is at higher levels of abstraction, however, where HPC could possibly help with circuit optimization, transpilation, and routing, working with circuit representations involving data types digestible by the DSL of the different processing units.

\section{Classical simulation of quantum systems\label{sec:ClassicalSimulationOfQSystems}}

\subsection{Why use classical simulators}

For near-term quantum computers, one might question the necessity of classical simulators to emulate quantum computations. After all, the promise of quantum computing lies in its ability to perform certain tasks exponentially faster than classical computers. However, classical simulators remain invaluable for several compelling reasons \cite{10.1038/s41598-019-47174-9, 9910084,doi.org/10.48550,Fedorov2022unitaryselective,preopt-0,preopt-1,preopt-2,IBMQuantumSummit2023}.

\begin{enumerate}
    \item {\bf Resource Efficiency:} Quantum computer time is a limited and valuable resource. When developing and testing new quantum circuits or algorithms, researchers often need to execute them numerous times to validate their functionality and robustness. Running experiments on a real quantum machine can be time-consuming and costly, particularly when waiting for quantum computer availability after each circuit modification. Classical simulators offer an efficient alternative, allowing researchers to rapidly iterate through experiments without waiting for quantum resources.
    \item {\bf Noise and Error Analysis:} Real quantum machines are susceptible to noise and errors due to environmental factors, making it challenging to control and maintain the desired quantum state fidelity. Classical simulators provide a controlled environment for introducing and analyzing various noise scenarios. Researchers can simulate quantum noises to assess how robust their algorithms are under different conditions, as well as explore error correction techniques. This is a crucial step in building fault-tolerant quantum systems.
    \item {\bf Scalability:} Current quantum machines have limitations in terms of the number of qubits and the noise levels they exhibit. Classical simulators, on the other hand, can be adapted to simulate quantum systems with a larger number of qubits, enabling researchers to explore complex quantum algorithms and states that are beyond the capabilities of existing quantum hardware.
    \item {\bf Versatility:} Classical simulators offer flexibility regarding what they can simulate. Researchers can use them to capture "snapshots" of quantum states during computations, perform measurements, and make decisions based on measurement outcomes. Additionally, they can obtain more comprehensive information, such as the density operator of the system, rather than just specific measurement results.
    \item {\bf Efficiency Trade-Offs:} Building an efficient quantum simulator involves addressing trade-offs between computational efficiency and the range of quantum states it can represent. The choice of simulator depends on the specific needs of the quantum computation being emulated.
    \item {\bf Connectivity:}
    When executing a circuit on quantum hardware, the transpilation step involves mapping logical qubits to physical ones, in a fashion that may one-to-many. This is due to the constraint of the physical connection of qubits, which is not usually all-to-all on true QPUs. When simulating, all qubits can be entangled with all others, thus keeping down the `physical' qubits needed and easing the computational burden of solving the minor-embedding problem.
  \item {\bf Pre-optimization:}
    Variational algorithms can be expensive and require many iterations of a quantum circuit.  Pre-optimization with classical simulators can be used to find approximate circuits parameters that can be used or refined with quantum hardware.\cite{preopt-0,preopt-1,preopt-2}
    \item {\bf HPC-assisted quantum computing:}
    There is a way in which HPC can substantially help to extend the depth of quantum circuits, as demonstrated in the recently developed method called "Operator Backpropagation."~\cite{IBMQuantumSummit2023} The idea is to compute a part of a quantum circuit on a quantum device, and compute the second part of a circuit classically by using a quantum circuit simulator on an HPC system, and then stitch the results together. 
\end{enumerate}
In summary, classical simulators allow for efficient development, testing, and analysis of quantum circuits, algorithms, and states, while also providing a controlled environment for exploring quantum noise and errors. While quantum computers hold immense promise, classical simulators remain a vital component of the quantum computing toolkit, enabling progress and innovation in the field. Furthermore, simulators will likely continue to serve as yet another type of circuit execution environment among multi-node compute clusters for executing distributed workloads, which will be adept at handling (sub-)circuits that fall within their scope.

\subsection{Quantum circuits simulators}

In this section, we describe different types of classical quantum circuit simulators. In particular, we describe their capabilities in terms of what the maximum size systems (in both qubits and gates) that can be simulated using HPC. The advantages and disadvantages of quantum circuit simulators are also discussed, as well as use cases.

Before we start, we note that not every quantum circuit is difficult to simulate on a classical computer. Trivially, a quantum circuit that carries out a purely classical computation is not difficult to simulate on a classical computer. Indeed, some of the quantum circuits that are not difficult to simulate include Clifford circuits~\cite{gottesman1998heisenberg,aaronson2004improved,garcia2014simulation} and the quantum Fourier transform~\cite{aharonov2006quantum,yoran2007efficient}, to name a couple. Some quantum circuits are known to be classically simulable for specific input states as well~\cite{garcia2014simulation}. If the simulation does not need to produce the full spectrum of the output state of a given quantum computation but rather a sampling of such, as it would be the case for obtaining a final, classical result out of a quantum computer upon measurements, the difficulty of the classical simulation can dramatically decrease~\cite{hillmich2020just}. Further decrease in the difficulty is obtained if the simulation is to mimic an error-prone quantum computer executing a quantum circuit~\cite{hillmich2022approximating}. We further note in passing that simulation of a quantum circuit that exhibits sparse couplings between densely connected sets of qubits can be more amenable to classical simulations~\cite{fatima2021faster}, using methodologies not unlike circuit cutting discussed in Sec.~\ref{sec:CNA}.

\subsubsection{State-vector simulators}

State-vector based simulation is to simulate the operations of applying a series of unitary operators $U_{m-1} \cdots U_1 U_0$ to the state-vector representation of the quantum states $\ket{\psi}=\sum_{i=0}^{2^n-1}\alpha_i\ket{i}$, where $n$ is the number of qubits and $m$ is the number of operations or gates. Typically, a complex-valued double or single precision floating-point vector $\vec{\alpha}$ of size $2^n$ is used to store the coefficients $\alpha_i$, which costs $16\times2^n$ bytes of memory for classical simulation. $U_i$ with $i\in[0,m-1]$ is a $2\times2$ (for one-qubit gate) or $4\times4$ (for two-qubit gate) complex matrix. It has been shown that an arbitrary quantum circuit can be decomposed into 1-qubit and 2-qubit gates \cite{barenco1995elementary}. In fact, most quantum devices internally run 1-qubit or 2-qubit basis gates. For example, IBM adopts 1-qubit gate \texttt{X}, \texttt{SX}, \texttt{RZ}, \texttt{ID} and 2-qubit gate \texttt{CX} (recently \texttt{ECR}) as the basis gates. To apply a gate $U$, the operation is $\ket{\psi}\to U\ket{\psi}$. For 1-qubit $U$ applying on qubit $q$ in a quantum register, $\vec{\alpha}$ is updated through the following expression where $s_i=\lfloor i/{2^q}\rfloor 2^{q+1}+(i \% 2^q)$ for every integer $i\in[0, 2^{n-1}-1]$:
\begin{equation}
\begin{bmatrix}
\alpha_{s_i} \\
\alpha_{s_i+2^q}  
\end{bmatrix}
\to
U_{2\times2}\cdot \begin{bmatrix}
\alpha_{s_i} \\
\alpha_{s_i+2^q}  
\end{bmatrix}
\label{eq:1q-op}
\end{equation}
Regarding 2-qubit unitary gate $U$ applying on qubit $p$ and $q$ (assuming $p<q$ without losing generality), $\vec{\alpha}$ is updated  through:
\begin{equation}
\begin{bmatrix}
\alpha_{s_i} \\
\alpha_{s_i+2^p} \\
\alpha_{s_i+2^q} \\
\alpha_{s_i+2^p+2^q}
\end{bmatrix}
\to
U_{4\times4}\cdot \begin{bmatrix}
\alpha_{s_i} \\
\alpha_{s_i+2^p} \\
\alpha_{s_i+2^q} \\
\alpha_{s_i+2^p+2^q}
\end{bmatrix}
\label{eq:2q-op}
\end{equation}
where $s_i{=}\lfloor \lfloor i/{2^p}\rfloor /2^{q-p-1} \rfloor  2^{q+1} + (\lfloor i/{2^p}\rfloor \% 2^{q-p-1})2^{p+1} +  (i \% 2^p)$ for every integer $i\in[0,2^{n-2}-1]$. 

Therefore, state-vector based quantum numerical simulation is to perform a sequence of $2\times2$ or $4\times4$ operations Eq.~(\ref{eq:1q-op}) and Eq.~(\ref{eq:2q-op}) over the large state-vector coefficient array of complex numbers in a classical system.

State vector simulators hold the full coefficients of pure quantum states within the classical system's memory, which scales exponentially with the number of qubits. Consequently, the state vector is often distributed across many nodes of a classical HPC~\cite{haner20170,pednault2017breaking,pednault2019leveraging,li2021sv}. Such simulators however exhibit linear scaling with respect to the circuit depth~\cite{fatima2021faster}. As such, state vector simulations are sensitive to the number of qubits but much less to the gate count or circuit depth. For example, a recent work shows that a low-energy nuclear quantum circuit with 115 million gates on 21-qubits can be effectively simulated within one hour using a GPU for state vector simulation~\cite{li2023deep}.

We note that various forms of approximate state-vector simulators can be tailored to problem specific applications.  For example, recent truncated state vector simulators have been used to simulate VQE for various chemistry problems on 64 qubits without significant loss of accuracy or the need for large computing resources~\cite{preopt-2,approxstatevec}. Combining the state vector simulators with Feynman path summation can trade the circuit depth with the number of qubits or memory usage~\cite{fatima2021faster}, rendering some hard to simulate circuits according to a naive state vector simulator to be simulable due to orders of magnitude improvement in their simulability~\cite{markov2020massively,fatima2021faster}.

\subsubsection{Density matrix simulators}

When dealing with a mixed system where noise is present, a pure state can no longer provide sufficient information about the system. In this case, a mixed state corresponds to a statistical ensemble, or probabilistic mixture of pure states, can be used to describe the condition where a system is entangled with another system, such as the environment from which the noise is imposed. A mixed state is represented by a density matrix or a density operator, which is defined by choosing the basis in the underlying space. The density matrix is given by:
\begin{equation*}
\rho=\sum{p_s\ket{\psi_s}\bra{\psi_s}}
\end{equation*}
where $\rho$ represents the fraction of the ensemble of each pure state and is generally an unknown real value. A density matrix contains all the information of a quantum system, allowing the calculation of the probabilities of the outcomes of any measurement performed.

Compared to state-vector simulation, the density operator requires the conservation of $4^n$ coefficients, where $n$ is the number of eigenstates for each pure state, i.e., the number of qubits. Therefore, the memory cost of a density matrix simulation is $2^n$ times that of a pure state simulation using state-vector. The system also evolves according to the operator or gate sequence. When a particular gate is applied, the density matrix evolves as follows~\cite{li2020density}:
\begin{equation}
\rho(t)=G(t)\rho(0)G(t)^{\dagger}
 \label{eq:onegate}
\end{equation}
where $G(t)$ is the gate at time $t$, and $G(t)^\dagger$ is its adjoint. The evolution of a density matrix is more complicated than that of a state vector, considering the size of $4^n$ and the extra adjoint operator per gate. Although $G(t)$ by itself is a unitary operator, it becomes a general matrix and is not necessarily unitary in the presence of noise. Depending on the noise model, the evolution can become an ensemble of evolutions of corresponding channels, each described by a Kraus operator.

The noisy density matrix quantum numerical simulation is to compute $\rho_\text{out}$ for a n-qubit quantum system, density matrix quantum circuit simulation is to compute $\rho_\text{out}$ for a n-qubit quantum register, given initial state $\rho_\text{in}$ and $m$ non-unitary transformations $G_0$, $G_1$, $\dots$, $G_{m-1}$:
\begin{equation}
 \rho_{\text{out}} = G_{m-1}\cdots (G_1 (G_0 \rho_{\text{in}} G^\dag_0) G^\dag_1)\cdots G^\dag_{m-1}
 \label{eq:dmsim}
\end{equation}
where $G$ and $\rho$ are $2^n\times 2^n$ matrices. Due to noise, $G$ is not necessarily a unitary matrix. $G^\dag$ is the adjoint of $G$ verifying $G^\dag=(G^*)^T$. As real quantum devices typically use 1-qubit or 2-qubit gates representing as $4\times4$ or $16\times16$ matrices, to obtain the matrix $G_i$ with a $2^n\times2^n$ size, Kronecker product or tensor product is used with the identity matrix $I$ for the other qubits.

\subsubsection{Tensor network simulators}

Quantum tensor networks can be used for the simulation of quantum circuits~\cite{markov2008simulating, lykov2021importance, lykov2021large, lykov2021performance, lykov2022tensor, berquist2022stochastic, shah2023gpu}. These simulators leverage the mathematical framework of tensor networks, which are graphical representations of quantum states and operations, as networks of interconnected tensors. Specifially, the nodes in these networks correspond to tensors, which encode gate operations, while the edges represent indices along which tensors are contracted, reflecting the entanglement and interactions between qubits.

The strength of tensor network simulators lies in their ability to compactly represent quantum states that would otherwise require a prohibitive amount of memory. This is particularly valuable for simulating shallow-depth quantum circuits with a large number of qubits, which are challenging for traditional state-vector simulators. These simulators are still limited by the amount of available distributed memory on a supercomputer. For the current generation of supercomputers, tensor network simulators can run quantum circuits up to approximately 200 qubits without approximations for certain quantum circuits~\cite{lykov2021large}. Larger simulations are possible with a truncated bond dimension as well as other applications with different resource requirements~\cite{preopt-1}.

\begin{figure}
\includegraphics[width=\columnwidth]{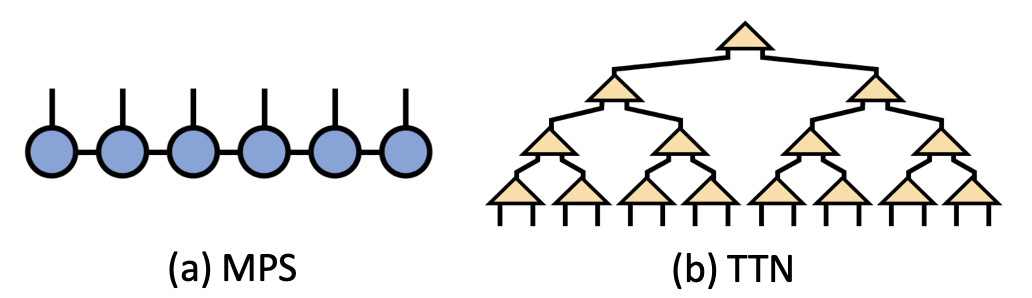}
\includegraphics[width=\columnwidth]{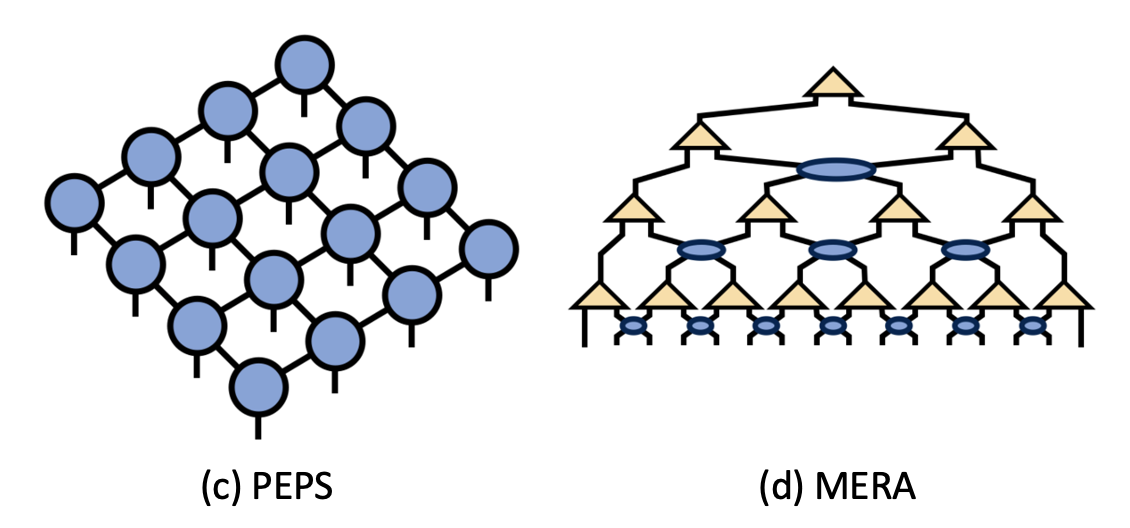}
\caption{Different types of tensor network simulators: (a) Matrix Product States (MPS), (b) Tree Tensor Networks (TTN), (c) Projected Entangled Pair States (PEPS), and (d) Multi-scale Entanglement Renormalization Ansatz (MERA).}
\label{fig:tensornetworks}
\end{figure}

There are different types of tensor network simulators: \gls{mps}, \gls{peps}, \gls{ttn}, \gls{mera}, and \gls{cvtn} to name a few as shown on Figure~\ref{fig:tensornetworks}. MPS is arguably one of the most popular simulators, and it is particularly useful for simulating one-dimensional quantum systems. In MPS, the tensors are arranged in a chain, with each tensor representing the state of a qubit and the connections between them capturing the nearest-neighbor entanglement. One of the MPS applications is an efficient tensor network simulation of lossy Gaussian boson sampling experiments, which has been recently demonstrated \cite{liu2023simulating, Oh2023}.

Each type of tensor network has its specific applications, advantages, and limitations, and the choice of which to use depends on the application, size, and structure of a quantum circuit:
\begin{itemize}
    
\item MPS simulators excel in one-dimensional quantum systems with short-range interactions. They are efficient for simulating ground and excited states, as well as dynamical properties, due to their ability to capture entanglement in a scalable manner with limited entanglement entropy.

\item  PEPS are generalizations of MPS to higher dimensions, making them suitable for two-dimensional quantum systems. They are adept at handling both short- and long-range interactions, but their computational complexity increases significantly with the system's size and entanglement.

\item  TTN simulators are structured in a hierarchical, tree-like manner, offering efficient computation for certain quantum circuits, especially those with hierarchical or layered structures. They are particularly useful for simulating states that exhibit hierarchical entanglement patterns and for providing insights into quantum many-body systems.

\item  MERA is designed for critical systems with long-range entanglement. It excels in representing ground states of quantum many-body systems near criticality, offering insights into scaling and universality in quantum phase transitions.

\item  CVTN simulators are tailored for quantum systems with continuous variables, like quantum fields or modes of light. They are adept at handling systems where particle number isn't conserved and are crucial in studying non-Gaussian states and processes in quantum optics and field theories.
\end{itemize}

Tensor network simulators continue to evolve, with ongoing research aimed at increasing their efficiency, scalability, and applicability to a broader range of quantum computing tasks.

\subsubsection{Open system Lindblad quantum simulators}

Open-system Lindblad quantum simulators are specifically designed to model the dynamics of quantum systems interacting with external environments. The applications include, for example, simulating quantum magnetism, topological materials, 
quantum phase transitions, and electron transportation. Unlike closed quantum systems, open systems are subject to environmental influences that lead to non-unitary processes such as decoherence and dissipation. These simulators leverage the Lindblad master equation, which blends the unitary evolution dictated by the system's Hamiltonian with the non-unitary aspects resulting from environmental interactions. This approach allows for a comprehensive simulation of open quantum systems, capturing the complexity of quantum noise and the environmental effects. Recently, a novel approach, called noisy quantum gates, has been proposed, as a classical simulation of the Lindblad dynamics~\cite{noisygates}. It is based on integrating the noise into the gates, rather than keeping gates and noise as two separate dynamics, an approach can be generalized to non-Markovian dynamics by using colored noises.

In general, there are a few ways to formulate, but the most common is to use the density matrix formalism, which is adept at representing mixed states, a critical aspect when dealing with open systems. These simulators provide researchers with the flexibility to define specific system-environment interactions, making them a versatile tool across various quantum research domains, especially in material science applications. Its ability to accurately model quantum noise induced by the environment is an especially valuable aspect of it. Overall, the open-system Lindblad quantum simulators stand as a very valuable tool in quantum research, enabling a deeper understanding of the environmental impacts on quantum systems and aiding in the advancement of practical quantum applications.

\subsection{Overview of classical simulators}

In summary, we described four main types of quantum circuit simulators. In the context of material science, the choice of a quantum circuit simulator depends on the specific characteristics of the system under study and the phenomena of interest. Here is how each type of simulator can be used:

State-vector simulators are ideal for systems that can be accurately represented by pure quantum states. In material science, they are particularly useful for studying the evolution of quantum states under Hamiltonians with relatively few degrees of freedom. They excel in scenarios where the full quantum state needs to be tracked, such as in the simulation of small, isolated quantum systems or systems with limited entanglement.

Density matrix simulators are well suited for studying systems where mixed states are prevalent, which includes most real-world scenarios in material science. They can handle decoherence and other noise effects, making them suitable for simulating open quantum systems or systems under non-ideal conditions. Density matrix simulators are ideal for investigating phenomena in quantum materials where environmental interactions play a significant role.

Different types of tensor network simulators have different use cases.
MPS and TTN are powerful in simulating one-dimensional and certain hierarchical quantum systems, respectively. They can efficiently model systems with short-range interactions, which are common in material science. These simulators are especially useful for studying ground state properties and low-energy excitations in materials.
PEPS and MERA are more suited for higher-dimensional systems. PEPS can handle both short- and long-range interactions in two-dimensional materials, making them valuable for exploring complex quantum materials, such as high-temperature superconductors. MERA is particularly effective in studying critical phenomena and phase transitions in materials.

Open system Lindblad quantum simulators are designed to handle non-unitary evolution, which is typical in open quantum systems, where the system is in contact with an external environment. In material science, they are crucial for studying decoherence, dissipation, and thermalization processes in materials. Lindblad simulators are particularly relevant for investigating quantum materials and devices operating under realistic, non-ideal conditions, where environmental interactions cannot be ignored. They are essential for understanding the behavior of materials in quantum information processing and quantum sensing applications, where control and mitigation of decoherence are critical.

Using HPC is critical for simulating large quantum circuits. Ultimately, supercomputers can simulate relatively small quantum circuits because of the exponential requirement of available distributed memory. The next generation of quantum simulators will probably use small quantum computers to simulate very large quantum circuits. The idea is to use small quantum computers to perform tasks that are inherently quantum in nature, such as the contraction of intermediate multi-dimensional tensors, which are central to tensor network simulations. This idea is closely related to Feynman's proposal to use quantum computers to simulate quantum systems~\cite{hey2018feynman}. One way is to use \gls{hhl} algorithm for the contraction of very high-dimension tensors, which is currently a bottleneck of classical tensor network simulators. The HHL algorithm can be adapted to perform tensor contractions by encoding the tensors as linear systems. This could potentially revolutionize tensor network simulations by dramatically reducing the computational complexity of these contractions. This approach promises a scalable pathway for quantum simulation, as improvements in quantum hardware, such as increased qubit counts and enhanced coherence times, would directly translate into an increased capacity for simulating larger and more complex quantum circuits.

\section{Use case identification\label{sec:UseCaseDiscovery}}

\subsection{Criteria for use case identification}

In this Section, we identify use cases relevant to real-world applications~\cite{2023arXiv230613126S}.
In such identification, we will rely on the following criteria:

\begin{enumerate}
\item
Potential impact: selecting real-world use cases requires identifying research topics at the frontier of materials science and engineering that could benefit from the deployment of quantum computing for solving one or more of the discovery steps involved. Representative examples include semiconductor integration and processing, electronics, and optoelectronics applications. Also, materials use cases in carbon capture and storage, catalysis, energy storage, materials replacement, and recycling are being intensely researched, along with pharmaceutical and biochemical applications. In some cases, impact arises from insight into fundamental phenomena, while in other cases it comes from screening a large number of candidate materials and selecting the highest-performing candidate.
\item
Physical scale: typically, realistic materials use cases involve computational workflows consisting of various simulation steps with length and time scales covering more than 10 orders of magnitude. Therefore, predicting a material’s performance, from the molecular scale to the process scale, requires a broad range of computational methodologies including classical and quantum mechanical molecular dynamics simulations as well as the thermodynamic and bulk continuum models. Recently, artificial intelligence and machine-learning techniques have been added to the physics-based simulations toolkit for applications including generative materials design, high-throughput materials screening, and accelerated simulation.
\item
Complexity: application-guided and performance-oriented computational discovery work streams feature complex materials, where important phenomena occur in volumes often consisting of thousands of molecular building blocks with millions of atoms.  These are challenging to represent adequately in any computational simulation environment. A real-world discovery use case requires, therefore, the identification of a minimum representative materials volume for conducting the performance analysis. This representation can be as small as the active space within a molecule, a surface at which a physical or chemical process occurs, or as large as an entire composite wing of an airplane.
\end{enumerate}

In principle, quantum computing could be applied in each discovery step with the potential of enhancing computational accuracy, reducing computational time, or delivering any other improvement of the materials discovery outcome. For selecting the proper application of quantum computing in a real-world use case, it is however important to first identify and compare the difficulty of the classical and quantum simulation tasks and to ensure that proper data representations and algorithms can be developed for performing the discovery task on a quantum computing system. To that end, in our analysis of use cases, we will consider:

\begin{enumerate}
\item the relevance to real-world problems
\item the hardness of simulating the use case under consideration on classical computers
\item the hardness of simulating the use case under consideration on quantum computers (near-term and fault-tolerant)
\item the hardness of simulating quantum circuits involved in the previous point on quantum hardware and with classical simulators 
\item the possibility of using classical HPC and quantum computers in concert
\end{enumerate}

\subsection{Overview of use cases}

\begin{figure}
\includegraphics[width=\columnwidth]{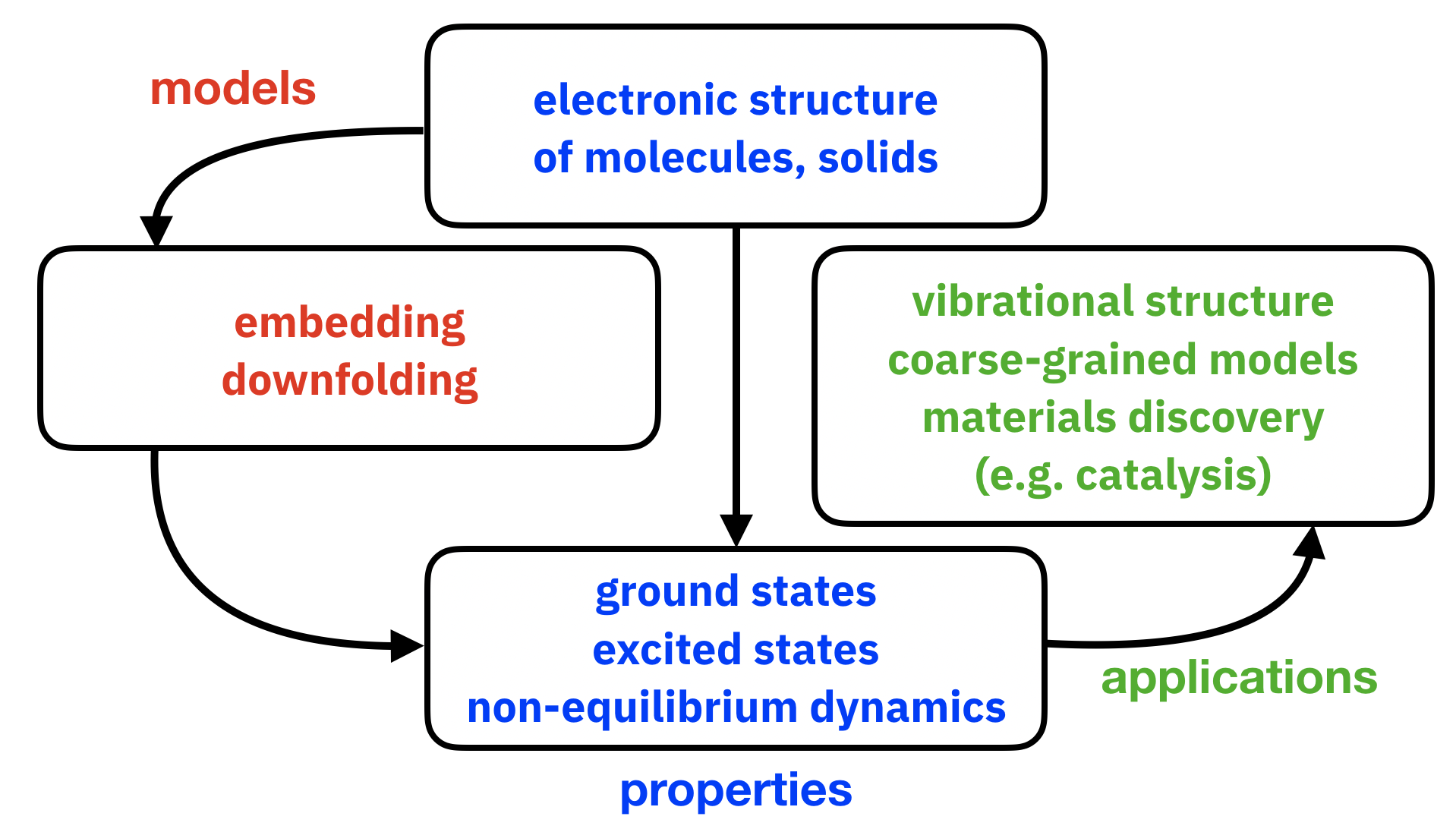}
\caption{Schematics overview of use cases in materials discovery.  A central task of quantum theory is describing the properties of molecules and solids (blue). This goal can be achieved by solving for the electronic Schrodinger equation to compute e.g. ground-state, excited-states, and non-equilibrium dynamical properties (vertical arrow). Due to computational limitations, one can introduce approximations to the fundamental electronic structure problem, that reduce the size of the quantum problem within the existing computational budget (red).
Accurate solutions of the electronic Schrodinger equation are the starting point (green) for important applications in materials science, including vibrational structure calculations, coarse-grained models (e.g. corrosion, deformation), and materials discovery (e.g. catalysis, metamaterials).
}
\label{fig:map}
\end{figure}

A central task of quantum theory, in the fields of quantum condensed matter physics and quantum chemistry, consists of describing the properties of interacting atomic, molecular, amorphous, and crystalline systems. Apart from advancing our fundamental understanding of quantum mechanics, a generic solution to the quantum problem would allow progress in application fields ranging from materials discovery (better magnets, solar cells, catalysts, or qubit hardware) to drug design.

The required operation count and the number of quantum degrees of freedom required for the accurate ab initio description of such systems puts them far outside of the accessible regime on current or near-term quantum hardware, except for very small systems. These requirements were elucidated by many authors including Reiher et al \cite{reiher2017elucidating},Tubman et al \cite{2018arXiv180905523T}, Elfving et al \cite{elfving2020will} and Goings et al \cite{goings2022reliably}, who considered the application of quantum phase estimation to challenging instances of the electronic structure problem (respectively nitrogenase cofactor, chromium dimer, and cytochrome P450) on fault-tolerant devices.

It is therefore advantageous to consider approximations to the fundamental electronic structure problem that reduce the size of the quantum problem both in terms of qubit and operation requirements (see Figure~\ref{fig:map}, left).

Two complementary approaches are presently pursued. First, the embedding route identifies a set of strongly quantum parts of the problem (typically ``orbitals'' or ``sites''), extracts the relevant physics to an auxiliary quantum problem, and solves this problem using a quantum processor. The remainder of the system is then treated classically with polynomial scaling approximations. The exact limit is recovered by enlarging the embedding subspace or increasing the accuracy of the classical method.

Second, the model systems route approximates the original quantum system by a minimal set of appropriately chosen effective low-energy degrees of freedom. In this scheme, the exact limit is recovered in a ``bottom-up'' approach by gradually adding degrees of freedom until the original model is recovered. A typical example of this approach is given by the Heisenberg spin model, which (in the appropriate limit) captures the low-energy limit of the so-called t-J model, which itself is an approximation of the low-energy physics of the Hubbard model.

For both of these approximation approaches, simulation on quantum computers will become feasible much earlier than the simulation of the full quantum electronic structure problem. Realistic embedding approaches for materials typically result in compact but dense quantum problems (requiring many operations on few qubits), whereas model systems generally yield larger but sparse systems (requiring many qubits but fewer operations). The sparsity allows us to take advantage of devices with limited qubit connectivity with a reasonable operation count.

A central question concerns the physical observables under study. Traditional electronic structure mainly investigates ground states and their properties, such as the ``band structure'' or the ground-state energy as a function of nuclear coordinates. Out of these states, information about structure, vibrational modes, and mechanical properties is inferred.

However, experiments are performed at non-zero temperatures and, additionally, probe aspects of the excitation spectra. Photoemission spectroscopy, for instance, obtains the single-particle excitation spectrum and the gap size \cite{10.1063/5.0044060}. Nuclear magnetic resonance and magnetic neutron spectroscopy obtain nuclear spin excitations, and optical conductivity measurements probe two-particle excitations related to a current-current correlation function. Standard ground state methodology such as \gls{dft} requires additional approximations to predict such observables. Famously the DFT ``band gap'' is not a reliable proxy for the gap in semiconductors \cite{bandgap,perdew1985density,cohen2012challenges}.
Quantum (or classical) computing methodologies that intrinsically contain excitation information present a valuable addition to our materials toolkit.

In many experimental setups, a weak external field (such as an incident light beam) probes the physics of the material under study. A central assumption is that this field leaves the material essentially unperturbed, such that linear response theory provides an appropriate description for its physics. However, once fields become strong, the system itself acquires an explicit time dependence which may often be modeled as a quench (an instantaneous change of the Hamiltonian such as a very short but very strong light pulse), a steady state (the time-evolution of a system in a translation-invariant long-time limit, such as a system with an applied voltage after initial transients have decayed), or a periodic drive (such as a continuous laser field).

While our classical toolkit for simulating ground-state physics is rather mature, the study of excitations within linear response is much more demanding and often requires high-performance computing resources. The study of time-dependent phenomena in real materials is still in its infancy. These science domains therefore offer promising application fields for future quantum computers. 

Reliable solutions to the electronic structure problem also allow us to define the nuclear Schrodinger equation, which in turn can be solved to determine molecular vibrations, and coarse-grained models, whose solution allows us to characterize complex phenomena including material stress, corrosion, and cracking (see Figure~\ref{fig:map}, right).

\subsection{Relevant use cases}

\subsubsection{Ground-state electronic structure}

One of the goals of electronic structure is to determine the ground and low-lying eigenstates of the Hamiltonian of a molecule, often within the Born-Oppenheimer approximation, i.e., for a fixed set of nuclear positions ${\bf{R}}$.
Determining the main features of the potential energy surface, i.e. the electronic energy as a function of nuclear positions $E({\bf{R}})$, is key to understanding chemical reactivity, reaction rates, and product distributions.
In particular, accurate estimation of reaction barrier heights (i.e. within $\sim 1$ kcal/mol from experiments) is important for many applications, such as catalyst development, battery electrode modeling, and manufacturing process design,
and a known challenge for electronic structure methods and in particular for DFT \cite{cohen2012challenges}.

As an example, in the design of manufacturing processes, the desired reaction(s) should dominate, e.g., have lower reaction barrier heights than undesired reactions.  One way to arrange this is through reactant selection, however, there may be many possible reactants.  Experimental screening to identify the best reactant candidates is often time, cost, and labor-intensive due to various factors.  Reactants may be expensive, hazardous, or rare.  Computational screening methods offer the possibility of screening candidate reactant barrier heights orders of magnitude faster, safer, and cheaper than experimental methods.  However, current state-of-the-art classical computational methods are often insufficiently accurate to compare reactant barrier heights.

\paragraph{Static and dynamic correlation}

Within electronic structure, wavefunctions are often classified as single-reference and multi-reference. In the former case, a single electronic configuration dominates the target wavefunction.
While the ground states of the majority of molecules have a single-reference character, one often encounters a multi-reference character in molecular excited states, at stretched bond geometries, in transition-metal chemistry, and in the presence of small HOMO-LUMO gap (e.g. large $\pi$-conjugated systems).

Single-reference problems can be accurately solved with various quantum chemistry methods for classical computers, from DFT methods (that can routinely treat $O(10^3)$ atoms~\cite{jones2015density})
to high-level wavefunction methods such as coupled-cluster theory (that can attain kcal/mol accuracy on systems of $O(10)$ atoms~\cite{sparta2014chemical}).
Extending quantum chemistry methods to multi-reference situations is an active and challenging research field where, notwithstanding steady progress, the accuracy attainable in molecules with more than a $O(1)$ atoms is significantly lower than in the single-reference case.

\begin{figure}
\includegraphics[width=\columnwidth]{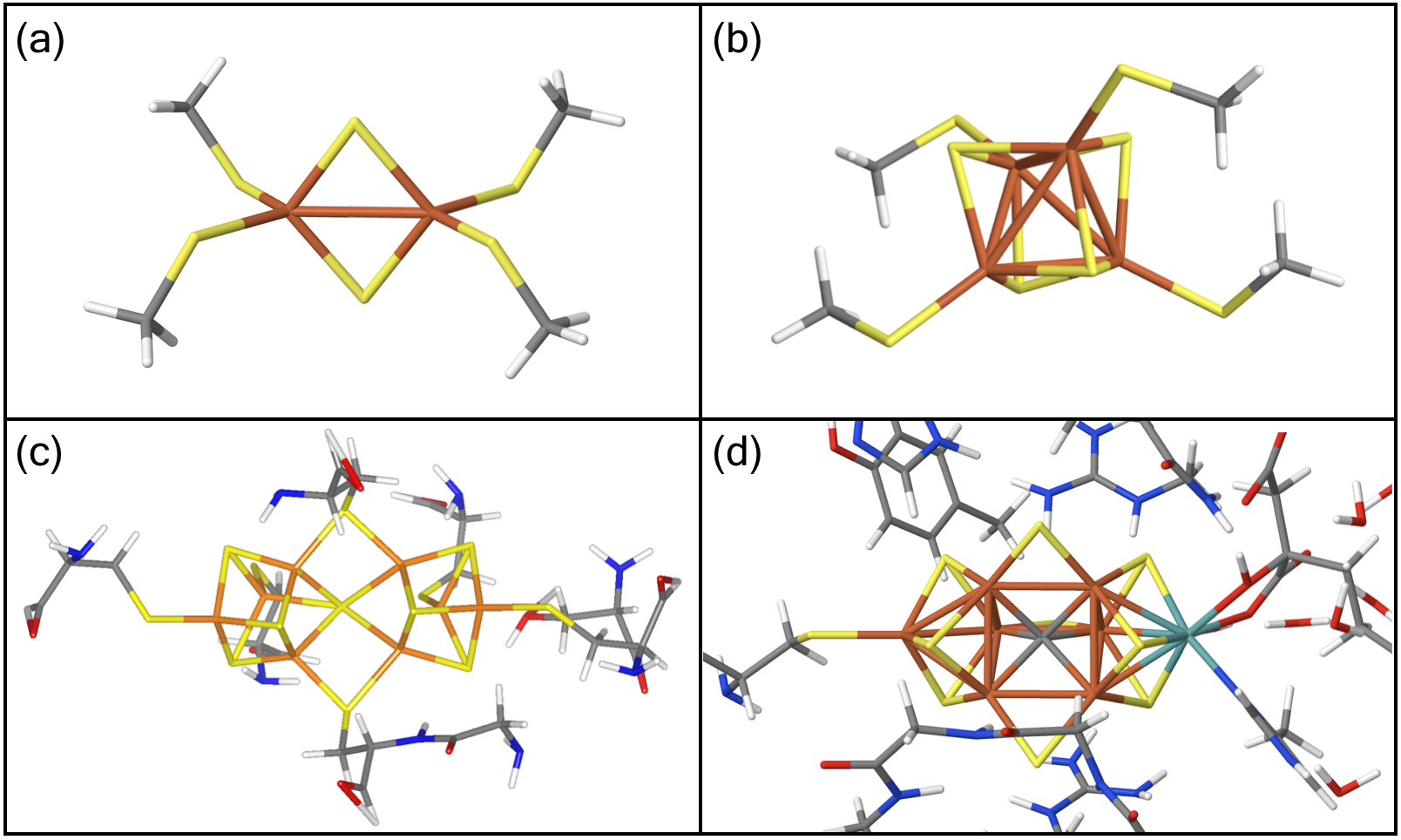}
\caption{The \ce{Fe2S2} (a) and \ce{Fe4S4} (b) clusters of ferredoxins, the P-cluster of nitrogenase in the resting state \ce{P^N} (c), and the FeMo-cofactor of nitrogenase (d). Orange, yellow, teal, red, blue, gray, and white sticks denote Fe, S, Mo, O, N, C, and H atoms respectively.}
\label{fig:fes}
\end{figure}

Important examples of multi-reference quantum-chemical problems are found in transition-metal chemistry~\cite{doi:10.1021/acs.jctc.9b00674,ciapprox-1}. Transition-metal oxides are an important class of materials that are used in catalysts, semiconductors, pigments, and many other applications.  Benchmarking studies of DFT for bond dissociation enthalpies of transition metal oxides and their cations demonstrate mean absolute errors of 30 kJ/mol (7 kcal/mol) or greater \cite{moltved2019performance}, identifying the energetics of reactions involving transition metals as a challenge for DFT \cite{moltved2019performance}. Examples of challenging instances of the electronic structure problem connected with transition-metal chemistry are:

\begin{enumerate}
\item Active sites of enzymes containing transition metals. These often involve multiple coupled transition metals, exemplified by the systems in Figure \ref{fig:fes}:
\begin{itemize}

\item the \ce{Fe2S2} and \ce{Fe4S4} clusters of ferredoxins \cite{sharma2014low}, where multi-reference character can be investigated in (30e,20o) and (36e,54o) active spaces respectively \cite{li2008github,li2017spin}

\item the iron-molybdenum cofactor (FeMo-cofactor) of nitrogenase, which catalyzes the 6-electron reduction of N$_2$ to ammonium in biological nitrogen fixation~\cite{noodleman2002insights,reiher2017elucidating}, and can be modeled by a (113e,76o) active space \cite{li2008github,li2019complexity}

\item the P-cluster of nitrogenase~\cite{li2019electronic} and oxygen-evolving complex of photosystem II \cite{cady2008functional,kurashige2013entangled}. In the context of nitrogenase, the hypothesized function of the P-cluster is to mediate electron transfer from the Fe-cluster to the FeMo-cofactor where nitrogen reduction takes place. The relevant states are the resting state \ce{P^N}, the one-electron oxidized state \ce{P^{1+}}, and the two-electron oxidized state \ce{P^{OX}}, which can be modeled by (114e,73o), (117e,75o) and (120e,77o) active spaces respectively.
\end{itemize}
Active sites of enzymes containing transition metals present highly complex multi-reference quantum chemistry problems motivated by biological applications.
Combined theoretical and experimental studies, primarily at the level of DFT, have illustrated many structural and electronic features of such active sites~\cite{siegbahn2009structures,lancaster2011x,batool2019magnetic}.
On the other hand, to interpret aspects of experimental spectroscopy, it is necessary to achieve a detailed understanding of the interplay between spin-coupling and delocalization between metals, which in turn requires accurate multi-reference quantum chemistry~\cite{sharma2014low,kurashige2013entangled,li2019electronic,chilkuri2019ligand,cao2018protonation}.
\item Nanocatalysts and surface catalysts containing transition metals. Simulating the mechanism of action of synthetic heterogeneous catalysts is challenging, as in the case of enzymes.
DFT predictions of quantities such as the adsorption energy of small molecules are unreliable~\cite{schimka2010accurate,capdevila2016performance}. 
While not all such problems are expected to be multi-reference in character, even the single-reference modeling of such chemistry, at a level significantly beyond DFT, is currently challenging or impossible.
In addition, multi-reference effects are expected to play a role in certain catalysts, such as transition metal oxides, or at intermediate geometries in reaction pathways~\cite{norskov2011density,schimka2010accurate,wodtke2016electronically,norskov2009towards}. 
\end{enumerate}

For all these systems, the basic quantum simulation problem is the computation of the ground-state (or low-energy eigenstate) of the electronic Hamiltonians, 
and the basic metric is whether quantum algorithms yield more accurate energies (for comparable computational resources) than their classical counterparts.
Proof-of-principle demonstrations can be carried out in the active spaces listed above.
However, to describe the actual chemistry of these systems, one should improve over active-space calculations by (i) treating larger numbers of electrons and orbitals, 
(ii) accounting for environmental, dynamical, and solvent effects, and (iii) accounting for dynamical electronic correlation.  While multi-reference problems are a known computational challenge,
even single-reference problems become hard to model classically when large numbers of electrons and orbitals are involved~\cite{vandevondele2012linear}.

\paragraph{Relativistic effects}

A discussion of the advantages and present challenges posed by hybrid quantum-classical approaches to electronic structure calculations begs mention of the treatment of relativistic effects.  
Accurate modeling of the ground and excited states of many molecules, such as atomic gold and uranium dimers, are examples of systems whose precise computational treatment requires the inclusion of relativistic effects in the Hamiltonian.  
Without this inclusion, a fundamental precision limit to quantum computation remains in place \cite{stetina2022simulating}.

\paragraph{Correlated electronic structure in solids}

There is some overlap in methods and ideas between the electronic structure problem in materials, e.g. crystalline solids, and quantum chemistry.
When electron-electron interactions are weak, the low-energy properties of a material can be described by computing the band structure using DFT. However, in some materials (called strongly correlated), the electron-electron interaction fundamentally changes the behavior and the effective non-interacting ansatz underpinning DFT is no longer appropriate. Typical examples are Mott insulators, which appear as conductors in band-structure theory but are insulating due to electron-electron interaction.
While Mott insulators are now well understood, many other phenomena in strongly correlated systems lack a justification in terms of an underlying microscopic mechanism and a quantitative and predictive theory of the associated physics.
Some important examples include:

\begin{itemize}
\item \textit{High-temperature superconductivity}. The mechanism driving superconductivity is not yet fully elucidated. Two regimes close to superconductivity, the {pseudogap} and {strange-metal} phase remain elusive~\cite{keimer2015quantum}. The nature and relation to the superconducting phase of both are not understood. The strange-metal phase features behavior inconsistent with a simple weakly interacting metal even at high energies, hence the name of ``non-Fermi liquid phase'', and has motivated a whole area of research on exotic metallic systems~\cite{lee2018recent}. On the other hand, the pseudogap phase features several competing ordering tendencies~\cite{fradkin2015intertwined}, which are very challenging to resolve in numerical methods, because most methods naturally favor a particular ordering pattern, so discriminating physical effects from method biases is very difficult.
\item \textit{Frustrated spin systems} realize a host of high non-trivial phases, in particular topological and gapless spin liquids~\cite{balents2010spin}. They have historically been the test bed for computational methods such as tensor networks and variational methods and, as such, appear as good test cases for quantum simulations.
\end{itemize}

\subsubsection{Embedding}
\label{embedding_sec}

As seen in the previous Section, solving the electronic structure problem in situations that are relevant to real-world applications, e.g. catalysis, requires significant quantum resources.
That example illustrates how the goals of relevance to real-world applications and ease of execution on quantum devices are in conflict.
However, there are two ways of making feasible progress toward the solution of challenging instances of the electronic structure problem: quantum embedding theories and effective low-energy models. 
These are not only useful steps toward the electronic structure problem, but also very compelling applications for current and near-term devices. Examples and applications are described in the remainder of this and following Sections.

The idea of quantum embedding theories stems from the fact that in chemistry, physics, and materials science, phenomena of interest often take place within a small region, motivating the separation of the whole system into a so-called active space surrounded by a host environment. The electronic structure problem is solved separately for the environment at a low level of theory and for the active space at a higher, i.e., more accurate and more expensive, level of theory. As an active space only accounts for a small portion of the whole system with a greatly reduced Hilbert space, simulating it using quantum chemistry methods becomes feasible on quantum computers and emulators. Examples of active spaces include point defects in semiconductors and insulators and active sites of catalysts at interfaces or on surfaces.

Various quantum embedding theories have been proposed, categorized by the level of theory used to describe different parts of the system.
\begin{itemize}
\item Density-based embedding theories define an active space in real space and partition the density of the system into the active space and the environment. The latter is described at the DFT level, and the former is described either using DFT with a higher-level exchange-correlation functional (DFT-in-DFT) or with wavefunction-based quantum chemistry methods (WF-in-DFT)~\cite{rossmannek2021quantum,rossmannek2023quantum,gao2023onsurface,roncevic2023aromaticity}.
\item Density matrix embedding theories (DMET)~\cite{knizia_density_2012,knizia_density_2013,wouters_practical_2016,pham_can_2018,hermes_multiconfigurational_2019,pham_periodic_2020,cui_efficient_2020,lau_regional_2021,cui_systematic_2022,mitra_density_2023}, as introduced previously, also define an active space in real space. The active space is treated at a higher level of theory. The interaction with the environment is accounted for through a set of bath orbitals obtained from a calculation of the full system at a lower level of theory, with an additional one-body term that ensures the density matrices at both levels of theory coincide.
\item Embedding theories based on Green's function define an active space by a set of single-particle electronic states. The non-local, frequency-dependent self-energy of the active space is expressed as a sum of terms evaluated at high and low levels of theory, with an additional term to remove the double counting. Examples of such theories include \gls{dmft} (DMFT+DFT~\cite{anisimov_first-principles_1997,lichtenstein_ab_1998} or DMFT+GW~\cite{sun_extended_2002,biermann_first-principles_2003,biermann_dynamical_2014,boehnke_when_2016,choi_first-principles_2016,nilsson_multitier_2017}), self-energy embedding theory (SEET)~\cite{lan_generalized_2017,zgid_finite_2017,rusakov_self-energy_2019}, and \gls{qdet}~\cite{ma_first-principles_2020,ma_quantum_2020,ma_quantum_2021,sheng_greens_2022}. They mainly differ in the choice of the high-level and low-level theories and the partitioning of the total self-energy into the active space and the environment.
\end{itemize}

In practice, a quantum embedding calculation typically starts with a low-level calculation of the full system, the results of which are subsequently employed to determine parameters of the many-body effective Hamiltonian of the active space. In QDET, for example, the effective Hamiltonian comprises a one-body term and a two-body term. In the one-body term, an exact double counting correction has been derived for the case where G$_0$W$_0$ is the low-level theory. In the two-body term, the effect of the environment is incorporated through an effective screening. The QDET Hamiltonian can be computed without any explicit summation over empty states, enabling its application to large systems. Leveraging GPU-accelerated classical supercomputers, it is feasible to simulate systems consisting of over a thousand atoms and active spaces consisting of over a hundred orbitals. The effective Hamiltonian, written in second quantization, can be diagonalized on classical computers using FCI, or on quantum computers~\cite{ma_quantum_2020,rungger_dynamical_2020,keen_quantum-classical_2020,kawashima_optimizing_2021,yao_gutzwiller_2021,tilly_reduced_2021,vorwerk_quantum_2022,huang_simulating_2022,huang_quantum_2023} by mapping it to qubits and quantum gates and finding its lowest eigenstates using a suitable quantum eigensolver.
Challenges remain in facilitating the communication of data from classical (super)computers to quantum computers, accommodating the effective Hamiltonian with the available number of gates and circuit depth, and mitigating noises present on near-term computers which may lead to unphysical results. Intriguingly, quantum computers themselves are employed to investigate the properties of spin defects in solids. These defects hold promise as prospective candidates for the implementation of improved quantum computers, thus forming a positive feedback loop at the intersection of computational materials science and quantum computing.

Additionally, CAS-DMET and NEVPT2-DMET described before have been effectively used to examine the adsorption and emission spectra of the negatively charged nitrogen-vacancy (NV$^{-1}$) and the neutral silicon vacancy (SiV$^{0}$) defects in diamond, as well as the neutral oxygen vacancy (OV$^{0}$) defect in both bulk and surface magnesium oxide, as reported in recent literature. \cite{mitra2021excited,haldar2023local,Verma2023} In particular, for the NV$^{-1}$ defect in diamond, NEVPT2-DMET with a significantly reduced embedding space and a complete active space (CAS) configuration of 10 electrons in 9 orbitals (denoted as 10e, 9o) accurately predicts a triplet-triplet excitation energy of 2.31 eV. This is in close agreement with the experimentally observed value of 2.18 eV. Additionally, it predicts an optically inactive singlet-singlet transition at 1.02 eV, closely approximating the experimental value of 1.26 eV. For the neutral OV$^{0}$ defect in MgO bulk, NEVPT2-DMET, with a CAS configuration of 2e, 5o, estimates the excitation energy to be 5.24 eV, aligning well with the experimental absorption maximum of 5.03 eV. Studies of the optically allowed $S_0\rightarrow S_1$ transition for the OV$^{0}$ defect on the MgO surface have also been conducted using CAS-DMET, NEVPT2-DMET, and DME-PDFT methods. However, varying experimental results ranging from 1-5 eV have been reported, which makes direct comparison difficult. Using MP2 and CCSD solvers for the DMET embedding subspace, the adsorption energies for CO on MgO have been calculated using the largest embedding subspaces, and these have been found to be within 1.2 kcal/mol of the non-embedding reference values. \cite{Mitra2022} So far, the LASUCC and LASSI methods have primarily been tested on strongly correlated molecular systems and has been utilized to study the dissociation of (H$_2$)$_2$ into two H$_2$ molecules, the dissociation of the two double bonds in trans-butadiene, and the $J$-coupling in tris-(\(\mu\)-hydroxo)-bridged chromium compound model~\cite{Pandharkar2022,Otten2022}. The accuracy of these studies has been confirmed through excellent agreement with the results from corresponding CASCI calculations performed using LAS orbitals.

\subsubsection{Model systems}

Low-energy models introduce an effective Hamiltonian describing the low-energy degrees of freedom of a quantum system. The effective Hamiltonian is then solved for the ground- and low-energy excited states.
How to propose a low-energy effective Hamiltonian, and how to propose reliable and precise parameters for that Hamiltonian, are important questions that affect the predictive power of that model for real-world applications.
On the other hand, accurate solutions of a given model allow us to capture trends in physical properties as the free parameters of the model vary (e.g. temperature, interaction strength, number of electrons).
Therefore, the value of a model as it relates to real-world application lies both in its definition and in its computational solution: the physical properties predicted by a model should be equally robust to perturbations in model parameters and solutions of the Schrödinger equation, for robust inferences about the physical properties of a real system to be drawn from the use of effective low-energy models.

Although band structure and Fermi liquid theory are used as the primary models for quantum particles in materials, these descriptions can fail to capture actual physical properties, for example in high-T$_c$ cuprates and other transition-metal compounds like spin liquids.
This offers a challenge in how to improve their performance for such systems.
Various models have been suggested for these systems, such as the Emery and Heisenberg models reviewed below.
However, the ability of such models to capture actual physics of interest is again often unclear, even though they have been widely studied and have greatly advanced our understanding of correlated electron physics.
Furthermore, these models can themselves be challenging to solve.

A typical approach to connecting ab initio electron-level models and large-scale models is downfolding using DFT.
In this method, DFT bands are modeled by projecting them onto localized Wannier functions, which yields one-particle hopping terms~\cite{pavarini2001band}.
Interactions are added by modeling screening, for example using constrained DFT or random phase approximations (RPA).
Finally, since the original DFT bands also already incorporated interactions, this double-counting must be corrected.
This approach and related downfolding techniques continue to be actively developed and broadly used~\cite{haule2015exact,aryasetiawan2004frequency,jeschke2013first,haule2015exact,zhou2010construction,white2002numerical,yanai2006canonical}.

In the remainder of this Section, we review some important model systems, namely spin Hamiltonians and lattice models of correlated electrons. While the relationship between these models and real materials is very delicate, both due to the choice of the effective Hamiltonian and its parameterization, studying computationally the phase diagram of these models as a function of their free parameters can reveal trends and phase transitions that are relevant for the physics of real materials. Furthermore, it constitutes a promising research area for simulations on current and near-term quantum devices.

\paragraph{Spin Hamiltonians}

Spin Hamiltonians are frequently used to model the magnetic properties of materials as well as other systems where there is local correlation or frustration. Consider a lattice with $n$ sites occupied by spin-$s$ particles that are allowed to interact with each other.
In such a lattice, there is an allowed number of $2s+1$ microstates at each lattice site, giving a total of $(2s+1)^n$ spin configurations for a lattice with $n$ sites. 
At an infinitely high temperature, spins point randomly in all directions, agitated by thermal motion. 
On the other hand, as the temperature approaches $T=0K$, in many situations spins display the tendency to become part of a long-range magnetically ordered state under the effect of the interaction between them. The two extreme limits are antiferromagnetic order, where adjacent spins point in opposite directions, and ferromagnetic order, where all spins point in the same direction.
If we consider paramagnetism (namely the absence of magnetic order in the absence of an externally applied magnetic field) as analogous to an ideal gas (where particles interact weakly with each other), and magnetic order as analogous to a solid (where particles form a static and regularly repeating lattice), it is natural to inquire if there exists an analogous liquid of spins.

Spin liquids \cite{savary2016quantum,zhou2017quantum,chamorro2020chemistry,sears2017phase,PhysRevA.106.022434,li2023benchmarking} are exotic materials where spins do not order at any finite temperature but interact strongly with each other, unlike in a paramagnet where spins are weakly coupled to each other.
Normally, strong spin interactions result in a single spin configuration (or a small subset thereof) having lower energy than the others, thereby driving magnetic order.  In a spin liquid, interactions conspire to make all spin configurations (or a large subset thereof) nearly equivalent in energy, thereby evading magnetic order.
In a spin liquid state, the ground state consists of a superposition of spin configurations that is not dominated by a single configuration (or a small subset thereof). A spin liquid thus does not possess magnetic order, as individual spins fluctuate, but possesses strong entanglement in the ground state, so that spins are not randomly mutually oriented.

The key ingredient to obtain a ground state that is a superposition of a large number of spin configurations is to have strong interaction between neighboring spins and have a large number of configurations with near-identical energies, i.e. to frustrate the spins. This often arises due to the underlying geometric frustration of the lattice in which spins sit and/or the nature of the interaction between spins. Furthermore, particles with higher spin $s$ tend to show more classical behavior as opposed to quantum behavior, because the energy barrier between different microstates scales as $s^2$, which strongly suppresses quantum fluctuations. Therefore, a spin as low as possible, e.g. $s=1/2$, greatly favors spin liquid behavior.

\begin{figure}[htbp]
\centering
\includegraphics[width=\columnwidth]{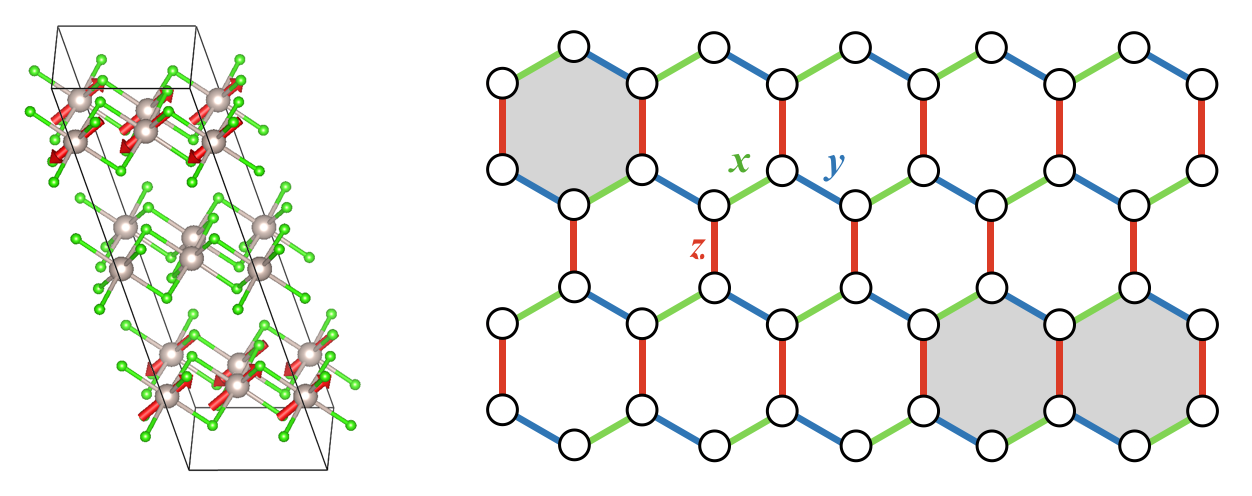}
\caption{Left: crystal structure of $\alpha$-RuCl$_3$, illustrating the van der Walls gap between the RuCl$_6$ honeycomb layers (with Ru/Cl atoms represented as gray/green circles) and magnetic Ru$^{3+}$ ions aligned as in the zigzag phase (red arrows). Right: schematic representation of the two-dimensional Heisenberg-Kitaev model with spins represented as white circles and $x/y/z$ bonds as green/blue/red links respectively. Quantum simulations to date have focused on 1 and 2 hexagons, corresponding to 6 and 10 spins respectively \cite{tazhigulov2022simulating}.}
\label{fig:rucl3}
\end{figure}

A concrete example of a material of interest as a proximate spin liquid \cite{chamorro2020chemistry} is ruthenium trichloride, $\alpha$-RuCl$_3$ (see Fig.~\ref{fig:rucl3}). This transition metal halide has a crystal structure made up of stacked honeycomb layers of edge-sharing RuCl$_6$ octahedra. Plumb et al \cite{plumb2014alpha} found that spin-orbit coupling in this material is substantial, leading to a $j=1/2$ description of the Ru$^{3+}$ valence electrons. 
Since this material is built up with edge-sharing RuCl$_6$ octahedra, it is believed \cite{kitaev2006anyons,jackeli2009mott} that its low-energy excitations are parametrized by the Kitaev-Heisenberg Hamiltonian
\begin{equation}
H_{kh} = \sum_{\langle ij \rangle_\gamma} J \vec S_i \cdot \vec S_j + K S_{i \gamma} S_{j \gamma} \;,
\end{equation}
where $\langle ij \rangle_\gamma$ denotes summation over the bonds $\gamma \in \{x,y,z\}$. This feature makes $\alpha$-RuCl$_3$ a material of great interest in the ongoing search for a Kitaev spin-liquid ground state \cite{koitzsch2016j,nasu2016fermionic,banerjee2016proximate}. Although $\alpha$-RuCl$_3$ orders magnetically at low temperature with zigzag magnetic order \cite{banerjee2016proximate,sears2015magnetic}, this material has shown some signatures of spin-liquid physics, such as a broad continuum of magnetic excitations identified in both Raman scattering \cite{sandilands2015scattering} and inelastic neutron scattering measurements \cite{banerjee2016proximate}.
Therefore, the degree of similarity to the Kitaev model, as well as the interpretation of spectroscopic and thermal measurements in terms of modifications to the Kitaev model, is currently intensely debated. We remark that, in the literature \cite{maksimov2020rethinking}, the Kitaev-Heisenberg Hamiltonian is augmented by additional couplings, 
\begin{equation}
\begin{split}
H &= H_{kh} + \sum_{\langle ij \rangle_\gamma} \Gamma ( S_{i\alpha} S_{j\beta} + S_{i\beta} S_{j\alpha} ) \\
&+ \sum_{\langle ij \rangle_\gamma} \Gamma^\prime ( S_{i\gamma} S_{j\alpha} + S_{i\gamma} S_{j\beta} + S_{i\alpha} S_{j\gamma} + S_{i\beta} S_{j\gamma} )  \;,
\end{split}
\end{equation}
with $\{\alpha,\beta,\gamma\} = \{x,y,z\} , \{z,x,y\}, \{y,z,x\}$ for $\gamma = z,y,x$ respectively. Several parameter configurations have been proposed to describe the Hamiltonian $H$, ranging from first-principle methods \cite{yadav2016kitaev,maksimov2020rethinking} to phenomenological analyses \cite{banerjee2016proximate,maksimov2020rethinking}. Recent literature has focused on $J = -1.53$, $K= -24.4$, $\Gamma = 5.25$, $\Gamma^\prime=-0.95$ \cite{tazhigulov2022simulating}. The coupling that is believed to be the leading one is the (negative) Kitaev term $K<0$. The off-diagonal exchange term $\Gamma>0$ is also believed to be significant, while the ferromagnetic exchange $J$ is believed to be subleading \cite{winter2017models}.
However, it should be noted that the parameters of $H$ vary quite significantly between the studies \cite{winter2016challenges,maksimov2020rethinking}. The exactly solvable Kitaev model corresponds to $J=\Gamma=\Gamma^\prime=0$, and it is argued that, in the parameter regime of $\alpha$-RuCl$_3$, both the excitations and the heat capacity show echoes of the two kinds of Majorana fermions that exist at the solvable point \cite{gohlke2017dynamics,laurell2020dynamical}.

The combination of rich physical behavior and challenging ground- and thermal-state preparation makes quantum spin liquid models like the Kitaev-Heisenberg Hamiltonian approximating $\alpha$-RuCl$_3$ interesting and challenging problems for classical and quantum simulations. The presence of $2$-local spin-$1/2$ Hamiltonians makes them compelling targets for simulations on quantum devices, focused on the calculation of specific heats and dynamical structure factors relevant to understanding the low-energy excitation spectrum of the Hamiltonian.

\paragraph{Lattice models}

A central feature of correlated electron materials is the competition between different inhomogeneous orders. Important examples are metal-insulator \cite{imada1998metal}, paramagnet-ferromagnet \cite{spaldin2010magnetic}, and conductor-superconductor phase transitions \cite{orenstein2000advances,dagotto1994correlated}.
Lattice models of correlated electrons serve as the canonical microscopic physical models for the computational description of such competition phenomena in materials.
A paradigmatic example of a fermionic lattice model is the one-band Hubbard model \cite{hubbard1964electron,kanamori1963electron,gutzwiller1963effect,arovas2022hubbard},
\begin{equation}
\label{eq:hubbard}
\hat{H} = -t \sum_{\langle ij \rangle \sigma} ( \hat{a}^\dagger_{i\sigma} \hat{a}_{j\sigma} + \mathrm{h.c.} ) + U \sum_i \hat{n}_{i \uparrow} \hat{n}_{i \downarrow} \;.
\end{equation}
In Eq.~\eqref{eq:hubbard} where $\hat{a}/\hat{a}^\dagger$ denote the usual fermion annihilation/creation operators, $\hat{n} = \hat{a}^\dagger\hat{a}$ is the number operator, $t$ and $U$ are the kinetic and repulsion energies,
and $\langle ij \rangle$ denote summation over nearest neighbors sites $ij$ on a lattice (e.g. the 2D square lattice).
The Hubbard model is one of the simplest models of interacting fermions, but despite its simplicity, it exhibits a wide range of correlated electron behavior including interaction-driven metal-insulator transitions, superconductivity, and magnetism. 
The precise behavior depends delicately on parameters, creating an interesting challenge for theory and computation.
Exact solutions are available for one-dimensional \cite{lieb1968absence} and infinite-dimensional cases \cite{metzner1989correlated}. 
High-temperature series expansions provide numerically exact results but only for temperatures too high to be relevant for physically interesting situations \cite{oitmaa2006series}. 
In general dimensions at relevant temperatures, only approximate numerical solutions are available, including wavefunction-based, diagrammatic, and embedding methods \cite{leblanc2015solutions}.
These numerical methods have been steadily developed and carefully benchmarked \cite{leblanc2015solutions,zheng2017stripe}, providing illuminating insights into the physics of this model.

\begin{figure}[h!]
\centering
\includegraphics[width=\columnwidth]{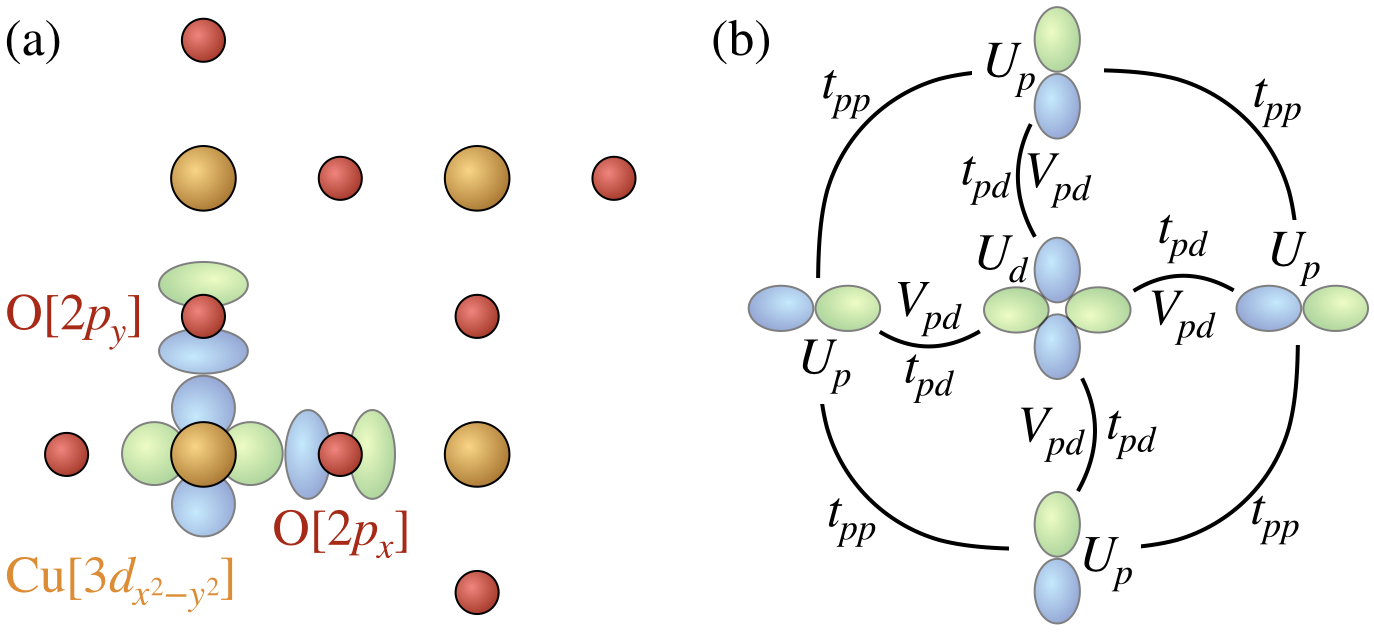}
\caption{(a) schematic view of a [CuO$_2]_4$ cell in a cuprate CuO$_2$ plane. Cu/O atoms are represented with orange/red circles and orbitals by blue and green contour plots (blue/green for positive/negative values). (b) definition of the three-band Hubbard model parameters (the curve connectors represent the kinetic energy coefficients).
}
\label{fig:lattice}
\end{figure}

One concrete example of a fermionic lattice model that can be connected to a technological application is the three-band Hubbard model in two dimensions, also known as the Emery model \cite{emery1987theory}. 
This model is generally believed to contain the essential properties of high-T$_c$ cuprate superconductors that arise from the interplay between the copper $d_{x^2-y^2}$ and oxygen $p_{x,y}$ orbitals in the CuO$_2$ layers (see Figure \ref{fig:lattice}). It is defined, in the so-called ``phase representation'', by the Hamiltonian
\begin{equation}
\label{eq:emery}
\begin{split}
\hat{H} &= 
t_{pd} \sum_{\langle ij \rangle \sigma} (\hat{d}^\dagger_{i\sigma} \hat{p}_{j\sigma} + \mathrm{h.c.} )
+ t_{pp} \sum_{\langle ij \rangle \sigma} (\hat{p}^\dagger_{i\sigma} \hat{p}_{j\sigma} + \mathrm{h.c.} ) \\
&- \Delta_{pd} \sum_{i\sigma} \hat{n}^d_{i \sigma} \\
&+ U_d \sum_i \hat{n}^d_{i\uparrow} \hat{n}^d_{i\downarrow} 
+ U_p \sum_i \hat{n}^p_{i\uparrow} \hat{n}^p_{i\downarrow} 
+ V_{pd}  \sum_{\langle ij \rangle \sigma \tau}  \hat{n}^p_{i\sigma} \hat{n}^p_{j\tau} \, \\
\end{split}
\end{equation}
where $\langle \rangle$ denotes nearest neighbors, $d^{(\dagger)}_{i\sigma}/p^{(\dagger)}_{i\sigma}$ destroys (creates) a hole with spin $\sigma$ on the Cu[3d]/O[2p] orbital, $\hat{n}^d_{i\sigma}/\hat{n}^p_{i\sigma}$ is the number of holes on a Cu[3d]/O[2p] orbital with spin $\sigma$, and the charge transfer gap $\Delta_{pd}$ is defined as the difference between the O[2p] and Cu[3d] orbital energies.

On classical computers, the Emery model has been investigated with several numerical methods, including direct simulations of lattices by exact diagonalization of up to 24 sites \cite{greiter2007no}, 
quantum Monte Carlo and DMRG of 80 and 144 sites \cite{huang2017numerical,vitali2019metal}, and embedding methods like density matrix embedding theory with 12 sites \cite{cui2020ground}.
These simulations have been primarily directed at simulating ground-state spin-spin, and $d$-wave correlation functions, to establish e.g. the presence of stripe or $d$-wave superconducting order \cite{greiter2007no,huang2017numerical,vitali2019metal,cui2020ground}.

Due to the complexity of the model, unlike in the one-band case, a consensus on much of the physics has yet to be reached, 
providing an opportunity for high-accuracy quantum computing algorithms to elucidate the properties of high-T$_c$ cuprate superconductors by simulating lattices of O(100) sites.

The Emery model is often simplified considering a one-band Hubbard model. While much of the physics of high-T$_c$ cuprate superconductors,
e.g. $d$-wave pairing, density waves, pseudogap phase, stripe order, is described by the one-band Hubbard model Hamiltonian,  there are still important reasons to go beyond the one-band picture to study the original three-band model directly. For instance, (a) some important physics may be lost in the reduction to the one-band approximation (such as a role for the oxygen degrees of freedom in the pseudogap phase \cite{fauque2006magnetic}), (b) near degeneracies of competing states seen in the one-band case \cite{zheng2017stripe} may be resolved with the additional degrees of freedom of the three-band model, and (c) the three-band model retains the atomic structure of the CuO$_2$ layer and thus has a direct link to the structure of real materials as well as experimental measurements of orders at the atomic scale \cite{hybertsen1989calculation,martin1996electronic,hanke20103}.

\subsection{Excited states}

Many experiments on condensed-matter systems probe dynamical properties, rather than equilibrium properties.
For example, material properties are often explored through scattering experiments, such as neutron and photon scattering or angle-resolved photoemission spectroscopy (ARPES)~\cite{Damascelli_2004}, 
which probe the structure factor and spectral function respectively~\cite{10.1063/1.5114818,2023arXiv231004333W}.
In these experiments, a system is initialized at equilibrium at inverse temperature $\beta$ under a Hamiltonian $\hat{H}$, and a time-dependent perturbation $\hat{V}(t)$ is applied. The response of the system is characterized by linear response theory, computing frequency-dependent correlation functions of the form
\begin{equation}
\label{eq:cabw}
C_{AB}(\omega) = \int_{-\infty}^{\infty} \frac{dt}{2\pi} \, e^{i\omega t} \, \mbox{Tr}[ \hat{\rho}(\beta) e^{it \hat{H}} \hat{A} e^{-it \hat{H}} \hat{B} ] \;.
\end{equation}
where $\hat{A},\hat{B}$ are suitable operators, e.g. components of the dipole moment operator and density operators at two different spatial points in UV/visible light and neutron scattering experiments respectively.
It should be noted that $C_{AB}(\omega)$ can be computed knowing the eigenpairs of the unperturbed Hamiltonian $\hat{H}$ and/or its time evolution operator $e^{-it \hat{H}}$.
In other words, by virtue of the linear response approximation, the perturbation $\hat{V}(t)$ does not appear in the function $C_{AB}(\omega)$.

Going beyond spectral properties, the {non-equilibrium real-time dynamics} of quantum systems have increasingly come into focus, both because of experiments that can probe quantum dynamics at atomic scales, 
{and because of fundamental interest in studying the equilibration of quantum systems, which serves as a bridge between the theories of quantum mechanics and statistical mechanics}. 
Experimental setups now exist that can probe ultra-fast dynamics in materials including, for example, free-electron lasers~\cite{patterson2010coherent,weathersby2015mega} as well as other pulsed laser systems. 
These allow the application of experimental techniques, such as pump-probe spectroscopy~\cite{fischer2016invited,buzzi2018probing}, to provide novel insights into the behavior of correlated quantum systems.
Within non-equilibrium real-time dynamics, one is often interested in the calculation of time-dependent expectation values, $B(t) = \mbox{Tr}[ \hat{\rho}(\beta) \hat{U}(t,0)^\dagger \hat{B} \hat{U}(t,0) ]$.
The time-evolution operator $\hat{U}(t,0)$ under $\hat{H}+\hat{V}(t)$ can be written as a time-ordered exponential, yielding the following expression for $B(t)$, based on a Keldysh contour integral,
\begin{equation}
\begin{split}
B(t) &= \sum_{nm=0}^{\infty} i^{n-m} \int_0^t dr_n \dots dr_1 \int_0^t ds_m \dots ds_1 \\
&\mbox{Tr}[ \hat{X}(r_n \dots r_1)^\dagger \hat{B} \hat{X}(s_m \dots s_1) ]
\end{split}
\end{equation}
where $r_1 < \dots <r_n$, and $s_1 < \dots <s_m$, and 
\begin{equation}
\begin{split}
\hat{X}(s_m \dots s_1) &= \hat{U}_0(t ,s_m) \hat{V}(s_m) \hat{U}_0(s_m , s_{m-1}) \dots \\
&\dots \hat{U}_0(s_2 ,s_1) \hat{V}(s_1) \hat{U}_0(s_1 ,0) \;, \\
\end{split}
\end{equation}
with $\hat{U}_0(t,t^\prime) = e^{-i (t-t^\prime) \hat{H}}$.
Linear-response frequency-dependent correlation functions and non-equilibrium real-time correlation functions are very relevant physical quantities, they are challenging to compute on classical computers because 
they require access to the time dynamics of a system (or equivalently to its excited states), and they are very suitable tasks for fault-tolerant quantum computers, due to their ability to simulate complex unitary transformations
like time evolution operators at polynomial cost and with controllable approximations only.

\subsection{Applications of electronic structure}

\subsubsection{Vibrational structure calculations}

Obtaining accurate vibrational spectra of molecules is a costly task on conventional computers. While uncovering the electronic structure of molecules stands as a fundamental challenge in quantum chemistry and material design, to truly make an impact in both scientific research and practical applications, it is vital to go beyond the electronic structure. This requires creating a kinetic model that relies on a deep understanding of a molecule's vibrational structure.  Knowing a molecule's vibrational structure enables the prediction of thermodynamic properties that are key in many fields, such as atmospheric science, catalysis, and fuel combustion modeling.  Although classical computers often handle simulation of the electronic structure of small molecules reasonably well, they struggle with calculating vibrational structures beyond the harmonic approximation, even for small molecules. Computational challenges emerge when higher-order terms are involved due to deviations from harmonicity and also the interplay between different bosonic modes. This can be described, e.g, by quartic force field Hamiltonian~\cite{ollitraultCS2020}
\begin{equation}
\begin{split}
\label{eq:vibrational}
H_{\rm{anharm}}&=\frac{1}{2} \sum_i^M \omega_i\left(\, q_i^2\, + p_i^2\, \right) +\sum_{\{ijk\}} h_{ijk}\, q_iq_jq_k \, +
\nonumber \\
&+  \sum_{\{ijk l \}} h_{ijkl}\, q_iq_jq_k q_l +\cdots \, , 
\end{split}
\end{equation}
where the first term  indicates the harmonic approximation and the remaining terms are the 
anharmonic corrections, with $h$ being the force constants.

Applications of quantum computing in calculating molecular vibrational structures have not been extensively explored, see the references in~\cite{QuantumAlgorithmsSrvey_AWS2023,VibResourceEstimate_IBM2023,2023arXiv230613126S}. While there have been a few proposals for both near and long-term quantum algorithms to address the vibrational structure problem, the implementation on currently available devices requires developing efficient quantum algorithms for near-term devices as well as a better understanding of possible optimizations to minimize the qubit numbers and circuit depth. Recent studies on quantum resources required for vibrational structure calculations on quantum computers indicate that the combined resources needed for achieving quantum advantage in vibrational structure problems might be lower than those for electronic structure problems~\cite{NSawaya2021,VibResourceEstimate_IBM2023}. The vibrational qubit Hamiltonian's Pauli strings are more localized, making them potentially easier and faster to simulate compared to those in the electronic qubit Hamiltonian. It is noted that the number of Pauli strings simultaneously executed in a quantum circuit during a single Trotter step is also larger for the vibrational structure problem compared to the electronic one. 

Enhancing vibrational structure calculations can be achieved through two possible approaches. One is centered on refining the accuracy of the electronic \gls{pes} using quantum computers. This method proves effective when dealing with rigid molecules, especially those where the vibrational structure is well-described by harmonic approximations or classical algorithms. The second approach focuses on elevating vibrational structure calculations through quantum computing, assuming a sufficiently accurate electronic energy surface. This becomes particularly crucial in characterizing floppy molecules and capturing anharmonic effects.
One pertinent example of the latter case, with industrial significance, is polyyne molecules. These organic compounds, characterized by alternating carbon single and triple bonds, are challenging to stabilize. They belong to high symmetry point groups and feature numerous silent modes, undetectable through infrared or Raman spectroscopy. A precise depiction of their low-frequency vibrations is crucial for understanding their reactivity. Polyynes play a pivotal role in floating catalytic chemical vapor deposition and the emerging technology of carbon nanotubes, ultimately contributing to reducing the carbon footprint. Precise numerical results on the quantum computational cost for polyyne molecules have been provided in the recent study~\cite{VibResourceEstimate_IBM2023}.
Another recent study has explored the first approach, examining how a hybrid quantum and classical computation could improve the vibrational structure simulations~\cite{StoberPRA2022}. For a model lithium hydride system, the PES is computed on quantum hardware, and then classical computation is used to both fit analytic forms to the PES and from those generate the vibrational energy levels that can be used to predict thermodynamic properties.  This hybrid algorithm was demonstrated with actual calculations on quantum hardware, though the authors point out that extensions to systems with more degrees of freedom may require innovations to avoid the combinatorial complexity of sampling the entire space of molecular deformations.

One more classic example of a multiscale problem that couples electronic, vibrational, and transport phenomena is photosynthesis.  The core process is the conversion of electromagnetic energy to a stored chemical form, which involves photon absorption plus exciton transmission, transduction, and dissipation.  This is all done by biological systems with optimized electronic-vibronic coupling to move energy, with estimated photochemical yields for some subsystems of up to 85\% \cite{DEMMIGADAMS2003707}.  It has been shown recently that these systems may even include alternate pathways to dissipate energy and avoid damage to the key molecular components ~\cite{doi:10.1073/pnas.2018240118}.  
Multi-scale modeling of photosynthesis has been a long-standing area of interest in biological simulation \cite{ritz2002quantum}.  Simplified molecular simulations can provide a model for vibrational modes, but they need to be combined with exciton transfer calculations as well as detailed simulations of reactive chemistry to create the full picture.  The exciton transfer calculations are a good example of a quantum process with a relatively simple Hamiltonian, as the number and geometry of donor and acceptor species are known.  Fluctuations in their geometry, however, require coupling to a simpler simulation model.  The last chemical step then requires accurate quantum chemical modeling of complex systems.  In photosystem II, for example, the key chemical step is using the excited electron for water oxidation.  This happens in a small active site complex of ten mixed magnesium, calcium, and oxygen atoms, and the process is still incompletely understood \cite{Cox2020ARB}.

\begin{figure}
    \centering
    \includegraphics[width=\columnwidth]{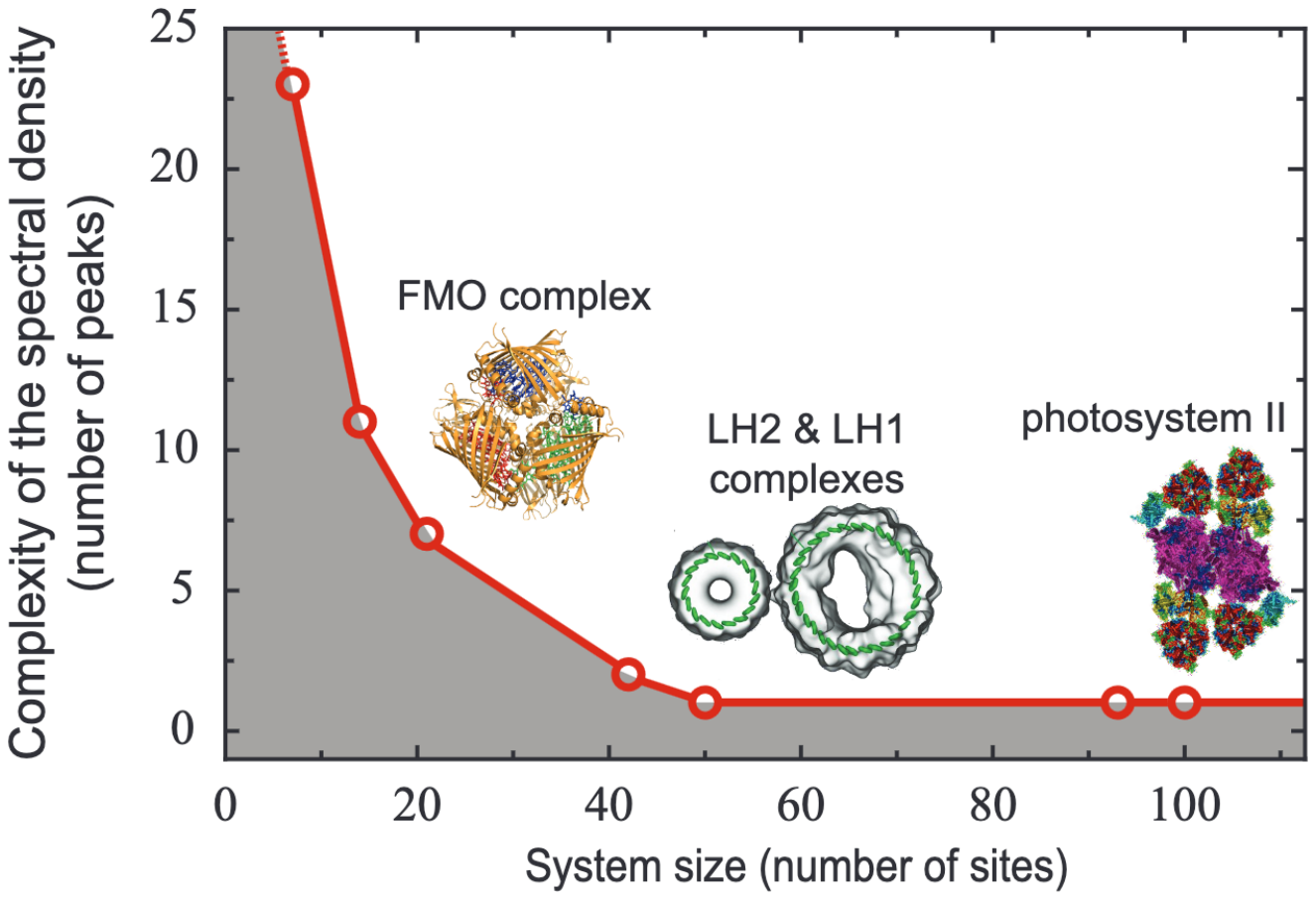}
    \caption{The shaded region in the graph illustrates the estimated sizes of treatable systems when simulating Frenkel exciton Hamiltonians using current classical supercomputing resources.  Three photosynthetic systems are presented: the Fenna-Matthews-Olson complex of Green-Sulfur Bacteria, the Light Harvesting I and II complexes of Purple Bacteria, and Photosystem II of higher plants. The simulation was conducted utilizing the hierarchical equations of motion approach on 64 AMD Opteron cores, utilizing a total of 250 GB of RAM. Image from~\cite{Mostame2016}.}
    \label{fig:Frenkel}
\end{figure}

Photosynthetic complexes, modeled as open quantum systems interacting with a vibrational environment, have been the focus of various analytical and numerical studies. However, the computational resources required by most classical methods increase exponentially as the number of particles in the simulation grows. This challenge intensifies significantly for open quantum systems with structured environments, such as photosynthetic complexes. While numerically exact solutions are achievable for only small systems (fewer than 20 particles) and under restrictions concerning bath modes, the exploration of more complex open quantum systems remains a challenging computational barrier. This limitation is illustrated in Figure \ref{fig:Frenkel}, indicating the current upper limits of computational capabilities on classical computers for simulating the dynamics of such open quantum systems. As it is shown, there is a trade-off between the complexity of the spectral density and the system size \cite{Mostame2016}.
Various proposals have surfaced for simulating these complex systems on quantum hardware, including gate-based simulations, a hybrid quantum–classical approach, as well as analog quantum simulators, see the references in \cite{Jaderberg_2022}, for example. Nevertheless, conducting large-scale simulations on existing quantum hardware poses a challenge, due to the limitations, e.g., circuit depth, in currently available quantum devices, so near-term progress will likely depend on hybrid quantum-classical algorithms or better optimization and resource reduction techniques.

\subsubsection{Coarse-grained simulations}

The commercialization of new, advanced materials is often a cost-, time-, and labor-intensive process.  Typical times to bring a new material to market are on the order of 10-20 years from initial research to first use \cite{national2011materials}.
The long, expensive commercialization process is partially due to the experimental nature of materials development, e.g., many material samples must be physically prepared and then evaluated.  Challenges in the process include the sample preparation, which for new materials has a steep learning curve, with many iterations of learning required to identify the best methods to create high-quality samples for evaluation.  The precursors and by-products for sample preparation may be expensive, hard to obtain, and/or unsafe, thus requiring specialized equipment to handle.  Further, evaluation of the samples may be laborious with many data points to be taken under carefully controlled conditions.  Finally, more often than not, once a sample is finally prepared and evaluated, the conclusion is that the material does not meet the desired specifications, and the entire process must be repeated for a different material candidate.

The materials commercialization process may be accelerated by conducting material exploration computationally.  New materials may be evaluated quickly and efficiently by virtually modeling a material and predicting its properties.  The role of experimentation is moved from exploration to confirmation of the computational results. 

In computational materials discovery, the process often has two steps.  The first identifies or generates material candidates, which may be done, for example, with machine learning methods.  In the second step, the material candidates are filtered by computationally predicting their material properties.  Methods to compute material properties are discussed here.

One type of material property in demand is those that are both light and strong. The main modeling for these concerns two main areas. The first is materials discovery, in which computation is used to test predicted materials' stability. This includes stability against spontaneous transition to alternative undesirable phases with the application of small amounts of pressure or temperature change; or indeed stability against phase separation, in which constituent atoms cluster together and do not form compounds or layers as desired. The computational tests for this include perturbing atomic positions and seeing minute energy changes in order to infer reaction barriers. These are DFT calculations using often nudged elastic band techniques. Additional tests are done by calculating the phonon spectra. A state-of-the-art set of such tests appears in \cite{long2023two}, pages 6-8.

   The second main area of modeling for advanced materials is defect studies: crack propagation and corrosion are two key areas. Both of these defects can be either across a surface in the form of failure of protective coatings and surface cracks; or deep into a surface, whether a crack or deep pit corrosion. Failure of mechanical properties occurs on all length scales, starting at the atomic and spreading to macroscopic, visible effects on the scale or centimeters or longer. For this, the main modeling approach is a type of embedding theory as introduced above. The essence of embedding techniques is a set of calculations that are done at much higher accuracy, usually at the cost of both time and resources, in a key important region chosen as the active space. The results of the embedding calculations then need to be matched to those done at a coarser level, a challenging task to do well, and which difficulty depends to some degree on the type of embedding done, see the paragraph below. For example, if the type of embedding is by calculating quantum wave functions in the active space, connecting to classical more macroscopic descriptions of phonon and other vibrational levels can be a very challenging task. However, if instead the active space consists of precise calculations of atomic positions at zero temperature, then these can be connected to those at finite temperature by a process of ramping up vibrations, and this can work quite well. When the defect to be modeled, such as a crack, occurs at multiple length scales, then each scale working up from the smallest to the largest can be considered the embedded active space of the one that comes at a larger scale, and matching must be done at each stage. This is termed multi-scale quantum embedding. 
   
   Both for crack propagation and defect modeling such as corrosion of the key material or failure of a coating, often first-principles molecular dynamics is also used, with DFT used to calculate the forces and energies. These calculations are used both to test the matching at the different scales and also as a study tool in itself, to try to infer mechanisms for materials failure. With a crack, the action can spread very quickly from the atomic to the macroscopic, if forces are strong enough, and MD is key for measures of the speed of such scale transitions. For other kinds of defects such as wear and corrosion, processes typically happen on a much slower time scale, and deeper calculations can be done at the atomic level to study the origins of the failure. However, in almost all cases, these, too, are done with quantum embedding methods of detailed fully quantum calculations, often extending to full-configurational chemistry, for the active regions.

   Other types of material challenges are in handling creep, a slow deformation when a material is exposed to long-term stress (pressure, force, twist). These types of slower macroscopic changes are usually treated with differential equations, which have optimal algorithms in quantum computing, particularly if the problem can be linearized. Additional observables such as Young's modulus and Poisson's ratio measure the properties of materials and can reveal defects if the material is not up to the expected level. Methods to compute these properties are also on a macroscopic scale and use both differential and integral equations. 

   When materials are exposed to magnetic or electrical fields, we term these open quantum problems. Systems in magnetic or electrical fields are known as driven systems, and those at finite temperatures as dissipative systems. Thus one can have a driven, dissipative system. These types of systems are usually handled with lattice models, as discussed in a previous Subsection.

\subsubsection{An example use case: quantum metamaterials}

The above use cases show how quantum and classical computing can effectively work together to answer key questions in materials design.  As noted in our criteria, such modeling must be relevant to real-world problems, challenging to simulate classically, include feasibly solvable sub-problems on quantum computers, and potentially involve classical HPC and quantum computers in concert.  The overall evolution of simulation in materials discovery generally follows the three phases of phenomenological simulation for insight, property prediction for candidate screening, and finally direct optimization of materials for target properties.  To illustrate this further, we explore an example use case of metamaterials in more detail.

The design of metamaterials, characterized by their unique properties like negative refractive index and selective spectral reflection, holds immense promise for diverse real-world applications \cite{doi:10.1126/science.1096796,10.1038/natrevmats.2016.1,10.1038/natrevmats.2017.66,10.1038/s41928-019-0257-7}. These artificially structured materials, functioning at sub-wavelength scales, offer a novel path to enhance the performance of conventional electronic devices. This improvement, in turn, has the potential to revolutionize energy-saving in fields such as thermophotovoltaics, heating, ventilation, and air conditioning systems. Consequently, the pursuit of optimal metamaterial designs has become an area of intense research interest.

\begin{figure}
\includegraphics[width=\columnwidth]{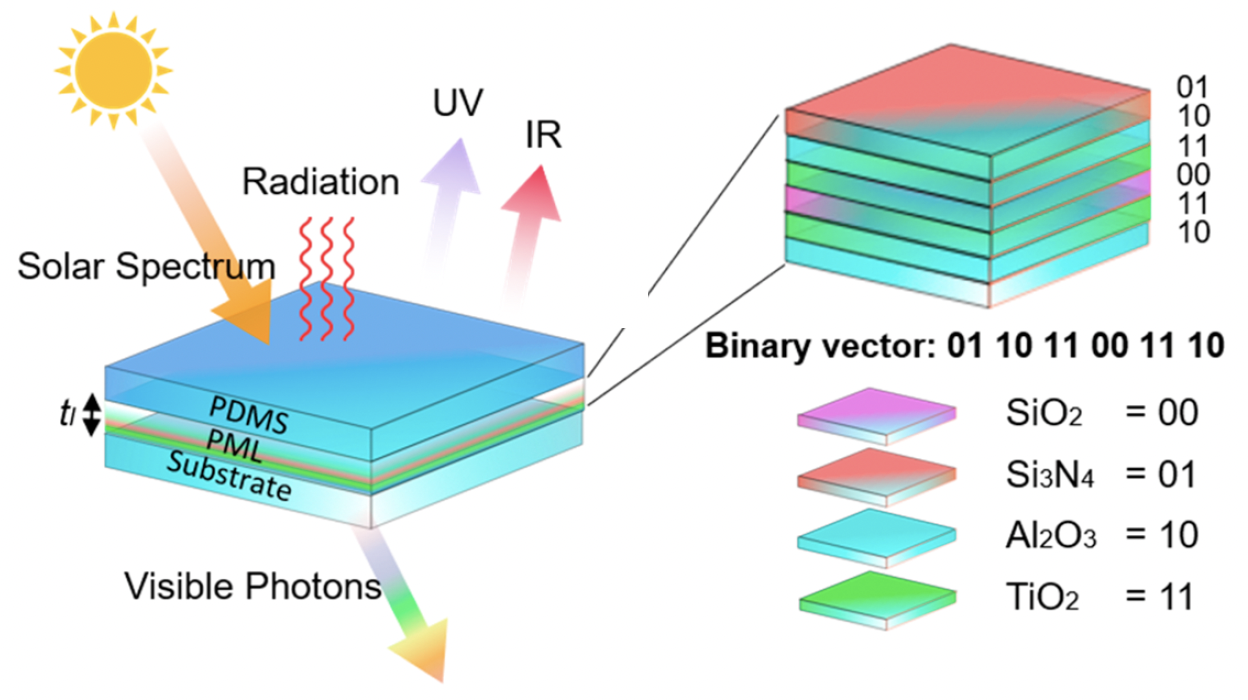}
\caption{The schematic structure of the planar multi-layer in the transparent radiative cooling metamaterial that can be mapped into a binary vector (from \cite{10.1021/acsenergylett.2c01969}).}
\label{fig:metamaterial}
\end{figure}

The properties of metamaterials hinge not only on their inherent material characteristics but also on their geometric parameters. Therefore, determining the ideal geometrical parameters is fundamental to achieving high-performance metamaterials. However, the challenge arises from the multi-dimensional nature of these parameters, creating vast, non-convex design spaces. As such, finding the optimal metamaterial structures through intuition, trial-and-error, exhaustive searches, or random approaches becomes a complex and laborious task.

This leads to essential questions in metamaterial design:
How can we identify high-performance metamaterials that are close to the global optimal design?
How can we optimally design metamaterials with complex geometric parameters that lead to significantly large design spaces?
To address these questions, a novel strategy is proposed - an active learning algorithm that harnesses the power of machine learning, quantum computing, and HPC to efficiently design metamaterials, even in cases with complex target properties and expansive search spaces. Metamaterials, often consisting of discretized design concepts, necessitate effective solutions for combinatorial optimization problems. Quantum computing, exploiting the unique principles of quantum physics like superposition, entanglement, and tunneling, offers exponential computational speedup over classical computing \cite{10.1038/nphys2900}.

To solve optimization problems using quantum computing processors, an objective function is formulated as a Hamiltonian, describing the relationship between metamaterial states and their properties. Factorization machine, a supervised machine learning model, can effectively capture the relationship between input variables and output values, providing model parameters such as bias, linear terms, and quadratic terms. These parameters can be directly utilized to construct a Hamiltonian matrix \cite{5694074}. Consequently, an iterative algorithm that combines factorization machines \cite{PhysRevResearch.2.013319} and quantum computing can efficiently discover optimal metamaterial designs.

Despite the efficiency of quantum computing in solving Hamiltonian objective functions, current quantum devices have a limited number of qubits. Thus, their utility for numerical simulations is restricted. Parallel computing, particularly HPC, holds the potential to provide substantial acceleration for such tasks. Therefore, the integration of machine learning, quantum computing, and HPC emerges as a powerful approach for the efficient design of complex metamaterials.
One more question emerges: How can continuous variables be optimized using a quantum computing-based algorithm, which typically returns binary variables? To address this limitation, another quantum computing-based algorithm is proposed \cite{PhysRevResearch.4.023062}, capable of simultaneously optimizing discrete and continuous spaces, making it a more versatile tool in metamaterial design.

\section{Conclusions and Outlook\label{sec:conclusion_outlook}}

This work can be summarized as a perspective on quantum computing in the field of materials science. Rather than an exhaustive review of the existing literature, this paper aims at pointing out the challenges for a quantum-centric supercomputing architecture to achieve quantum advantage for materials science use cases. For that, we have presented a comprehensive view of topics at different level of abstraction, ranging from hybrid quantum-classical architectural problems to the the final use cases. We believe this is the major contribution of this paper.

The different sections have highlighted the challenges at various levels. The first two sections closely tie together, listing fundamental algorithms for simulation of quantum systems and specific hurdles that arise in implementing them at scale both for noisy quantum devices and fault-tolerant architectures, with an emphasis on the classical workloads. Then, we moved our focus closer to the architecture design arguing in details about the difficulties that are faced when optimizing quantum and classical workloads coming from hybrid algorithms. We discussed more general topics around quantum-centric supercomputing, namely, workload management and programming models, and how extensions to existing classical computing frameworks can help to address quantum computing device requirements. Before tying up all the threads in the section about use cases, we have discussed how classical simulations of quantum systems can be used both for approximate verification and to identify hard use cases. Finally, we have provided criteria for the identification of good use cases in materials science, drawing upon the previous sections. Based on these criteria, we have discussed a few exemplary cases, representative of the variety of applications of materials science in mind.   

One major takeaway from this perspective is to suggest how we can lay the grounds to think about quantum computing together with classical high-performance computing, in quantum-centric supercomputing centers. Materials science provides a great setting for use cases, which have potential for quantum advantage in scientific and industrial applications. 

\begin{acknowledgments}

Y.A. acknowledges support from the U.S. Department of Energy, Office of Science, under contract DE-AC02-06CH11357 at Argonne National Laboratory.
A.D.M., M.G, and S.V.\ are supported by CERN through the CERN Quantum Technology Initiative (CERN QTI). A.F.I. acknowledges financial support from the Natural Sciences 
and Engineering Council of Canada (NSERC).
The work at the DIPC was funded by the Gipuzkoa Provincial Council (project QUAN-000021-01), the European Union (project NextGenerationEU/PRTR-C17.I1), as well as by the IKUR Strategy under the collaboration agreement between Ikerbasque Foundation and DIPC on behalf of the Department of Education of the Basque Government.
Y.A., G.G., L.G., A.M. M.H. and B.H. acknowledge funding support from the Next Generation Quantum Science and Engineering (Q-NEXT), supported by the U.S. Department of Energy, Office of Science, National Quantum Information Science Research Centers. A.L. and T.S.H. acknowledge support from the U.S. Department of Energy, Office of Science, National Quantum Information Science Research Centers, Quantum Science Center (QSC). W.A.d.J. acknowledges funding support from the Quantum Systems Accelerator (QSA), supported by the U.S. Department of Energy, Office of Science, National Quantum Information Science Research Centers.  N.M.T. acknowledges support from by U.S. Department of Energy, Office of Science,
National Quantum Information Science Research Centers, Co-Design Center for Quantum Advantage under
Contract No. DE-SC0012704 (C2QA).
L.G., A.M. and M.H. acknowledge partial support from the NSF QuBBE Quantum Leap Challenge Institute (Grant No. NSF OMA-2121044). 
The work at the Oak Ridge National Laboratory used resources of the Oak Ridge Leadership Computing Facility which is supported by the Office of Science of the U.S. Department of Energy under Contract No. DE-AC05-00OR22725.
\end{acknowledgments}

\setglossarystyle{list}
\pagestyle{plain}
\printglossary[type=\acronymtype]
\bibliography{bibliography}

\end{document}